\documentclass[%
 reprint,
superscriptaddress,
longbibliography,
showkeys,
 amsmath,amssymb,
 aps,
pre, 
]{revtex4-2}

\usepackage{graphicx}  
\usepackage{dcolumn}   
\usepackage{bm}        
\usepackage{amssymb}   
\usepackage{amsmath}   
\usepackage{mathtools} 
\usepackage{xcolor}
\usepackage{ulem}
\usepackage[colorlinks,linkcolor=blue,urlcolor=blue,citecolor=blue]{hyperref}
\usepackage{algorithm}
\usepackage{algpseudocodex}

\usepackage{enumitem}
     
\usepackage{dcolumn}
\usepackage{bm}

\usepackage{hyperref}
\hypersetup{
    colorlinks=true,
    linkcolor=blue,
    filecolor=magenta,      
    urlcolor=cyan,
    citecolor = purple,
    pdftitle={Overleaf Example},
    pdfpagemode=FullScreen,
    }
\usepackage{array}
\usepackage{float}

\newcommand{\ee}{\end{equation}} 
\newcommand{\be}{\begin{equation}}

\usepackage[mathscr]{euscript}  
\usepackage{xcolor}  

\usepackage{comment}
\usepackage{float}
\usepackage{xr}
\makeatletter

\def\rmd{{\rm d}}

\def\rmd{{\rm d}}

\begin{document} 

\preprint{APS/123-QED}

\title{Entropy Production Rate in Stochastically Time-evolving Asymmetric Networks}

\author{Tuan Minh Pham}
\email[Contact author: ]{m.t.pham@uva.nl} 
\thanks{Equal contributions}
\affiliation{Dutch Institute for Emergent Phenomena, 1090 GE, Amsterdam, The Netherlands}
\affiliation{Complexity Science Hub Vienna, Metternichgasse 8, A-1030, Vienna, Austria}
\affiliation{Institute for Advanced Study,
Oude Turfmarkt 147, 1012 GC Amsterdam, The Netherlands}
\affiliation{Institute of Physics, University of Amsterdam, Science Park 904, Amsterdam, The Netherlands}
\author{Deepak Gupta}
\email[Contact author: ]{phydeepak.gupta@gmail.com} 
\thanks{Equal contributions}
\affiliation{Institute of Physics, Polish Academy of Sciences, Al. Lotników 32/46, 02-668,
Warsaw, Poland}
\affiliation{Institut für Physik und Astronomie, Technische Universität Berlin, Hardenbergstraße 36, D-10623 Berlin, Germany}

\date{\today}

\begin{abstract}
\textcolor{black}{Networks that characterise the interactions between units composing complex
systems are typically treated as fixed. Yet, such networks often
stochastically evolve over time, shaping} 
the collective behavior of complex systems.
To date, we lack a general non-equilibrium thermodynamic treatment of such  time-dependent networks. 
In this Letter, to address this problem, we  model fluctuating interactions between units of nonlinear network systems  as uncorrelated colored noise (i.e., annealed disorder) with a correlation time. This approach enables us to quantify how the entropy production rate (EPR) depends on both the  time-scale  and the strength of the disorder. Using {\it dynamical mean field theory}, we derive an exact expression for EPR  at {\it any} transient time that is validated by simulations of the full  dynamics and establish a  relation between EPR and autocorrelation at stationarity. 
{\color{black}We find that annealed disorder shifts the transition from the fixed-point to chaos toward a higher variance of the interactions, thereby suppressing chaos but at the cost of greater  dissipation.}
\end{abstract}

\maketitle

Finding  effective representations 
of large-scale complex systems is \textcolor{black}{one of the central topics in complexity science} 
because their direct simulations 
are computationally very expensive. This \textcolor{black}{has led} 
to the resurgence of interest in {\it Dynamical Mean Field Theory} (DMFT)--a powerful method offering  low-dimensional effective descriptions for  systems \textcolor{black}{of many statistically equivalent agents},  with applications ranging from {\color{black}neural~\cite{Sompolinsky, Kadmon2015, Keup, Pereira-Obilinovic, Clark2025}}
and ecological systems \cite{De_Giuli, Bunin, Altieri2021} to evolutionary \cite{pham2024dynamical} and social dynamics \cite{Baron2021, Stamatescu, Garnier}, and to machine learning \cite{Cakmak2020, Bordelon_2023, Gerbelot}. See
~\cite{Buice_2013, Hertz2017, Crisanti2018, Marti2018, helias2020statistical, galla2024generatingfunctionalanalysisrandomlotkavolterra, huang2022statistical, Cugliandolo2024, Zuniga-Galindo_2025} for detailed expositions  of DMFT.

In addition, such complex systems are typically out of equilibrium because they require a constant energy input for their operation~\cite{Lucente2025}. Within the framework of stochastic thermodynamics,  how far these systems are from equilibrium 
can be quantified by the entropy production rate (EPR)~\cite{VANDENBROECK2015, Seifert_2012}. 
Therefore, recent research 
has been devoted to quantifying EPR in neural \cite{Kaluarachchi, Lynn2021} and living \cite{Battle, Fang2019, Yang} systems. 
{\color{black}Apart from being a measure of dissipation, EPR also plays a key role in generative diffusion models~\cite{SosukePRX, Whitelam}. Moreover, EPR correlates with consciousness levels in  human brains~\cite{Perl,Gilson} and reflects the cognitive load demanded by tasks~\cite{Lynn2021}.}

Recent studies show a promising direction \textcolor{black}{of using 
DMFT to quantify  EPR in large-scale  systems with assumed random interactions between units~\cite{pham2024irreversibilitynonreciprocalchaoticsystems, Aguilera2023, rooke2026}.} While these studies have focused on out-of-equilibrium processes with {\color{black}{\it quenched} disorder}~\footnote{Quenched disorder implies that some relevant random parameters (drawn from some distribution at the initial time) do not evolve in time.}, there are many situations where the underlying network structures temporally vary. These include synaptic plasticity in the brain \cite{Abbott},  temporal socio-economic networks~\cite{holme2019temporal}, the adaptation of gene-regulatory networks~\cite{KanekoPloSOne2007}, \textcolor{black}{interaction-mediated fluidization in dense biological systems~\cite{Sezik},} and stochastically evolving ecological networks~\cite{Suweis2024, davide-eco}. 
 This setting is referred to as an {\it annealed} disorder according to  the spin-glass nomenclature~\cite{Spin-glass}. 
\textcolor{black}{Unlike the case of quenched disorder, it remains 
an open question of how  annealed disorder affects the dynamics and  nonequilibrium thermodynamics of non-linear networked systems.  Motivated by this  gap in the literature,} 
in this Letter, we use DMFT to calculate EPR in both transient dynamics and non-equilibrium steady state (NESS) for such systems. 
Here, we model \textcolor{black}{the network couplings as uncoupled Ornstein-Uhlenbeck processes}~\cite{haunggi1994colored}. 
In active matter systems, the Ornstein-Uhlenbeck process appears as a stochastic driving force  (active noise) that can lead to an improvement in system performance compared to thermal  noise~\cite{behera2022}.

{\color{black}To illustrate our approach, we consider $N$ interacting units (neurons), labeled $x_i$ for $1\leq i\leq N$, with interactions governed by a general nonlinear function $F[\cdot]$. Then, the dynamics of the system are given as follows}
\begin{align}
\dot {x}_i(t)= -x_i(t) + F\left[\sum_{j=1,j\neq i}^N J_{i,j}(t) x_j(t)\right] + \zeta_i(t) \ , \label{eq:eqn1}
 \end{align}
where the dot denotes a time-derivative and $\zeta_i(t)$ is a Gaussian white noise with zero mean and delta-correlation in time $\langle \zeta_i(t)\zeta_j(t')\rangle = \sigma^2\delta_{ij}\delta(t-t')$. Here, $\sigma$ is the parameter that controls the strength of the noise. In Eq.~\eqref{eq:eqn1}, $J_{i,j}(t)$ is the time-dependent coupling term that describes the interaction of the (pre-synaptic) $j$th unit/neuron with the (post-synaptic) $i$th unit/neuron [we {\color{black}exclude} self-loops, i.e., $J_{i,i}(t) =0$]. Beyond this specific application to neural networks, where it is referred to as the firing-rate \textcolor{black}{(activity-based) model~\cite{Miller2012, Bressloff2010}} and corresponds to the so-called nonlinear integration scheme of information processing~\cite{Nicoletti2005}, the model~\eqref{eq:eqn1} has also been widely used in modeling cellular systems~\cite{Cellbox, Zecchina2013, Paczko}.

\textcolor{black}{In this Letter, we extend previous studies~\cite{pham2024irreversibilitynonreciprocalchaoticsystems,martorell2025ergodicitybreakinghighdimensionalchaos, martorell2023dynamically} by considering the interaction matrix $\mathbf{J}(t)$ as time dependent, where} each $J_{i,j}(t)$ stochastically evolves as a function of time. Specifically, we assume that the annealed interaction $J_{i,j}(t)$ has mean $\mu/N$ and variance $g^2/N$, so that  $J_{i,j}(t)\equiv \mu/N + g/\sqrt{N}Z_{i,j}(t)$, where   $Z_{i,j}(t)$ is an active noise obeying the Ornstein-Uhlenbeck process: 
 $\dot Z_{i,j}= -Z_{i,j}/\tau_0 + \sqrt{1+2\tau_0}/\tau_0~\xi_{i,j}(t)$~\cite{Suweis2024, davide-eco, Gupta_2025}. Here $\xi_{i,j}(t)$ is
 Gaussian white noise  with zero mean and delta-correlation in time $\langle \xi_{i,j}(t)\xi_{k,\ell}(t')\rangle = \delta_{ik}\delta_{j\ell}\delta(t-t')$. \textcolor{black}{Since in our model $\mathbf{J}(t)$ is treated as an externally prescribed stochastic driving protocol rather than what generated from ``environmental variability'', it has different origin from the thermal fluctuation $\boldsymbol{\zeta}(t)$}. Therefore, the  white noises $\zeta_a(t)$ and $\xi_{b,c}(t)$ can be \textcolor{black} {assumed to be}  uncorrelated for all time $t,t'$: $\langle \zeta_a(t)\xi_{b,c}(t')\rangle = 0$ for all $a,b,c$. 
 {\color{black}In what follows, we take $\sigma^2 = 2 T$, where $T$ is the temperature of the environment (Boltzmann's constant $k_{\rm B}=1$). }
 
In the long-time limit, the process $Z_{i,j}(t)$ approaches a stationary Gaussian distribution {\color{black}with} zero mean and two-time correlation:
    $\langle Z_{i,j}(t)Z_{i',j'}(t')\rangle = \delta_{i,i'}\delta_{j,j'} q_{\tau_0}(|t-t'|)$,
where $q_{\tau_0}(|\tau|)\equiv (1+2\tau_0)/(2\tau_0)e^{-|\tau|/\tau_0}$ is the memory kernel and $\tau_0$ is the correlation-time parameter. {\color{black}In the limiting cases $\tau_0\to0$ and $\tau_0\to\infty$, the memory kernel $q_{\tau_0}(|\tau|)$ approaches $\delta(\tau)$ and $1$, respectively. Accordingly, $J_{i,j}(t)$, through $Z_{i,j}(t)$, reduces to Gaussian white noise as $\tau_0\to0$, whereas it approaches Gaussian quenched noise as $\tau_0\to\infty$. By tuning the correlation time $\tau_0$, we can interpolate between the quenched case, in which $J_{i,j}$ is drawn once from a Gaussian distribution and then held fixed in time, and the annealed case, in which $J_{i,j}$ evolves in time, starting from $Z_{i,j}(0)\sim\mathcal{N}\big(0,q_{\tau_0}(0)\big)$.

We stress that the current form of the correlation-time kernel $q_{\tau_0}$  is just a convenient choice to interpolate between the quenched and annealed disorders.
The following theoretical development also applies to other forms of $q_{\tau_0}$ as long as $Z_{i,j}$ and $Z_{j,i}$ are uncorrelated. Furthermore, because of the time-dependent nature of $\mathbf{J}(t)$, our model~\eqref{eq:eqn1} is not equivalent to an \textit{annealed} extension of the classical model by Sompolinsky et al.~\cite{Sompolinsky} that would be obtained by moving the sum inside the nonlinear function $F(\cdot)$ outside (see Sec.~\ref{Inequivalence}~\cite{SM} for a detailed discussion on this point). Nevertheless, our methodology straightforwardly applies to the latter model \footnote{Detailed calculations of the EPR for the extension of the Sompolinsky et al.~\cite{Sompolinsky} model with annealed disorder will be published elsewhere}.}


Apart from being in contact with a thermal bath at temperature $T$, the system~\eqref{eq:eqn1}  is simultaneously subjected to stochastic driving via the {\it  nonreciprocal}  coupling matrix ${\bf J}$, i.e., $J_{ij}(t)\neq J_{ji}(t)$ for every pair $(i,j)$.  Consequently, the system breaks the detailed balance condition and produces entropy even in NESS~\cite{EP-NRI, Kreienkamp_2026}. 
Our setting hence contrasts with previous work on partially-annealed couplings in equilibrium spin-dynamics with reciprocal couplings \cite{Pham2022,Penney1993, Allahverdyan1998, Poderoso2007, Uezu2009}.
{\color{black} This setting shall allow us to show how EPR could serve as a characteristic signature to distinguish different dynamical regimes of the model~\eqref{eq:eqn1}.}

The total entropy production rate along a single stochastic trajectory is decomposed as $\dot{S}_{\rm tot} = \dot{S}_{\rm sys}+\dot{S}_{\rm res}$, where $\dot{S}_{\rm sys}(t)$ and $\dot{S}_{\rm res}(t)$ are the rates of change of the system entropy and the environmental entropy, respectively. These are, respectively, defined as~\cite{Seifert_2012}
\begin{subequations}
   \begin{align}
     \dot{S}_{\rm sys}&\equiv  -\sum_i \partial_{x_i}\ln p(\mathbf{x},t) \circ \dot x_i(t) - \partial_t\ln p(\mathbf{x},t)\ , \\  
     \dot{S}_{\rm res}&\equiv \ T^{-1}\sum_{i} \bigg[F\bigg(\sum_{j\neq i}J_{i,j}(t) x_j(t)\bigg) -x_i(t) \bigg]\circ \dot x_i(t)\ , \label{reservoir_entropy}
\end{align} 
\end{subequations}
where the symbol ``$\circ$" denotes the Stratonovich convention~\cite{vanKampen1981},  and the probability density function  $p(\mathbf{x},t)$ is  the time-dependent solution of the Fokker-Planck equation corresponding to Eq.~\eqref{eq:eqn1} for $\mathbf{x}\equiv (x_1, x_2, \dots x_N)$ at time $t$.
The average of the system entropy production rate vanishes in NESS, i.e., $\langle \dot{S}_{\rm sys}\rangle_{\rm NESS} = 0$~\cite{Seifert_2012}. Thus, the average rate of total entropy production becomes equal to the average rate of environmental entropy production, i.e., $\big\langle \dot{S}_{\rm tot}\big\rangle_{\rm NESS} =\big\langle \dot{S}_{\rm res}\big\rangle_{\rm NESS}$, where the  brackets $\langle\cdot\rangle_{\rm NESS}$ represent the average over noise realizations in the NESS. In the following, we hence focus only on the EPR of the environment given by Eq.~\eqref{reservoir_entropy} 
\footnote{Notice that if one considers $\mathbf{\xi}_{i,j}(t)$ in the dynamics of $Z_{i,j}$, as another thermal noise, the averaged dissipation due to the dynamics of $Z_{i,j}$ via this channel is zero since $\mathbf{Z}(t)$ is an equilibrium OU process.}.


{\color{black}On the one hand, finite-$N$ simulations of the full dynamics~\eqref{eq:eqn1} may not fully capture the behaviors that arise in the infinite-size limit. On the other hand, increasing $N$ can alleviate these finite-size effects, but at the expense of rapidly increasing computational costs, making simulations challenging for $N\gg1$. Therefore, an alternative approach is desirable for understanding the dynamics and thermodynamics of system~\eqref{eq:eqn1} in the limit $N\to\infty$.} {\color{black} 
To this end, we derive the following DMFT equation for the effective process associated with Eq.~\eqref{eq:eqn1}, in which the coupling $J_{i,j}(t)$ are stationary OU process  with correlation-time $\tau_0$, using the moment generating functional approach similar to \cite{Suweis2024} (see Sec.~\ref{app-eff-eqn}~\cite{SM} for its detailed derivation)}
\begin{align}
    \dot x(t) = -x(t) +F[\mu M(t) + g\eta(t)] + \zeta(t)\ ,\label{effective}
\end{align}
where $x(t)$ is a representative unit of an infinite population, in which all units with
statistically similar connections and dynamical properties become identical after taking the ensemble
average \cite{Touboul2012}. {\color{black} In Eq.~\eqref{effective}, $\zeta(t)$ is again zero mean Gaussian thermal white noise with delta correlation in time;}  $M(t)\equiv \langle x(t) \rangle^{\rm (MF)}$; and $\eta(t)$ Gaussian noise with zero mean and two-time correlation \begin{align}
C_\eta(t,t')\equiv \langle \eta(t)\eta(t')\rangle^{\rm (MF)} = q_{\tau_0}(|t-t'|) C_x(t,t')\ , \nonumber
\end{align}
where $C_x(t,t')\equiv \langle x(t)x(t') \rangle^{\rm (MF)}$ is the  positional autocorrelation.  Here, the angular brackets with superscript ``MF'' represent the average performed over realizations of the effective (mean-field) dynamics~\eqref{effective}. {\color{black}Thus, in Eq.~\eqref{effective}, the mean $M(t)$ and the noise correlation $C_\eta(t,t')$ are to be solved in a self-consistent manner}~\footnote{Note that in NESS with time-translational invariance $C_x(t,t') =C_x(t-t')\rightarrow Q$ as $t\rightarrow t'^+$,  we could also formulate the above dynamics using a Fokker-Planck equation for the joint distribution of both $x$ and $\eta$~\cite{Wio1989}.}.

Since the original dynamics~\eqref{eq:eqn1} can be  \textit{exactly} represented by the effective dynamics~\eqref{effective} as $N\rightarrow \infty$, the average rate of environmental entropy production~\eqref{reservoir_entropy} can be  shown to be equivalent to that  of this effective process. Specifically, denoting $\dot{s}_{\rm res}(t):= \dot{S}_{\rm res}(t)/N$ as the average rate of environmental entropy production per unit of the full dynamics~\eqref{eq:eqn1} and $\langle \dot{s}_{\rm res}(t) \rangle^{\rm (MF)}$ as that of the effective dynamics~\eqref{effective}, we show that as $N\rightarrow \infty$ (Sec.~\ref{app:EPR}~\cite{SM})
\begin{align}
\langle \dot{s}_{\rm res}(t) \rangle =\langle \dot{s}_{\rm res}(t) \rangle^{\rm (MF)}\ .
\end{align} 
The average rate of entropy production at time $t$ can be further expressed in terms of correlations (Sec.~\ref{app:EPR}~\cite{SM})
\begin{align}
\langle \dot{s}_{\rm res}(t) \rangle^{\rm (MF)}=1/T[C_x(t,t) +C_{F}(t,t) - 2C_{xF}(t,t)] -1 \ ,
\label{EPR}
\end{align}
for the equal-time correlation $C_{AB}(t,t)\equiv \langle A(t)B(t) \rangle^{\rm(MF)}$ evaluated using Eq.~\eqref{effective}. This is our first main result. In  the $\tau_0\rightarrow 0$ limit for the linear model $F[z]=z$, an effective temperature $T_{\rm eff}\neq T$  can be defined such that the dynamics become equilibrium, and hence there is no dissipation \cite{Caprini_2019, Martin2021}.
However, following~\cite{Verley_2014,Semeraro_2021,SS-1,ss-2,Gomez-Solano_2010}, here we attribute dissipation only to the thermal channel, as long as the active colored noise is considered a stochastic driving that cannot be combined with the thermal noise. 

{\color{black}We compare the full description~\eqref{eq:eqn1} and the DMFT~\eqref{effective} in Fig.~\ref{TD-comp-FD-and-DMFT}~\cite{SM} and it shows excellent agreement between them. Further, Fig.~\ref{TD-comp-FD-and-DMFT} validates our Eq.~\eqref{EPR}.
Additionally, Figs.~\ref{fig:DMFT-SP-tau1}, \ref{fig:DMFT-SP-tau10}, \ref{fig:DMFT-SP-tau100} show their comparison 
at long-time. Therefore, in what follows, we shall use only the DMFT description to discuss the results.}

  \begin{figure}
    \centering
    \includegraphics[width=\linewidth]{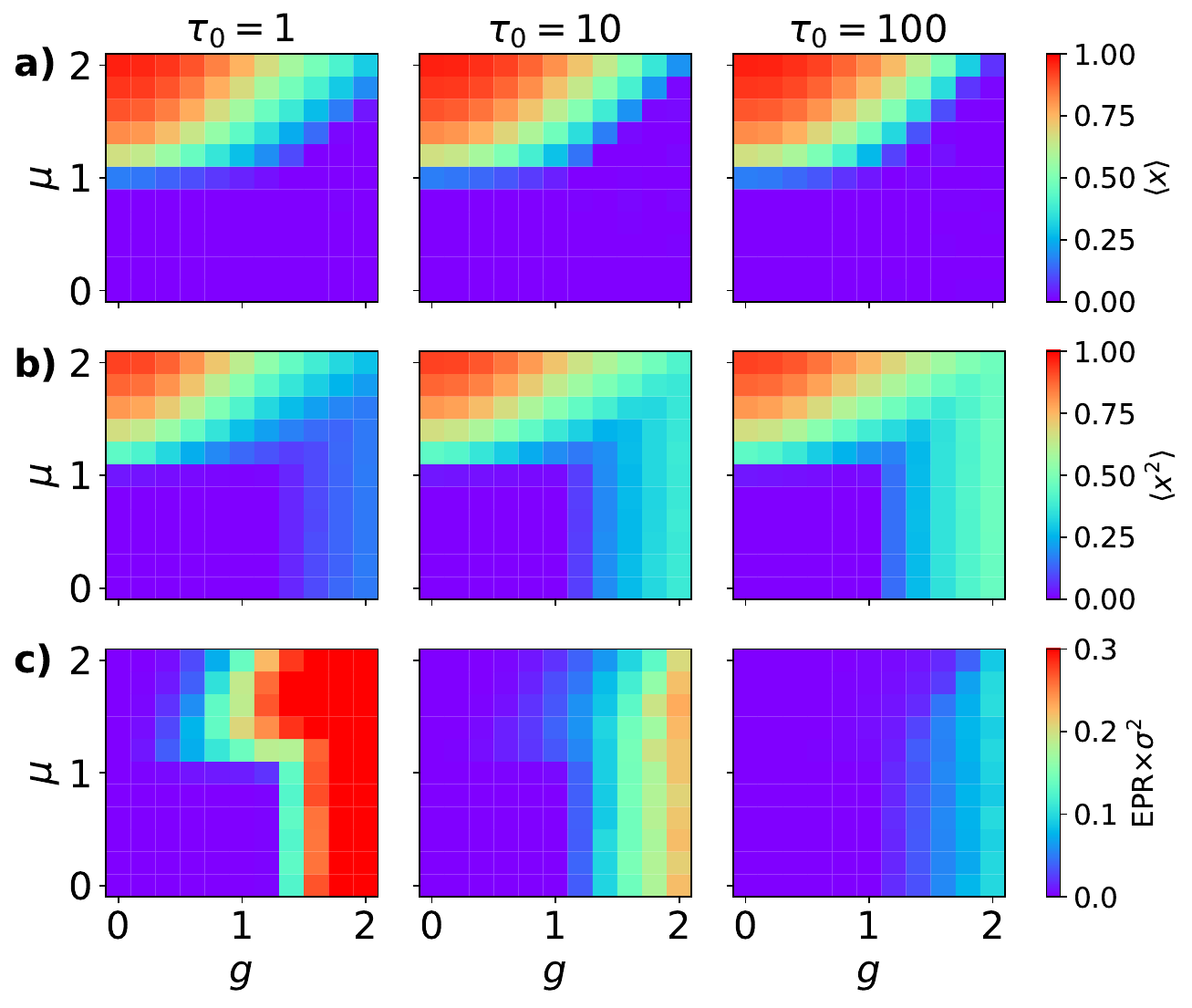}
    \caption{Long-time behavior of the mean $\langle x \rangle$ (a), second moment $\langle x^2\rangle$ (b), and entropy production rate (EPR) (c) in the parameter space $(g,\mu)$ obtained from the dynamical mean-field theory (DMFT) Eq.~\eqref{effective}. Comparison of DMFT results with results obtained from the full dynamics~\eqref{eq:eqn1} is shown in Figs.~\ref{TD-comp-FD-and-DMFT},~\ref{fig:DMFT-SP-tau1}, \ref{fig:DMFT-SP-tau10}, and \ref{fig:DMFT-SP-tau100}. Here and in what follows, unless specified, we focus on $F(z) = \tanh(z)$, a standard  form of activation function in modeling neural networks. Parameters: noise strength $\sigma = 0.01$, time step $dt = 10^{-2}$, total simulation time $t=50$, DMFT iterations 1000, and each point is averaged over 2000 number of realizations.}
    \label{fig:DMFT-PDF}
\end{figure}

Let us first focus on the long-time behavior of the system to characterize its different dynamical regimes.
To this end, we plot the mean and second moment, $M\equiv \langle x\rangle^{\rm (MF)}$ and $Q\equiv \langle x^2\rangle^{\rm (MF)}$, both obtained using DMFT Eq.~\eqref{effective} in the $(g,\mu)$ plane, respectively, 
in Figs.~\ref{fig:DMFT-PDF}a and b. Here, we find (i) two quasi-static solutions: paramagnetic 
$(M=0, Q=0)$ and ferromagnetic or persistent
activity  
$(M>0,Q>0)$; and (ii) two chaotic regions. {\color{black}Both of these behaviors will be explained later.

Figure~\ref{fig:DMFT-PDF}c displays the EPR. At fixed $\mu$, EPR increases with  $g$ for all  $\tau_0$, but decreases with $\tau_0$ for fixed $(\mu,g)$. The latter behavior  can be intuitively understood as follows: time-dependent interactions give rise to higher activity compared to the quenched ``frozen'' ones (i.e., higher the annealed disorder, more active the system is), resulting in an increase of EPR as $\tau_0$ decreases \footnote{Although  we do not have a rigorous proof to support this intuitive picture, it has been shown in a much simpler setting, a 2D OU model with fluctuating parameters, that the faster the fluctuations of the parameters, the higher the entropy production rate~\cite{Alston_2022}.}.

\begin{figure}
    \centering
    \includegraphics[width = \linewidth]{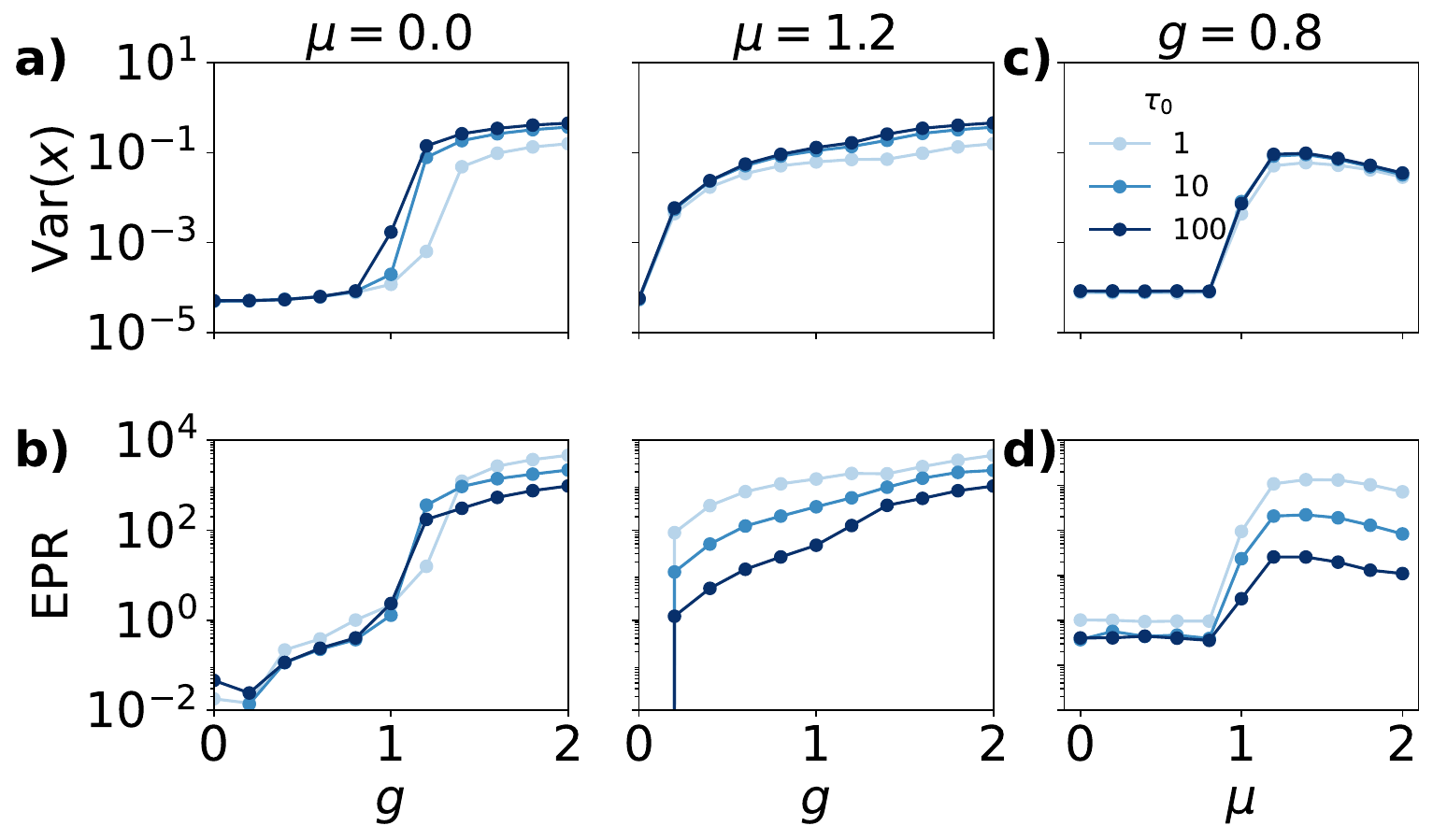}
    \caption{Var$(x)\equiv \langle x^2\rangle -\langle x\rangle^2$ and entropy production rate (EPR). a) and b): Var$(x)$ and EPR, each as a function of $g$ for $\mu = 0, 1.2$. c) and d) Var$(x)$ and EPR as a function of $\mu$ for $g=0.8$. The connecting lines are guides to the eye. }
    \label{fig:EPR-var}
\end{figure}

Figure~\ref{fig:EPR-var} shows how the variance of $x$ [Var$(x)$] and EPR behave as functions of $g$ and $\mu$. In Fig.~\ref{fig:EPR-var}a and b, at small $\mu$ ($\mu=0$), Var$(x)$ and EPR both start increasing from negligible to a finite value beyond a critical $g_c$, while at large  $\mu=1.2$, the variance smoothly increases without undergoing any sharp transition. (See Fig.~\ref{fig:compare1} for more parameter combinations.)  In particular, we find that stronger annealed disorder reduces the variance, pushing its change to occur at a much higher $g_c$, above which the curves for different $\tau_0$ become distinguishable. 

We further observe a non-monotonic dependence of Var$(x)$ and EPR on $\mu$ at $g>0$ (Fig.~\ref{fig:EPR-var}c and d). Additionally, higher $\tau_0$ enhances quenched disorder, thus reducing EPR and its non-monotonic behavior. (See Fig.~\ref{fig:compare2} for more parameter combinations.) This phenomenon of non-monotonicity may  originate from  a competition between $\mu$ and $g$, where the former promotes static order (a broken symmetry of local activity $x_i$) and the latter drives dynamic disorder (persistent fluctuations).} 

\begin{figure}
    \centering
    \includegraphics[width=0.8\linewidth]{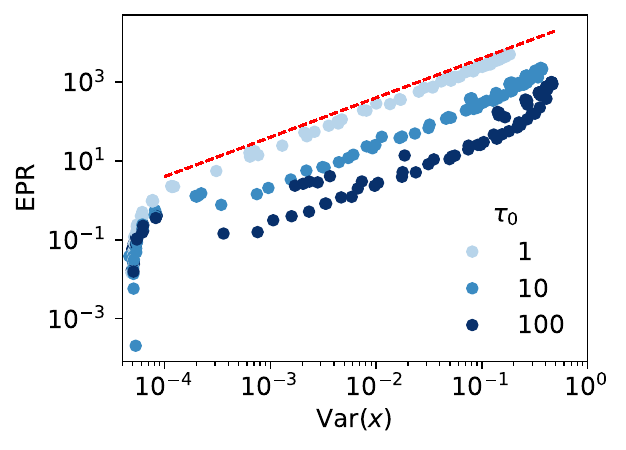}
    \caption{EPR as a function of Var$(x)$. Symbols corresponding to the same color (or $\tau_0$) are obtained from Fig.~\ref{fig:DMFT-PDF} for all combinations of {\color{black}$\theta\equiv (\mu, g, \tau_0)$}. The red straight dashed line is a guide to the eye.}
    \label{fig:epr-var-para}
\end{figure}

While Figs.~\ref{fig:DMFT-PDF} and~\ref{fig:EPR-var} give us an overview of the model's dynamic and thermodynamic properties, we want to  
extract a direct relationship between these two properties. Therefore, using the data shown in Figs.~\ref{fig:DMFT-PDF}a, b and c, we make a parametric plot for the behavior of EPR as a function of Var($x$) in Fig.~\ref{fig:epr-var-para}, where each data-point  corresponds to  a fixed set of parameters $\theta\equiv(\mu, g, \tau_0)$ and has its $(x,y)$-coordinates specified by  variance$(\theta)$ and EPR$(\theta)$. Note that the $x$-axis for the variance   ranges from $O(\sigma^2/2)$ to an $O(1)$ value.  As the system undergoes a  transition upon crossing the critical manifold $\theta_c\equiv (\mu_c, g_c)$ (Fig.~\ref{fig:EPR-var}), we find a change in the functional relationship between EPR and variance. Specifically, below this manifold, all three different $\tau_0$ curves collapse into a single one, suggesting that there is no visible difference between them. {\color{black}Thus, annealed and quenched disordered systems are indistinguishable below the critical manifold.} However, above this manifold, the exact relation between EPR and variance begins to depend on $\tau_0$. Interestingly, despite this dependence on $\tau_0$, all lines have the same slope on the log-log scale, as shown by the dashed red line in Fig.~\ref{fig:epr-var-para}. {\color{black}We note a few points that deviate from the $\tau_0=100$ ensemble, for which the DMFT dynamics may not yet have reached a stationary state at $t=50$. We expect this discrepancy to be eliminated with increased computational resources.}

The above data analysis suggests an intriguing relationship between the EPR and 
Var$(x)$ in the NESS. Therefore, \textcolor{black}{similar to what been established for quenched interactions~\cite{pham2024irreversibilitynonreciprocalchaoticsystems} and for overdamped dynamics without disorder~\cite{Terlizzi},} the general formula for the EPR~\eqref{EPR} can be expected to reduce to a simpler relation between the time-independent value of the EPR, $\dot{s}_{\rm NESS}\equiv \lim_{t\rightarrow \infty}\langle \dot{s}_{\rm res}(t) \rangle^{\rm (MF)}$, and the autocorrelation function under time-translational invariance $C_x(\tau)$, where  $C_x(\tau)\equiv C_x(t-t')$ for $\tau=t-t'\geq 0$ and $t,t'\rightarrow \infty$. 
Here, we generalize such a relation for  annealed interactions with arbitrary  $\tau_0$. To do so,  
we first derive the following equation [see~\ref{EOM_for_Cx}~\cite{SM} for more details]
\begin{equation}
\partial_\tau^2\bar{C}_x(\tau) = \bar{C}_x(\tau) - \bar{C}_F(\tau) -\sigma^2\delta(\tau)\ ,
\label{main:EOM_for_Cx}
\end{equation}
where $\bar{C}_x(t,t') \equiv C_x(t,t') - \langle x(t) \rangle \langle x(t') \rangle$; $\bar{C}_{xF}(t,t') \equiv  \big\langle \delta x(t)\delta F(t')]\big\rangle$; $\bar{C}_F(t,t') \equiv C_F(t,t) - \langle F(t) \rangle \langle F(t') \rangle$, for $\delta x(t) = x(t) - \langle x(t)\rangle^{\rm (MF)}$; and $\delta F[\cdot]\equiv F[\cdot]  - \langle F[\cdot]\rangle^{\rm (MF)}$.
Equations similar to Eq.~\eqref{main:EOM_for_Cx} have been derived for the quenched case~\cite{Cabana, Schuecker2018, martorell2023dynamically}. 
{\color{black}The physical solution of Eq.~\eqref{main:EOM_for_Cx} must obey the following boundary conditions: $\bar{C}_x(0)$ is finite, $\bar{C}_x(\tau)\leq \bar{C}_x(0)$ for $\tau>0$, $\dot{\bar{C}}_x(\tau\to 0^+)=-\sigma^2/2$~\footnote{The relation $\dot{\bar{C}}_x(\tau\to 0^+)=-\sigma^2/2$ follows by integrating both sides of Eq.~\eqref{main:EOM_for_Cx} over the interval $[-\epsilon,\epsilon]$ and taking the limit $\epsilon\to 0$. To derive this result, we use the continuity of $\bar{C}_x(\tau)$ and $\bar{C}_F(\tau)$ in the limit $\tau\to 0$, together with the evenness property of the correlation function $\bar{C}_x(\tau)=\bar{C}_x(-\tau)$, which implies $\dot \bar{C}_x(0^+)=-\dot \bar{C}_x(0^-)$ in the limit $\tau\to 0$.}, and $\bar{C}_x(\tau\rightarrow \infty)\rightarrow 0$.}
Furthermore, as derived in Sec.~\ref{EOM_for_Cx}~\cite{SM}, we have
\begin{equation}
    \partial_\tau \bar{C}_x(\tau) = -\bar{C}_x(\tau) + \bar{C}_{xF}(\tau)\ ,
    \label{EOM_for_Cx2}
\end{equation}
Multiplying 2 on both sides of Eq.~\eqref{EOM_for_Cx2}, and then adding it to Eq.~\eqref{main:EOM_for_Cx}, we arrive at 
\begin{equation}
    \ddot{\bar{C}}_{x}(\tau) +  2\dot{\bar{C}}_x(\tau) = -\bar{C}_x(\tau) - \bar{C}_F(\tau) + 2\bar{C}_{xF}(\tau)\ ,
\end{equation}
where a dot indicates a time-derivative with respect to $\tau$.
Applying the above to the NESS restriction of Eq.~\eqref{EPR} which works for any arbitrary time, we obtain the following exact relation connecting the EPR with the correlation function in the stationary state
\begin{equation}
  \dot{s}_{\rm NESS}=-\ddot{\bar{C}}_x(0^+)/T +1\ .
  \label{NESS_entropy}
\end{equation}
Here, we remark that special care needs to be taken in using this formula in the limit $\tau_0\rightarrow 0$. Specifically, according to our physical picture, where we  attribute dissipation only to the thermal bath $\zeta(t)$~\cite{Verley_2014,Semeraro_2021,SS-1,ss-2,Gomez-Solano_2010}, we  need to compute the EPR for finite $\tau_0$ 
and then take $\tau_0\rightarrow 0$ afterwards. {\color{black}A direct consequence of Eq.~\eqref{NESS_entropy} is that the EPR can  be analytically calculated once $\bar{C}_x(\tau)$ is known.
We illustrate this point in Fig.~\ref{fig:Bessel_solution} (Sec.~\ref{sec:linear-entropy}), using an  important application of model~\eqref{eq:eqn1} when the underlying force is linear, i.e., $F(z)=z$~\cite{Morales}. There, we show that the EPR for this case diverges in the limit  $\tau_0\rightarrow 0$ due to the average $\langle \eta(t)^2\rangle\sim\delta(0)$ appearing in the EPR calculations for the effective white noise $\eta(t)$. We confirm this divergence by direct integration of the DMFT Eq.~\eqref{effective} in Fig.~\ref{fig:EPRdt}. Physically, this means that the housekeeping cost of maintaining an infinitely fast stochastic protocol diverges, despite the long-time linear dynamics of 
$x$ converging to a well-defined stationary distribution. A similar divergence has been observed for the stationary heat flow rate in \cite{Gupta_2025} for a harmonic oscillator with fluctuating stiffness modeled by white noise. }

{\color{black}
We are now in a position to address the physics underlying the results shown in Fig.~\ref{fig:DMFT-PDF}a and b. We first discuss the stability of the fixed point $M_*=0$. As discussed in Sec.~\ref{sec:fixed-point-m-0}, the condition for its instability is
\begin{align}
\mu\left\langle F'\left[z~g\sqrt{q_{\tau_0}(0)~\bar{C}_x(0)}\right] \right\rangle = 1 \ ,
\label{eq:m-eq0stab}
\end{align}
where $\bar{C}_x(0)$ is the solution of Eq.~\eqref{main:EOM_for_Cx}, and the angular brackets denote an average over a Gaussian random variable $z$ with zero mean and unit variance. Figure~\ref{fig:stability-plot} shows the dot-dashed contour given by Eq.~\eqref{eq:m-eq0stab}. This contour,  which separates the regions in which the fixed point $M_*=0$ is stable and unstable, qualitatively explains Fig.~\ref{fig:DMFT-PDF}a.

Next, to understand Fig.~\ref{fig:DMFT-PDF}b, we need to establish a transition from the quasi-static regime to a chaotic one.
Here, we only sketch the key steps of calculating the largest Lyapunov exponent $\lambda_{\rm max}$ through the temporal evolution of 
the distance $d(t)$ between two stochastic trajectories, referred to as two replicas (see Sec.~\ref{lyapunov_analysis-main} for detailed analysis). We generalize the approach in Ref.~\cite{Schuecker2018} to annealed disorder to obtain the distance $d(t)$. This distance can be expressed in terms of  the cross-correlations  between the two replicas, $\bar{C}^{12}_x(t,s)$, and their respective autocorrelation $\bar{C}^{11}_x(t,s),  \bar{C}^{22}_x(t,s)$, where $\bar{C}^{\alpha\beta}_x(t,t') \equiv   \left \langle \delta x_\alpha(t)\delta x_\beta(t')\right\rangle$ for $\alpha, \beta \in\{1,2\}$. In NESS, under time-translational invariance,
$\bar{C}^{11}_x(t,s) =  \bar{C}^{22}_x(t,s) = \bar{C}_x(t-s)$ for $t,s\rightarrow \infty$, where $\bar{C}_x(t-s)\equiv \bar{C}_x(t,s)$ denotes the  solution of Eq.~\eqref{main:EOM_for_Cx}.
\color{black}A small perturbation around this solution of the form $\bar{C}^{12}_x(t,s) = \bar{C}_x(t-s) + \epsilon k^{(1)}(t,s) $, for $ \epsilon \ll 1$, allows us to derive the distance $d(t) = -2\epsilon k^{(1)}(t,t)$ from the growth rate $k^{(1)}(t,t)$. 
Using the ansatz $k^{(1)}(t,t')= e^{\lambda (t+t')}\psi(\tau)$ for  $\tau = t- t'$,  we arrive at the following Schrödinger equation for $\psi(\tau)$ that generalizes the one in~\cite{Schuecker2018} for the annealed case: 
\begin{align}
    [1 -\partial_\tau^2  - g^2q_{\tau_0}(\tau)K(\tau)]\psi(\tau) = \left[1- (\lambda +1)^2\right]\psi(\tau)\ ,
    \label{Schrodinger_main}
\end{align}
where $K(\tau) \equiv \langle F'(0)F'(\tau) \rangle$.
The Lyapunov exponent $\lambda_{\rm max}$ is then computed from the asymptotic expression
$k^{(1)}(t,t) \propto e^{\lambda_{\rm max} t}$  as $t\rightarrow \infty$ {\color{black}(see numerical method to solve Eq.~\eqref{Schrodinger_main} and to compute $\lambda_{\rm max}$ in  Sec.~\ref{sec:algorithm}).}

\begin{figure}
    \centering
    \includegraphics[width = \linewidth]{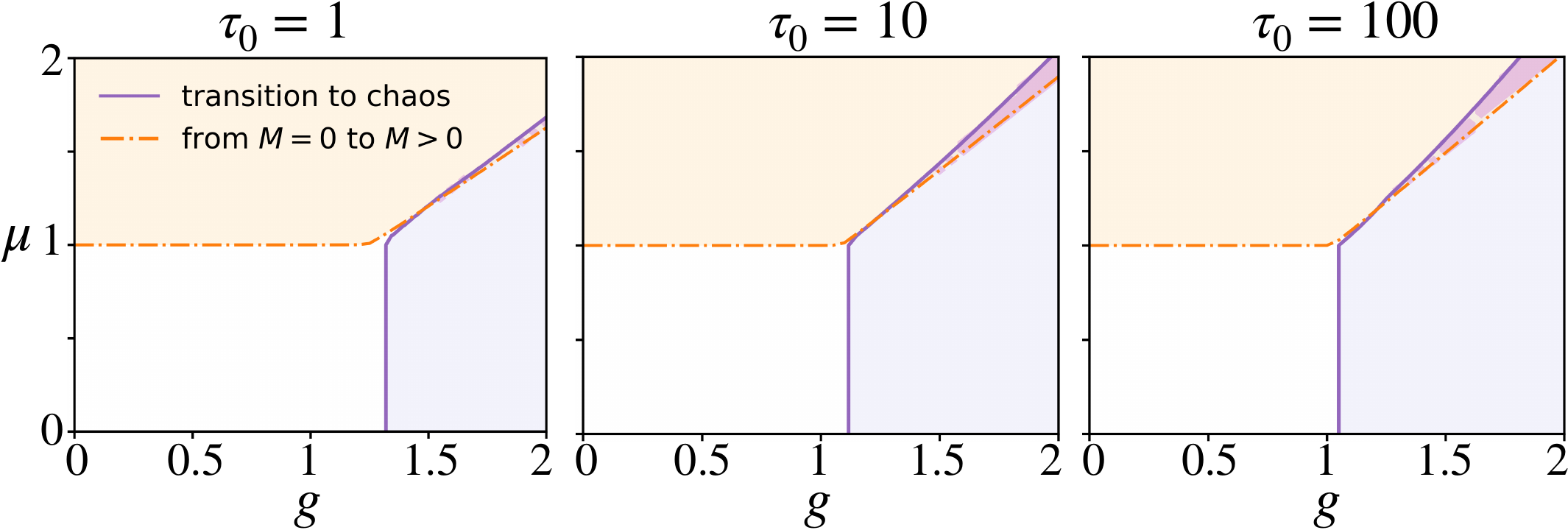}
    \caption{Effect of annealed disorder.
    (i) Transition (bottom to top) from $M=0$ to $M>0$ [dot-dashed contour, Eq.~\eqref{eq:m-eq0stab}]. (ii) Transition (left-to-right) from quasi-static regions to chaotic ones (solid contour, largest Lyapunov exponent $\lambda_{\rm max}=0$). The overlap area between the orange dot-dashed line and purple one is so-called synchronous chaos regime. Here $\sigma=0.01$.}
    \label{fig:stability-plot}
\end{figure}

Finally, the  quasi-static region is characterized by $\lambda_{\rm max}  < 0$, while the chaotic one is characterized by $\lambda_{\rm max} > 0$. The boundary between these regions is given by $\lambda_{\rm max} = 0$, see the solid contour in Fig.~\ref{fig:stability-plot}. 
Furthermore, Fig.~\ref{fig:stability-plot} shows the overlap between the fixed point stability contour~\eqref{eq:m-eq0stab} and the $\lambda_{\rm max}=0$ contour. 
{\color{black}(Notice that the linear stability analysis provides the lower bound on this contour, see Fig.~\ref{fig:region-plot}.)} 
The overlapping region corresponds to synchronous chaos, with $M>0$ and $\lambda_{\rm max}>0$, in contrast to the ferromagnetic region where $M>0$ and $\lambda_{\rm max}<0$.
Therefore, we note that (i) higher the annealed disorder (i.e., the smaller $\tau_0$), smaller the synchronous chaos region; (ii) annealed disorder pushes the transition from fixed-point to chaos toward higher $g_c$ -- so it suppresses chaos (Fig.~\ref{fig:stability-plot}), but at the cost of dissipating higher entropy~(Fig.~\ref{fig:EPR-var}b).} 

{\color{black}
In light of the results in Fig.~\ref{fig:stability-plot}, the transition of $\langle x^2\rangle$ from negligible to a finite value in Fig.~\ref{fig:DMFT-PDF}b can be understood as follows. The first transition in Fig.~\ref{fig:DMFT-PDF}b (moving rightward) is due to the fact that the asynchronous-chaos phase exhibits larger fluctuations than the paramagnetic phase, although both have zero mean. The second transition in Fig.~\ref{fig:DMFT-PDF}b (moving upward) arises because, although both the ferromagnetic and paramagnetic phases have small variance, the former develops $M^2>0$ upon crossing the dashed instability line in Fig.~\ref{fig:stability-plot}, leading to an increase in $\langle x^2\rangle={\rm Var}(x)+M^2\simeq M^2$. This explains the behavior of $\langle x^2\rangle$ in Fig.~\ref{fig:DMFT-PDF}b.
Then, given our identification of the two distinct instabilities in Fig.~\ref{fig:stability-plot}, the aforementioned non-monotonicity in Fig.~\ref{fig:EPR-var}c and d can be understood as the consequence of two successive transitions as $\mu$ increases: first, from asynchronous to synchronous chaos (purple to overlap region), and then from synchronous chaos to persistent activity (overlap to orange region), analogous to the behavior observed in the quenched limit \cite{martorell2025ergodicitybreakinghighdimensionalchaos}.}


In this Letter, we offer an exact quantification of the dissipation of infinite non-linear systems with annealed interactions~\eqref{EPR}. 
\textcolor{black}{A $\tau_0$-independent scaling relation between the entropy production rate and the population-averaged variance of the neural activity  is established in NESS. Our finding further reveals that the annealed disorder pushes the chaotic regime, but at the expense of greater dissipation.  These results can be expected to hold  for other non-linear stochastic dynamical systems with annealed disorder.} In the future, it would be interesting to extend our results to sparse networks~\cite{Metz2025}, {\color{black}to networks with tunable non-reciprocity~\cite{Fournier2026}}, to non-Gaussian interaction matrices~\cite{Azaele}, to different non-linear functions exhibiting discontinuous transitions~\cite{Diego}, to 
dynamics featuring synchronization~\cite{Pazo}, spatially-extended systems with traveling waves~\cite{Salvatore}, systems with feedback~\cite{aslyamov}, and neural dynamics with plasticity~\cite{clark2023theory, Pereira-Obilinovic}. One could also apply our framework to optimal control problems for active matter systems~\cite{Luke2024} (via tuning $g$ and $\mu$), making use of their thermodynamic cost associated with dynamical realizations of $J_{ij}(t)$. 
By designing an optimal protocol for $g=g(t)$ and/or $\mu=\mu(t)$, we can achieve the desired distribution of activity $p(\mathbf{x})$ at any given cost, paving the way to address various trade-offs between dissipation and performance in information-processing systems \cite{Lan2012, Proesmans2023}. 
{\color{black}Lastly, though there was  no evidence of aging behavior in our model as long as $Z_{ij}(t)$  and $Z_{ji}(t)$  remain uncorrelated all the time,   aging behavior is expected  to exist when they are strongly correlated \cite{Cugliandolo1997, Sezik}. The thermodynamics of this scenario will be studied in future work.} 


\textit{Acknowledgment.} We are grateful to Amos Maritan for carefully reading the manuscript and for suggesting relevant references. We also thank Anjali Sharma, Alessio Zaccone, Antony Speranza, and Miguel A. Muñoz for their valuable comments. {\color{black}We further thank the reviewers for their critical feedback on our manuscript, which has improved its quality and accessibility.}
TP is supported  by the Dutch Institute for Emergent Phenomena (DIEP) cluster at the University of Amsterdam under the Research Priority Area Emergent Phenomena in Society: Polarization, Segregation, and Inequality. DG gratefully acknowledges support from the
Alexander von Humboldt Foundation. DG thanks  DIEP for hosting his research stay, during which  part of the work was carried out.
TP and DG acknowledge the satellite meeting of StatPhys29 ``Fluctuations in Self-Interacting and Learning Processes (2025)'' at NORDITA, where this collaboration started.

%


\widetext
\newpage
\pagebreak

\setcounter{equation}{0}
\setcounter{figure}{0}
\setcounter{table}{0}
\setcounter{page}{1}
\setcounter{section}{0}
\setcounter{subsection}{0}
\makeatletter
\renewcommand{\theequation}{S\arabic{equation}}
\renewcommand{\thefigure}{S\arabic{figure}}
\renewcommand{\thesection}{S\Roman{section}} 
\renewcommand{\bibnumfmt}[1]{[#1]}
\renewcommand{\citenumfont}[1]{#1}

\begin{center}\Large{Supplemental Material for ``Entropy Production Rate in Stochastically Time-evolving Asymmetric Networks'' }\end{center}

\section{Comment on the difference between two dynamics}
\label{Inequivalence}
As discussed in \cite{Nicoletti2005}, there  are two distinct dynamics, both of which are commonly used in the neural network literature, though they correspond to different information-processing schemes: nonlinear integration and nonlinear summation. Our choice in Eq.~\eqref{eq:eqn1} is the former:
\begin{equation}
   \frac{\rmd x_i}{\rmd t} = -x_i(t) + F\left( \sum_{j\neq i} J_{ij}(t) x_j(t)\right) + \zeta_i(t)\,,\qquad \langle \zeta_i(t)\zeta_j(t')\rangle =\sigma^2 \delta_{ij}\delta(t-t')\ .
\label{tanh_of_sum}
\end{equation}
The latter is given by the following form 
\begin{equation}
   \frac{\rmd x_i}{\rmd t} = -x_i(t) + \sum_{j\neq i}  J_{ij}(t) F\big(x_j(t)\big) + \zeta_i(t) \ .
   \label{sum_of_tanh}
\end{equation}
i.e., here, the signals generated by each node are first nonlinearly transformed, and then linearly projected by means of the interaction matrix $\mathbf{J}(t)$. 

Before going to demonstrate the difference between the two processes Eq.~\eqref{tanh_of_sum} and  Eq.~\eqref{sum_of_tanh} due to time-dependent $\mathbf{J}(t)$, let's us recapitulate the exact equivalence between the two forms of non-linearity (see Appendix~A in~\cite{martorell2023dynamically}) for the \textit{quenched limit} of $q_{\tau_0} \rightarrow 1$, i.e., time-independent $\mathbf{J}$ once the noise strength is rescaled. Specifically, for $\langle \zeta_i(t)\zeta_j(t')\rangle =\sigma^2 \delta_{ij}\delta(t-t')$, and as long as interactions
are uncorrelated and zero-averaged, the two below dynamics are equivalent:
\begin{equation}
 \frac{\rmd x_i}{\rmd t} = -x_i(t) + F\left( \sum_{j\neq i} J_{ij} x_j(t)\right) + \zeta_i(t)\,,\qquad \Leftrightarrow\qquad  \frac{\rmd x_i}{\rmd t} = -x_i(t) + \sum_{j\neq i}  J_{ij} F\big(x_j(t)\big) + g^2\zeta_i(t) \ .
   \label{sum_of_tanh2}
\end{equation}

However, these two dynamical processes become different in our annealed case. Let us demonstrate this point. 
We introduce $y_{\ell} = \sum_i J_{\ell i} x_i$ which satisfy the following equation:
\begin{equation}
  \dot y_{\ell}(t) = \sum_i [\dot J_{\ell i}(t) x_i(t) +J_{\ell i}(t)\dot  x_i(t)] \ .
   \label{y_equation}
\end{equation}
Notice that 
\begin{align}
    \dot J_{\ell,i}(t) = \dfrac{g}{\sqrt{N}}\dot Z_{\ell,i}(t) = \dfrac{g}{\sqrt{N}}\bigg[-\dfrac{Z_{\ell,i}}{\tau_0} + \sqrt{\dfrac{1+2\tau_0}{\tau_0^2}} \xi_{\ell,i}(t)\bigg]\ .
\end{align}
Multiplying both sides $x_i$ and summing over index $\ell$, we have 
\begin{align}
    \sum_i x_i\dot J_{\ell,i}(t)  =\dfrac{g}{\sqrt{N}}\bigg[- \dfrac{1}{\tau_0}\sum_i x_iZ_{\ell,i}  + \sqrt{\dfrac{1+2\tau_0}{\tau_0^2}}  \sum_i x_i\xi_{\ell,i}(t)\bigg]\ .
\end{align}
Similarly, we have
\begin{equation}
   \sum_i J_{\ell,i}(t)\frac{\rmd x_i}{\rmd t} = - \sum_i J_{\ell,i}(t)x_i(t) +  \sum_i J_{\ell,i}(t) F(y_i) +  \sum_i J_{\ell,i}(t) \zeta_i(t) \ .
\end{equation}
Adding the two equations above, we get
\begin{align}
    \dot y_\ell(t) = -y_\ell(t) + \sum_i J_{\ell,i}(t) F(y_i) + \nu^{(1)}_\ell(t) +  \dfrac{g}{\sqrt{N}}\bigg[- \dfrac{1}{\tau_0} \nu^{(2)}_\ell(t)  + \sqrt{\dfrac{1+2\tau_0}{\tau_0^2}} \nu^{(3)}_\ell(t) \bigg]\ ,
    \label{difference}
    \end{align}
where we defined the quantities above for convenience 
\begin{subequations}
\begin{align}
    \nu^{(1)}_\ell(t)&\equiv \sum_i J_{\ell,i}(t) \zeta_i(t)\ ,\\
    \nu^{(2)}_\ell(t)&\equiv \sum_i Z_{\ell,i}(t)x_i(t)\ ,\\
    \nu^{(3)}_\ell(t)&\equiv \sum_i x_i(t)\xi_{\ell,i}(t)\ .
\end{align}    
\end{subequations}
Just a remark that, by its definition, it is straightforward to show that $\nu^{(1)}_\ell(t)$ has delta-correlated correlations: 
\begin{equation}
\begin{aligned}
   \langle \nu^{(1)}_\ell(t) \nu^{(1)}_{\ell'}(t') \rangle &= \sum_{i,j}\langle J_{\ell i}J_{\ell' j} \rangle_{\xi} \langle \zeta_i(t) \zeta_j(t') \rangle_{\zeta}     \\ &= \sigma^2 q_{\tau_0}(|t-t'|)\frac{g^2}{N}~\sum_{i,j}  \delta_{\ell \ell'}\delta_{i j} \times [\delta_{i j}\delta(t-t')]\\ &= \sigma^2 q_{\tau_0}(0)g^2~\delta_{\ell \ell'} \delta(t-t')\ ,
\end{aligned}
\end{equation}
The correlations {\color{black}$\big\langle \nu_{\ell}^{(2)}(t) \nu_{\ell'}^{(2)}(t')\big\rangle$, can be easily calculated as follows: 
\begin{equation}
    \big\langle \nu_{\ell}^{(2)}(t) \nu_{\ell'}^{(2)}(t')\big\rangle =\left\langle \sum_{i,j} x_i(t)x_j(t')Z_{\ell,i}(t)Z_{\ell',j}(t') \right\rangle \ . 
\end{equation}
As done in Eq.~\eqref{last-line} of the following section, the above can be factorized as follows 
\begin{equation}
    \big\langle \nu_{\ell}^{(2)}(t) \nu_{\ell'}^{(2)}(t')\big\rangle =\sum_{i,j}  \left\langle x_i(t)x_j(t')\right\rangle \left\langle Z_{\ell,i}(t)Z_{\ell',j}(t') \right\rangle  \ .
\end{equation}
Now, due to the stationary assumption of $\boldsymbol{Z}$, we have $\langle Z_{i,j}(t)Z_{i',j'}(t')\rangle = \delta_{i,i'}\delta_{j,j'} q_{\tau_0}(|t-t'|)$. Substituting this into the above equation, we get
\begin{equation}
    \big\langle \nu_{\ell}^{(2)}(t) \nu_{\ell'}^{(2)}(t')\big\rangle =\sum_{i,j}  \left\langle x_i(t)x_j(t')\right\rangle \delta_{\ell,\ell'}\delta_{i,j} q_{\tau_0}(|t-t'|) =  \delta_{\ell,\ell'}q_{\tau_0}(|t-t'|) \sum_i\left\langle x_i(t)x_i(t')\right\rangle \ .
\end{equation}
This expression indicates that the noise $\nu_{\ell}^{(2)}(t)$ is non-stationary, in general. When $N\rightarrow \infty$, we can replace the population average autocorrelation with the DMFT counterpart $\sum_i\left\langle x_i(t)x_i(t')\right\rangle = N C_{x}(t,t')$
\begin{equation}
    \big\langle \nu_{\ell}^{(2)}(t) \nu_{\ell'}^{(2)}(t')\big\rangle = N\delta_{\ell,\ell'}q_{\tau_0}(|t-t'|) C_x(t,t')\ .
\end{equation}
Since we have a factor $g/\sqrt{N}$ in front of the bracket in Eq.~\eqref{difference}, we can introduce a new noise $\tilde{\nu}_{\ell}^{(2)} =g/\sqrt{N} \nu_{\ell}^{(2)}(t)$ whose covariance is given by
\begin{equation}
    \big\langle \tilde{\nu}_{\ell}^{(2)}(t) \tilde{\nu}_{\ell'}^{(2)}(t')\big\rangle = g^2\delta_{\ell,\ell'}q_{\tau_0}(|t-t'|) C_x(t,t')\ .
\end{equation}
Following the same reasoning as above, i.e., making use of the same Eq.~\eqref{last-line} from the following section, we get 
\begin{equation}
    \big\langle  \nu^{(3)}_\ell(t)    \nu^{(3)}_{\ell'}(t')\big\rangle =  \sum_{i,j} \big\langle x_i(t)\xi_{\ell,i}(t) x_j(t')\xi_{\ell',j}(t')\big\rangle =  \sum_{i,j} \big\langle x_i(t) x_j(t')\big\rangle\big\langle \xi_{\ell,i}(t) \xi_{\ell',j}(t')\big\rangle = \delta_{\ell,\ell'} \delta(t-t')\sum_{i}\big\langle x_i(t) x_i(t')\big\rangle\ .
\end{equation}
Introducing a new noise $\tilde{\nu}_{\ell}^{(3)} =g/\sqrt{N} \nu_{\ell}^{(3)}(t)$ whose covariance is given by
\begin{equation}
    \big\langle \tilde{\nu}_{\ell}^{(3)}(t) \tilde{\nu}_{\ell'}^{(3)}(t')\big\rangle = g^2\delta_{\ell,\ell'} C_x(t,t')\delta(t-t')\ .
\end{equation}
In summary, the final effective DMFT equation for the model~\eqref{difference} reads
\begin{equation}
    \dot y_\ell(t) = -y_\ell(t) + \eta_\ell + \nu^{(1)}_\ell(t)  - \dfrac{1}{\tau_0} \tilde{\nu}^{(2)}_\ell(t)  + \sqrt{\dfrac{1+2\tau_0}{\tau_0^2}} \tilde{\nu}^{(3)}_\ell(t) \ ,
\end{equation}
where $\langle \eta_\ell(t)\eta_\ell(t') \rangle = q_{\tau_0}(|t-t'|) \langle F[y_\ell(t)] F[y_{\ell}(t)] \rangle$ is coming from the interaction term $\sum_i J_{\ell,i}(t) F(y_i)$. At this point, we can drop the index $\ell$. One can easily see that in the quenched limit ($\tau_0\to \infty$), the last two terms vanish, leading to the equivalence of the two models, as shown above in Eq.~\eqref{sum_of_tanh2}.}


\section{Derivation of effective equations}
\label{app-eff-eqn}
In this section, we derive the effective equation for the following equations of $N$ particle system~\eqref{eq:eqn1}, for which the interactions are modeled by the colored noise (and without self-loop  $J_{ii}=0$). The equations are 
\begin{align}
    \dot x_{i} &= -x_i(t)+  F\bigg[\sum_{j=1}^N J_{i,j}(t)~x_j(t)\bigg] + \zeta_i(t)\ ,\label{xeq-1}\\
     J_{i,j}(t) &\equiv \dfrac{\mu}{N} + \dfrac{g}{\sqrt{N}}~Z_{i,j}(t)\ , \label{J-t}\\
    \dot Z_{i,j}&= -\dfrac{Z_{i,j}}{\tau_0} + \sqrt{\dfrac{1+2\tau_0}{\tau_0^2}} \xi_{i,j}(t)\ ,~\label{Z-t}
\end{align}
where $i\in\{1,2,\dots,N\}$, $\zeta_i(t)$ is a Gaussian white noise with zero mean and correlation $\langle \zeta_i(t)\zeta_j(t')\rangle = \sigma^2 \delta_{ij}\delta(t-t')$, and $\boldsymbol{J}(t)$ is the  stochastic time-dependent interaction matrix. In addition, $\boldsymbol{J}(t)$ consists of two parts: the time-dependent part $\mu/N$ indicating the average value, and the time-dependent part
$\boldsymbol{Z}(t)$. The strength of the time-dependent part is controlled by $g$. We model $\boldsymbol{Z}(t)$ as an Ornstein–Uhlenbeck (OU) process with zero mean and stationary finite-time exponential correlations through
\begin{align}
   \langle Z_{i,j}(t) Z_{i',j'}(t')\rangle = \underbrace{\dfrac{1+2\tau_0}{2\tau_0} e^{-|t-t'|/\tau_0}}_{\displaystyle q_{\tau_0}(|t-t'|)}\delta_{j,j'}\delta_{i,i'}\ . \label{z-corr-t}
\end{align}
with $\tau_0$ quantifying the correlation time. (Notice that the above two-time correlation~\eqref{z-corr-t} can be shown using Eq.~\eqref{Z-t} in the long-time limit.)
In the short and long-$\tau_0$ limit, this correlation function behaves as 
\begin{align}
    \langle Z_{i,j}(t) Z_{i',j'}(t')\rangle = \delta_{j,j'}\delta_{i,i'}
    \times\begin{cases}
    \delta(t-t')       &\qquad \tau_0 \to 0\\
    1                   &\qquad  \tau_0 \to \infty
    \end{cases}\ .
\end{align}
By construction, we are not considering the cross-correlations of the colored noised $Z$ among different particles, i.e., only diagonal terms are contributing to the interactions. In principle, one can also consider the cross-correlation terms.

The effective equation corresponding to equation~\eqref{xeq-1} is obtained by performing the path-integral. To this end, we rewrite Eq.~\eqref{xeq-1} as follows
\begin{align}
\dot x_{i}(t) &= -x_i(t)+  F[y_i(t)] +\theta_i(t) + \zeta_i(t)\ ,\label{xeq-2}\\
 y_i(t) &= \sum_{j=1}^N  J_{i,j}(t)~x_j(t)  +\chi_i(t) ,\label{yeq-2}
\end{align}
where $\theta_i(t)$ and $\chi_i(t)$ are the axillary fields used in the end to compute the response function. In addition, these fields can also be considered as the input deterministic drive. In some other cases for which $x_i(t)$ denotes population growth, this can also be considered as a particle flux for an open system. For closed systems, this can be set to zero.

We write the dynamical generating function as 
\begin{align}
    \mathcal{Z}[\psi,\phi]\equiv \overline{\bigg\langle e^{i\sum_\sigma \int~dt~\psi_\sigma(t) x_\sigma(t)+i\sum_\sigma \int~dt~\phi_\sigma(t) y_\sigma(t)} \bigg\rangle_{\rm paths}},
\end{align}
where the angular brackets $\langle \cdots \rangle$ indicate the average over paths~\eqref{xeq-2} and the overline $\overline{[\cdots]}$ indicates the average over the random trajectories of the interaction matrix elements $J_{i,j}(t)$.
Also, $\mathcal{Z}[0,0]=1$, which ensures normalization, as expected.

Let us first write the average over the paths~\eqref{xeq-2} and \eqref{yeq-2}:
\begin{align}
    \langle (\cdots) \rangle_{\rm paths} &= \int \mathcal{D}[x,y]~e^{i\sum_\sigma \int~dt~\psi_\sigma(t)x_\sigma(t)+i\sum_\sigma \int~dt~\phi_\sigma(t)y_\sigma(t)}\prod_{\sigma,t}\delta\bigg(\dot x_{\sigma}(t) +x_\sigma(t)  -  F[y_\sigma(t)] -\theta_\sigma(t) -\zeta_\sigma(t)\bigg) \nonumber\\
    &\times\prod_{\sigma',t}\delta\bigg(y_{\sigma'}(t) -\sum_{j=1}^N  J_{\sigma',j}(t)~x_j(t) - \chi_{\sigma'}(t) \bigg)\ , \label{z-path}
\end{align}
where $\prod_{\sigma,t}$ indicates the product of all $\sigma$ indices for the entire duration of $t$. We use the integral representation of the Dirac delta function
$ \delta(y-y_0) \propto \int~d\hat{y}~e^{i\hat{y}(y-y_0)}$
in Eq.~\eqref{z-path}, and it gives:
\begin{align}
\langle (\cdots) \rangle_{\rm paths} &= \int \mathcal{D}[x,\hat{x},y,\hat{y}]~\exp\bigg[i\sum_\sigma \int~dt~\psi_\sigma(t)x_\sigma(t)+i\sum_{\sigma'} \int~dt~\phi_{\sigma'}(t)y_{\sigma'}(t)\bigg]\nonumber\\
&\times\exp\bigg[i\sum_\sigma \int~dt~\hat{x}_\sigma(t)\bigg(\dot x_{\sigma }(t) +x_\sigma(t)  -  F[y_\sigma(t)] -\theta_\sigma(t)-\zeta_\sigma(t)\bigg)\bigg] \nonumber\\
&\times\exp\bigg[i\sum_{\sigma'} \int~dt~\hat{y}_{\sigma'}(t)\bigg(y_{\sigma'}(t) -\sum_{j=1}^N J_{\sigma',j}(t)~x_j(t) - \chi_{\sigma'}(t)\bigg)\bigg],\\
&=\int \mathcal{D}[x,\hat{x},y,\hat{y}]~\exp\bigg[i\sum_\sigma \int~dt~\psi_\sigma(t)x_\sigma(t)+i\sum_{\sigma'} \int~dt~\phi_{\sigma'}(t)y_{\sigma'}(t)\bigg]\nonumber\\
&\times\exp\bigg[i\sum_\sigma \int~dt~\hat{x}_\sigma(t)\bigg(\dot x_{\sigma}(t) +x_\sigma(t)  -  F[y_\sigma(t)] -\theta_\sigma(t)\bigg)\bigg] \nonumber\\
&\times\exp\bigg[i\sum_{\sigma'} \int~dt~\hat{y}_{\sigma'}(t)\bigg(y_{\sigma'}(t) - \chi_{\sigma'}(t)\bigg)\bigg]\times\exp\bigg[-i\sum_\sigma \int~dt~\hat{x}_\sigma(t)\zeta_\sigma(t)\bigg]\nonumber\\
&\times \exp\bigg[-i\sum_{\sigma',j} \int~dt~\hat{y}_{\sigma'}(t) J_{\sigma',j}(t)~x_j(t)\bigg]\ ,\label{eq:second}
\end{align}
where in Eq.~\eqref{eq:second} we have separated terms that have annealed disorder $\zeta_\sigma(t)$ and $J_{\sigma',i}(t)$. Now we perform the averaging on an ensemble of trajectories of $\zeta_\sigma(t)$ and $J_{\sigma',i}(t)$. Since both $\zeta_\sigma(t)$ and $J_{\sigma',i}(t)$ are independent of each other, we can perform the average independently. This gives
\begin{align}
\mathcal{Z}[\psi,\phi]&\equiv \overline{\langle (\cdots) \rangle_{\rm paths}}\nonumber\\
&= \int \mathcal{D}[x,\hat{x},y,\hat{y}]~\exp\bigg[i\sum_\sigma \int~dt~\psi_\sigma(t)x_\sigma(t) + i\sum_\sigma \int~dt~\hat{x}_\sigma(t)\bigg(\dot x_{\sigma}(t) +x_\sigma(t) -  F[y_\sigma(t)] -\theta_\sigma(t)\bigg) \bigg] \nonumber\\
&\times\underbrace{\exp\bigg[i\sum_{\sigma'} \int~dt~\phi_{\sigma'}(t)y_{\sigma'}(t) + i\sum_{\sigma'} \int~dt~\hat{y}_{\sigma'}(t)\bigg(y_{\sigma'} - \chi_{\sigma'}(t)\bigg)\bigg]}_{B[y,\hat{y}]}\times~\overline{\exp\bigg[-i\sum_\sigma \int~dt~\hat{x}_\sigma(t)\zeta_\sigma(t)\bigg]}\nonumber\\
&\times \underbrace{\overline{\exp\bigg[-i\sum_{\sigma',j} \int~dt~\hat{y}_{\sigma'}(t) J_{\sigma',j}(t)~x_j(t)\bigg]}}_{\Delta[x,\hat{x},y,\hat{y}]}\\
&= \int \mathcal{D}[x,\hat{x},y,\hat{y}]~\underbrace{\exp\bigg[i\sum_\sigma \int~dt~\psi_\sigma(t)x_\sigma(t) + i\sum_\sigma \int~dt~\hat{x}_\sigma(t)\bigg(\dot x_{\sigma}(t) +x_\sigma(t) -  F[y_\sigma(t)] -\theta_\sigma(t) + iD\hat{x}_\sigma(t)\bigg) \bigg]}_{A[x,\hat{x},y,\hat{y}]} \nonumber\\
&\times\underbrace{\exp\bigg[i\sum_{\sigma'} \int~dt~\phi_{\sigma'}(t)y_{\sigma'}(t) + i\sum_{\sigma'} \int~dt~\hat{y}_{\sigma'}(t)\bigg(y_{\sigma'} - \chi_{\sigma'}(t)\bigg)\bigg]}_{B[y,\hat{y}]}\nonumber\\
&\times \underbrace{\overline{\exp\bigg[-i\sum_{\sigma',j} \int~dt~\hat{y}_{\sigma'}(t) J_{\sigma',j}(t)~x_j(t)\bigg]}}_{\Delta[x,\hat{x},y,\hat{y}]}\ ,
\label{decompo}
\end{align}
where $D = k_{\rm B}T$ is the diffusion constant, where $k_{\rm B}$ is the Boltzmann constant, and $T$ the temperature. 
Let us rewrite the last term as follows:
\begin{align}
    \Delta[x,\hat{x},y,\hat{y}] &\equiv\overline{\exp\bigg[-i\sum_{\sigma',j} \int~dt~\hat{y}_{\sigma'}(t) J_{\sigma',j}(t)~x_j(t)\bigg]}\\
    &=\exp\bigg[-i\sum_{\sigma',j} \dfrac{\bar{J}}{N} \int~dt~\hat{y}_{\sigma'}(t)~x_j(t)\bigg] \times \overline{\exp\bigg[-i\sum_{\sigma',j} \dfrac{g}{\sqrt{N}} \int~dt~\hat{y}_{\sigma'}(t) Z_{\sigma',j}(t)~x_j(t)\bigg]}\\
     &=\exp\bigg[-i\frac{\bar{J}}{N}\sum_{\sigma,j}\int~dt~\hat{y}_{\sigma}(t)~x_j(t)\bigg] \exp\bigg[-\frac{g^2}{2N}\sum_{\sigma,j}\int~dt~\int~dt'~\hat{y}_{\sigma}(t)~x_j(t)\hat{y}_{\sigma}(t')~x_j(t')q_{\tau_0}(|t-t'|)\bigg]\ , \label{last-line}
\end{align}
where in the last line we used the following result (see  Sec.~\ref{sec:lastline} for the proof):
\begin{align}
    \overline{\exp\bigg[\frac{ig}{\sqrt{N}} \int_0^t dt~A(t)Z(t)\bigg]}  = \exp\bigg[-\dfrac{g^2}{2N} \int_0^t dt_1\int_0^t dt_2 ~A(t_1)A(t_2)q_{\tau_0}(|t_1-t_2|)\bigg]\ . \label{pf-lastline}
\end{align}

Thus, Eq.~\eqref{last-line} becomes 
\begin{align}
\Delta[x,\hat{x},y,\hat{y}]&=\exp\bigg[-i \bar{J}N \int~dt~[\rho_{x}(t)\lambda_y(t)]\bigg]\exp\bigg[-\frac{Ng^2}{2}\int~dt~\int~dt'\big[Q_x(t,t')L_y(t,t') q_{\tau_0}(|t-t'|) \bigg] \ ,\label{eqn-sub}
\end{align}
where, arriving from Eq.~\eqref{last-line} to Eq.~\eqref{eqn-sub}, we defined the following quantities:
\begin{align}
    \rho_x(t) = \dfrac{1}{N} \sum_\sigma x_\sigma(t)\ , &\qquad\qquad  \lambda_y(t) = \dfrac{1}{N} \sum_\mu \hat{y}_\mu(t)\label{cond-1}\ ,\\
    Q_x(t,t') = \dfrac{1}{N} \sum_\sigma x_\sigma(t)x_\sigma(t') \ ,&\qquad\qquad L_y(t,t') = \dfrac{1}{N} \sum_\sigma \hat{y}_\sigma(t)\hat{y}_\sigma(t')\ .\label{cond-2}
\end{align}
Now we use the above equalities in $\mathcal{Z}$~\eqref{decompo} using the integral representation of the Dirac delta function (for all time), i.e., substituting expressions for \eqref{cond-1}-\eqref{cond-2} similar to the following
\begin{align}
    \prod_t\delta\bigg(\rho_x N - \sum_\sigma x_\sigma(t)\bigg) \propto \int~\mathcal{D}[\hat{\rho}] e^{iN \int~dt~\hat{\rho}_x(t)\rho_x(t)}e^{-i\sum_\sigma\int~dt~\hat{\rho}_x(t)~x_\sigma(t)}
\end{align}
in $\mathcal{Z}$~\eqref{decompo}. We define $\Pi = (\rho_x,\lambda_y,Q_x,L_y)$ and $\hat{\Pi} = (\hat{\rho}_x,\hat{\lambda}_y,\hat{Q}_x,\hat{L}_y)$ and rewrite $\mathcal{Z}$ as
\begin{align}
    \mathcal{Z}[\psi,\phi] &= \int~\mathcal{D}[\Pi,\hat{\Pi}]~e^{N(\Psi[\Pi,\hat{\Pi}]+\Phi[\Pi])}\nonumber\\
    & \times\int~\mathcal{D}[x,\hat{x},y,\hat{y}]\mathcal{A}[x,\hat{x},y,\hat{y}]~e^{-i\sum_\sigma\int~dt[\hat{\rho}_x(t)x_\sigma(t)]}e^{-i\sum_\sigma \int~dt\int~dt'~[\hat{Q}_x x_\sigma x_\sigma' ]}\nonumber\\
    &\times \mathcal{B}[y,\hat{y}]e^{-i\sum_\sigma\int~dt[\hat{\lambda}_y(t)\hat{y}_\sigma(t)]}e^{-i\sum_\sigma \int~dt\int~dt'~[\hat{L}_y \hat{y}_\sigma \hat{y}_\sigma']}\ ,\label{Z-eqn2}
\end{align}
where
\begin{align}
   \Psi[\Pi,\hat{\Pi}] &= i \int~dt~\big[\hat{\rho}_x(t)\rho_x(t)+\hat{\lambda}_y(t)\lambda_y(t)\big]+i\int~dt\int~dt'~[\hat{Q}_x ~Q_x+\hat{L}_y ~L_y](t,t')\ ,\\
   \Phi[\Pi]& = -i \bar{J} \int~dt~[\rho_{x}(t)\lambda_y(t)]-\frac{g^2}{2}\int~dt~\int~dt'\big[Q_x(t,t')L_y(t,t') q_{\tau_0}(|t-t'|) \ .
\end{align}

Now $\mathcal{Z}$~\eqref{Z-eqn2} can be written in compact form:
\begin{align}
   \mathcal{Z}[\psi,\phi] &= \int~\mathcal{D}[\Pi,\hat{\Pi}]~e^{N(\Psi[\Pi,\hat{\Pi}]+\Phi[\Pi]+\Omega_1[\hat{\Pi}]+\Omega_2[\hat{\Pi}])},\label{z-compact}
\end{align}
where we identify  ($e^{\log\int D[n] e^{\sum_\sigma (\dots)}} = e^{\log\int D[n] \prod_\sigma e^{(\dots)}} =  e^{\log\prod_\sigma\int D[n_\sigma] e^{(\dots)}}$, and this gives the following)
\begin{align}
    \Omega_1[\hat{\Pi}] &= \dfrac{1}{N}\sum_\sigma \ln \int~\mathcal{D}[\cdots] \exp\bigg[i \int~dt~\psi_\sigma(t)x_\sigma(t)+i \int~dt~\hat{x}_\sigma(t)\bigg(\dot x_{\sigma}(t) + x_{\sigma}(t) -F(y_\sigma)-\theta_\sigma(t) + i D \hat{x}_\sigma\bigg)\bigg]\nonumber\\
    &\times \exp\bigg[-i\int~dt[\hat{\rho}_x(t)x_\sigma(t)]-i\int~dt\int~dt'~[\hat{Q}_x x_\sigma x_\sigma']\bigg]\nonumber\\
    &=\ln\int~\mathcal{D}[\cdots] \exp\bigg[i \int~dt~\psi(t)x(t)+i \int~dt~\hat{x}(t)\bigg(\dot x(t) + x(t) -F(y)-\theta(t)+ i D \hat{x}\bigg)\bigg]\nonumber\\
    &\times \exp\bigg[-i\int~dt[\hat{\rho}_x(t)x(t)]-i\int~dt\int~dt'~[\hat{Q}_x x x']\bigg]\ .
\end{align}
Here we first write $e^{\sum_\sigma(\dots)} = \prod_{\sigma}e^{(\dots)}$, and then factorize the integral of the path over $\mathcal{D}$ for each $\sigma$. The integrals over $\sigma$ are not coupled and can be carried out independently. Finally, we take log, which converts the product into a summation. Since each term inside the summation is identical, the summation gives $N$. Similarly, we have 
\begin{align}
  e^{\Omega_2[\hat{\Pi}]} &=\int~\mathcal{D}[y,\hat{y}] \exp\bigg[i \int~dt~\phi(t)y(t)+i \int~dt~\hat{y}(t)\bigg(y(t)-\chi(t)\bigg)\bigg]\exp\bigg[-i\int~dt[\hat{\lambda}_y(t)\hat{y}(t)]-i\int~dt\int~dt'~[\hat{L}_y \hat{y} \hat{y}']\bigg]\ .
\end{align} 

Then, $\mathcal{Z}$ in Eq.~\eqref{z-compact} can be approximated using the saddle-point method (in the limit $N\to\infty$), 
\begin{align}
    \mathcal{Z}\approx e^{N(\Psi^*[\Pi,\hat{\Pi}]+\Phi^*[\Pi]+\Omega_1^*[\hat{\Pi}]+\Omega_2^*[\hat{\Pi}])}, \label{sad-z}
\end{align}
where the extrema solutions indicated by $*$ are the solutions of the following equations:
\begin{align}
    &\dfrac{\delta \Psi^*}{\delta \Pi}+\dfrac{\delta \Phi^*}{\delta \Pi} = 0\label{ex-1}\ ,\\
    &\dfrac{\delta \Psi^*}{\delta \hat{\Pi}}+\dfrac{\delta \Omega_1^*}{\delta \hat{\Pi}}+\dfrac{\delta \Omega_2^*}{\delta \hat{\Pi}} = 0~\label{ex-2}\ .
\end{align}
Equation~\eqref{ex-1} gives the following   
\begin{align}
    \hat{\rho}_x(t) = 
    \mu \lambda_y(t)  \ , && 
    \hat{\lambda}_y(t) = \mu \rho_x(t)\ ,\\
    i\hat{Q}_x =\frac{g^2}{2}L_y ~q_{\tau_0}(|t-t'|)  \ ,  && i\hat{L}_y =\frac{g^2}{2}Q_xq_{\tau_0}(|t-t'|)\ .
\end{align} 
Similarly, equation~\eqref{ex-2} gives the following
\begin{align}
    \rho_x(t) = \langle x \rangle_{\Omega_1}\ , &&\lambda_y(t) =\langle \hat{y}\rangle_{\Omega_2}\ , \\
    Q_x = \langle xx'\rangle_{\Omega_1}\ ,&&
    L_y = \langle \hat{y}\hat{y}'\rangle_{\Omega_2}\ .
\end{align}
By construction we have the normalization for $\mathcal{Z}$, namely, $\mathcal{Z}[\psi=0,\phi=0]= 1$. Therefore, 
\begin{equation}
   \langle \hat{y}(t) \rangle = i  \lim_{\psi\rightarrow 0, \phi\rightarrow 0} \frac{\delta \mathcal{Z}[\psi,\phi]}{\delta \chi(t)} =0 \,,\qquad  \langle \hat{y}(t) \hat{y}(t') \rangle = i  \lim_{\psi\rightarrow 0, \phi\rightarrow 0} \frac{\delta^2 \mathcal{Z}[\psi,\phi]}{\delta \chi(t)\delta \chi(t')} =0\ .
\end{equation}

Using this information, we first have  $\Psi^*[\Pi,\hat{\Pi}] +\Phi^*[\Pi]=i \int~dt~\hat{\lambda}_y(t)\lambda_y(t)+i\int~dt\int~dt'~\hat{L}_y (t,t')~L_y(t,t')=0 $ and
\begin{align}
    \Omega_1[\hat{\Pi}] &=\ln\int~\mathcal{D}[x,\hat{x}] \exp\bigg[i \int~dt~\psi(t)x(t)+i \int~dt~\hat{x}(t)\bigg(\dot x(t) + x(t) -F(y)-\theta(t) + i D \hat{x}\bigg)\bigg]\nonumber\\
    &\times \exp\bigg[-i\int~dt[\hat{\rho}_x(t)x(t)]-i\int~dt\int~dt'~[\hat{Q}_x x x' ]\bigg]\\
    &=\ln\int~\mathcal{D}[x,\hat{x}] \exp\bigg[i \int~dt~\psi(t)x(t)+i \int~dt~\hat{x}(t)\bigg(\dot x(t) + x(t) -F(y)-\theta(t) + i D \hat{x}\bigg)\bigg] \ . \label{omega_x-eqn}
\end{align}
Therefore, the above equation gives the first effective equation:
\begin{align}
    \dot x(t) = -x(t) +F(y) + \theta(t) +\zeta(t)\ ,
\end{align}
where $\langle \zeta(t)\rangle = 0$ and $\langle \zeta(t)\zeta(t')\rangle = 2 D \delta(t-t')$.
Similarly, we write 
\begin{align}
 \nonumber \Omega_2[\hat{\Pi}]&=\ln\int~\mathcal{D}[y,\hat{y}] \exp\bigg[i \int~dt~\Big[\phi(t)y(t)+\hat{y}(t)\big(y(t)-\chi(t)-\hat{\lambda}_y(t)\big)\Big]-i\int~dt\int~dt'~[\hat{L}_y \hat{y} \hat{y}']\bigg]\\
 \nonumber &=\ln\int~\mathcal{D}[y,\hat{y}] \exp\bigg[i \int~dt~\Big[\phi(t)y(t)+\hat{y}(t)\big(y(t)-\chi(t)-\mu\rho_x(t)\big)\Big]-\dfrac{g^2}{2}\int~dt\int~dt'~[Q_x \hat{y} \hat{y}'q_{\tau_0}(t-t')]\bigg]\\
    &=\ln\int~\mathcal{D}[y,\hat{y}] \exp\bigg[i \int~dt~\phi(t)y(t)+\hat{y}(t)\bigg(y(t)-\chi(t) -\mu\rho_x\bigg)-\frac{g^2}{2}\int~dt\int~dt'~Q_x\hat{y} \hat{y}'q_{\tau_0}(|t-t'|)\bigg]\ .
    \label{omega_r-eqn}
\end{align}
So, our second effective equation is 
\begin{align}
    y(t)=\chi(t) + \mu\rho_x(t) + g\eta(t)\ ,
\end{align}
with noise correlations:
\begin{align}
    \langle \eta(t)\eta(t')\rangle = Q_x(t,t') q_{\tau_0}(|t-t'|) = q_{\tau_0}(|t-t'|) \langle x(t)x(t') \rangle \ .
\end{align}

Let us write our effective equations:
\begin{align}
          \dot x(t) = -x(t) +F(y) + \theta(t) \label{eff-1}\ ,\\
         y(t)=\chi(t) + \mu\rho_x(t) + g \eta(t)\label{eff-2}\ ,
\end{align}
or (after setting the axillary fields $\theta(t)$ and $\chi(t)$ to zero) 
\begin{align}
    \dot x(t) &= -x(t) +F[\mu\langle x \rangle + g \eta(t) ] + \zeta(t)\ ,\label{si:eff-eqn}
\\
 \langle \eta(t)\eta(t')\rangle &= q_{\tau_0}(|t-t'|) \langle x(t)x(t') \rangle \, \label{eff-3}
\end{align}
where $q_{\tau_0}(|t-t'|)$ is given in Eq.~\eqref{z-corr-t}. Equation~\eqref{si:eff-eqn} is the effective equation~\eqref{effective} given in the main text.

\subsection{Calculation of Response function}
In this section, we calculate the response function
\begin{align}
   G_\theta(t,t')\equiv\dfrac{\delta \langle x(t)\rangle}{\delta (\theta(t'))} =\left \langle \dfrac{\delta x(t)}{\delta \zeta(t')}\right\rangle \ .
\end{align}
by taking the derivative of Eq.~\eqref{si:eff-eqn} with respect to $\zeta(t')$ on both sides, we get 
\begin{align}
\dot G_{\theta} = -G_{\theta} +\left \langle  F'[y]\frac{\partial y(t)}{\partial \zeta(t')}\right\rangle + \delta(t,t')\ ,
\end{align}
where $y(t)= \bar{J}\langle x \rangle + g \eta(t) + \chi(t)$. To this end, we write the dynamical moment generating function evaluated at the saddle-point [see Eqs.~\eqref{sad-z}--\eqref{ex-2}]. It turns out that $\Psi^*=0$ and $\Phi^*=0$ at this saddle point. Then, the dynamical moment generating function  becomes
\begin{align}
    \mathcal{Z}\approx e^{N\Omega^*[\hat{\Pi}]}\ , \label{sad-z-2}
\end{align}
where, for convenience, we define 
\begin{align}
   \Omega^*[\hat{\Pi}] \equiv \Omega_1^*[\hat{\Pi}] + \Omega_2^*[\hat{\Pi}]\ .
\end{align}
Notice that $\Omega_{1,2}$ are defined in Eqs.~\eqref{omega_x-eqn} and \eqref{omega_r-eqn}.

From Eq.~\eqref{sad-z}, we write the single particle moment generating function:
\begin{equation}
\begin{aligned}
   Z_1[\psi,\phi] = e^{\Omega^*} &= \int~\mathcal{D}[x,\hat{x},y,\hat{y}] \exp\bigg[i \int~dt~\psi(t)x(t)+i \int~dt~\hat{x}(t)\bigg(\dot x(t) + x(t) -F(y)-\theta(t) + i D \hat{x}\bigg)\bigg]\\
    &\times \exp\bigg[i \int~dt~\phi(t)y(t)+\hat{y}(t)\bigg(y(t)-\chi(t) -\bar{J}\rho_x\bigg)\bigg]\times \exp\bigg[-\int~dt\int~dt'~\frac{g^2}{2}Q_x\hat{y} \hat{y}'q_{\tau_0}(|t-t'|)\bigg]\\
  & = \int~\mathcal{D}[x,\hat{x},y,\hat{y}]D[\zeta]P[\zeta] D[\eta]P[\eta] \exp\bigg[i \int~dt~\psi(t)x(t)+i \int~dt~\hat{x}(t)\bigg(\dot x(t) + x(t) -F(y)-\theta(t) - \zeta(t)\bigg)\bigg]\\
    &\times \exp\bigg[i \int~dt~\phi(t)y(t)+\hat{y}(t)\bigg(y(t)-\chi(t) -\bar{J}\rho_x - \eta(t)\bigg)\bigg]\ .
\end{aligned}
\label{moment_generating_function_single_particle}
\end{equation}
Note that $Z_1[0,0]=1$, as expected. Differentiating $Z_1$ with respect to $\psi$ and setting both $\psi$ and $\phi$ equal to zero, we get
\begin{align}
   \dfrac{\delta Z_1[\psi,\phi]}{\delta \psi(t)}\bigg|_{\psi=0,\phi=0} 
  & = i\int~\mathcal{D}[x,\hat{x},y,\hat{y}]D[\zeta]P[\zeta] D[\eta]P[\eta]~x(t)~\exp\bigg[i \int~dt~\hat{x}(t)\bigg(\dot x(t) + x(t) -F(y)-\theta(t) - \zeta(t)\bigg)\bigg]\nonumber\\
    &\times \exp\bigg[i \int~dt~\hat{y}(t)\bigg(y(t)-\chi(t) -\bar{J}\rho_x - \eta(t)\bigg)\bigg] \\
    &= i\langle x(t)\rangle\ .
\end{align}
Differentiating $\langle x(t)\rangle$ (in the above equation) on both sides with respect to $\theta(t)$ gives the response function
\begin{align}
   G_\theta(t,t')\equiv\dfrac{\delta \langle x(t)\rangle}{\delta (\theta(t'))} = -i\langle x(t) \hat{x}(t')\rangle\ .\label{res-fun-1}
\end{align}

To evaluate the right-hand side of the above equation~\eqref{res-fun-1}, we write the following average 
\begin{align}
  \langle x(t) \zeta(t')\rangle &= \int~\mathcal{D}[x,\hat{x},y,\hat{y}]D[\zeta]P[\zeta] D[\eta]P[\eta]~x(t)\zeta(t')~\exp\bigg[i \int~dt~\hat{x}(t)\bigg(\dot x(t) + x(t) -F(y)-\theta(t) - \zeta(t)\bigg)\bigg]\nonumber\\
    &\times \exp\bigg[i \int~dt~\hat{y}(t)\bigg(y(t)-\chi(t) -\bar{J}\rho_x - \eta(t)\bigg)\bigg] \label{x-zeta-1}\\
    & = \int D[\dots] D[\zeta]~P[\zeta]~x(t)~\zeta(t')~e^{-i\int~dt~\hat{x}(t)\zeta(t)}~\mathcal{P} \ ,\label{x-zeta-2}
\end{align}
where, for convenience, we write $\mathcal{P}$, which contains all the other terms in Eq.~\eqref{x-zeta-1}. Now for a Gaussian noise with two-time correlation $\langle \zeta(t)\zeta(s)\rangle = C(t-s)$, we can show that
\begin{align}
 \zeta(t')~P[\zeta]  =  -\int~ds~C(t'-s) \dfrac{\delta P[\zeta]}{\delta \zeta(s)}  \ . 
\end{align}
Since the correlation function for noise $\zeta(t)$ is a delta function, this gives
\begin{align}
 \zeta(t')~P[\zeta]  =  -2D~\dfrac{\delta P[\zeta]}{\delta \zeta(t')} \ . \label{z-p-eqn}  
\end{align}
Then, substituting the product $\zeta(t')~P[\zeta]$~\eqref{z-p-eqn} in Eq.~\eqref{x-zeta-2}, we get
\begin{align}
  \langle x(t) \zeta(t')\rangle
    & = -2D\int D[\dots] D[\zeta]~x(t)~\int~ds~\delta(s-t')\dfrac{\delta P[\zeta]}{\delta \zeta(s)}   ~e^{-i\int~dt~\hat{x}(t)\zeta(t)}~\mathcal{P}\ .
\end{align}
Integrating by parts over the integration variable $s$ gives the following
\begin{equation}
\begin{aligned}
  \langle x(t) \zeta(t')\rangle
    =&~  -2Di\int~\mathcal{D}[x,\hat{x},y,\hat{y}]D[\zeta]P[\zeta] D[\eta]P[\eta]~x(t) \hat{x}(t')\exp\bigg[i \int~dt~\hat{x}(t)\bigg(\dot x(t) + x(t) -F(y)-\theta(t) - \zeta(t)\bigg)\bigg]\\
    &\times \exp\bigg[i \int~dt~\hat{y}(t)\bigg(y(t)-\chi(t) -\bar{J}\rho_x - \eta(t)\bigg)\bigg] = -2D i\langle x(t)\hat{x}(t')\rangle = 2 D G_\theta(t,t')\ ,
\end{aligned}
\label{x_linear_response_unequal_time}
\end{equation}
where in the last line we used Eq.~\eqref{res-fun-1}. Using effective dynamics~\eqref{si:eff-eqn}
\begin{align}
    \dot x(t) = - x(t) +F(y) + \theta(t) + \zeta(t)
\end{align}
we can obtain the response function $G_\theta(t,t')$, which in the Stratonovich convention \cite{Hertz2017}
\begin{align}
    G_\theta(t,t) = 1/2\ ,
\end{align}
for $t=t'$.
Thus, we have
\begin{equation}
  \langle x(t) \zeta(t)\rangle = T \ ,
  \label{x_linear_response}
\end{equation}
where $T$ is the temperature of the heat bath.

\subsection{Proof of Eq.~\eqref{pf-lastline}}
\label{sec:lastline}
Here we give the proof of Eq.~\eqref{pf-lastline}:
\begin{align}
    \overline{\exp\bigg[\frac{ig}{\sqrt{S}} \int_0^t dt~A(t)Z(t)\bigg]} &= 1 - \frac{g^2}{2!S} \int_0^t dt_1\int_0^t dt_2~A(t_1)A(t_2)\overline{Z(t_1)Z(t_2)}+\nonumber\\
    &+  \frac{g^4}{4!S^2} \int_0^t dt_1\int_0^t dt_2\int_0^t dt_3\int_0^t dt_4 ~A(t_1)A(t_2)A(t_3)A(t_4)\overline{Z(t_1)Z(t_2)Z(t_3)Z(t_4)} + \dots\ , \label{exp-fun}
\end{align}
where we have dropped odd ordered terms in $Z$ since these will be zero after averaging over the Gaussian distribution $P(Z)$. (Also, remember that $Z_\sigma^i$ are identically distributed; therefore, we drop the indices $\sigma$ and $i$ from $Z_\sigma^i$.) $Z$ is a stationary colored noise with two-time correlation: $\overline{Z(t_1)Z(t_2)} = q_{\tau_0}(|t_1-t_2|)$.

To evaluate the right-hand side of Eq.~\eqref{exp-fun}, we have the following result:
\begin{align}
   \overline{Z(t_1)Z(t_2)} &= q_{\tau_0}(|t_1-t_2|) \\
   \overline{Z(t_1)Z(t_2)Z(t_3)Z(t_4)} &= \overline{Z(t_1)Z(t_2)}~~\overline{Z(t_3)Z(t_4)} + \overline{Z(t_1)Z(t_3)}~~\overline{Z(t_2)Z(t_4)}+\overline{Z(t_1)Z(t_4)}~~\overline{Z(t_2)Z(t_3)}\ ,\label{wick}
\end{align}
where we have used the Wick's theorem for Gaussian random variables.

Since there are integrals in the last line of Eq.~\eqref{exp-fun} over $t_{i}$s (these are dummy variables), three contributions in Eq.~\eqref{wick} are identical. So we need to evaluate only one. Therefore, we get
\begin{align}
\overline{\exp\bigg[\frac{ig}{\sqrt{S}} \int_0^t dt~A(t)Z(t)\bigg]} &= 1 - \frac{g^2}{2!S} \int_0^t dt_1\int_0^t dt_2~A(t_1)A(t_2)q_{\tau_0}(|t_1-t_2|)+\nonumber\\
    &+  \frac{3g^4}{4!S^2} \int_0^t dt_1\int_0^t dt_2\int_0^t dt_3\int_0^t dt_4 ~A(t_1)A(t_2)A(t_3)A(t_4)q_{\tau_0}(|t_1-t_2|)q_{\tau_0}(|t_3-t_4|) + \dots\ , \nonumber\\
    &= 1 - \frac{g^2}{2S} \int_0^t dt_1\int_0^t dt_2~A(t_1)A(t_2)q_{\tau_0}(|t_1-t_2|)+\nonumber\\
    &+  \dfrac{1}{2!} \bigg(\frac{g^2}{2S}\bigg)^2\bigg[\int_0^t dt_1\int_0^t dt_2 ~A(t_1)A(t_2)q_{\tau_0}(|t_1-t_2|)\bigg]^2 + \dots\\
    & = \exp\bigg[-\dfrac{g^2}{2S} \int_0^t dt_1\int_0^t dt_2 ~A(t_1)A(t_2)q_{\tau_0}(|t_1-t_2|)\bigg]\ ,
\end{align}
where the above result is used to evaluate the last line of Eq.~\eqref{last-line}.

\subsection{Proof of $\langle \eta(t) \zeta(t')\rangle
      =0$}
      \label{no_correlation_between_noises}
We use Eq.~\eqref{moment_generating_function_single_particle} (with $\psi=\phi=0$) to write the following average 
\begin{align}
  \langle \eta(t) \zeta(t')\rangle &= \int~\mathcal{D}[x,\hat{x},y,\hat{y}]D[\zeta]P[\zeta] D[\eta]P[\eta]~\eta(t)\zeta(t')~\exp\bigg[i \int~dt~\hat{x}(t)\bigg(\dot x(t) + x(t) -F(y)-\theta(t) - \zeta(t)\bigg)\bigg]\nonumber\\
    &\times \exp\bigg[i \int~dt~\hat{y}(t)\bigg(y(t)-\chi(t) -\bar{J}\rho_x - \eta(t)\bigg)\bigg] \ ,\label{x-zeta-3}\\
    & = \int D[\dots]  D[\eta]~P[\eta]~\eta(t)~e^{-i\int~dt~\hat{y}(t)\eta(t)}~D[\zeta]~P[\zeta]~\zeta(t')~e^{-i\int~dt~\hat{x}(t)\zeta(t)}~\mathcal{P} \ ,\label{x-zeta-4}
\end{align}
where, for convenience, we write $\mathcal{P}$, which contains all the other terms in Eq.~\eqref{x-zeta-3}. Now for a Gaussian noise with two-time correlation $\langle \zeta(t)\zeta(s)\rangle = C(t-s)$, we can show that
\begin{align}
 \zeta(t')~P[\zeta]  =  -\int~ds~C_\zeta(t'-s) \dfrac{\delta P[\zeta]}{\delta \zeta(s)}  \ . 
\end{align}
The correlation function for noise $\zeta(t)$ is a delta function (which we will use later), 
but for 
\begin{align}
 \eta(t)~P[\eta]  &=  -\int~ds~C_\eta(t-s) \dfrac{\delta P[\eta]}{\delta \eta(s)} \ , \\
\end{align}
where $C_\eta(t-s)$ is given above. Then, substituting the product $\zeta(t')~P[\zeta]$~\eqref{z-p-eqn} in Eq.~\eqref{x-zeta-2}, we get
\begin{align}
  \langle \eta(t) \zeta(t')\rangle
    & = \int D[\dots] D[\eta]~\int~du~C_\eta(u-t)\dfrac{\delta P[\eta]}{\delta \eta(u)}   ~e^{-i\int~dt~\hat{y}(t)\eta(t)} D[\zeta]~\int~ds~C_\zeta(s-t')\dfrac{\delta P[\zeta]}{\delta \zeta(s)}   ~e^{-i\int~dt~\hat{x}(t)\zeta(t)}~\mathcal{P}\ .
\end{align}
Integrating by parts over the integration variable $s$ (first over $\zeta$ and then over $\eta$, both of them two $i$-s) gives the following
\begin{align}
  \langle \eta(t) \zeta(t')\rangle
    & = -\int~\mathcal{D}[x,\hat{x},y,\hat{y}]D[\zeta]P[\zeta] D[\eta]P[\eta]\exp\bigg[i \int~dt~\hat{x}(t)\bigg(\dot x(t) + x(t) -F(y)-\theta(t) - \zeta(t)\bigg)\bigg]\nonumber\\
    &\times \exp\bigg[i \int~dt~\hat{y}(t)\bigg(y(t)-\chi(t) -\bar{J}\rho_x - \eta(t)\bigg)\bigg]  \int~du~C_\eta(u-t)\hat{y}(u) \int~ds~C_\zeta(s-t')\hat{x}(s)\ .
\end{align}
Since $C_\zeta$ is a delta-function, we get
\begin{equation}
  \langle \eta(t) \zeta(t')\rangle
     = -2D \int~du~C_\eta(u-t)\langle \hat{y}(u)\hat{x}(t')\rangle  =0\ ,
     \label{most_important_identity}
\end{equation}
where the correlation$\langle \hat{y}(u)\hat{x}(t')\rangle$ on the right-hand side is zero, as we prove below.
Let us start with the  moment-generating function in Eq.~\eqref{moment_generating_function_single_particle}
\begin{equation}
\begin{aligned}
    Z[\psi, \phi, \theta,\chi] =&~ \int~\mathcal{D}[x,\hat{x},y,\hat{y}]~\exp\left\{ i\int~dt~\bigg[\psi(t)x(t)+ \hat{x}(t)\bigg(\dot x(t) + x(t) -F(y)-\theta(t) +i D\hat{x}\bigg)\bigg]\right\}\\ &\times \exp\left\{i \int~dt~\bigg[\phi(t)y(t)+\hat{y}(t)\bigg(y(t)-\chi(t) -\bar{J}\rho_x \bigg)\bigg]  \right\}\nonumber\\
    &\times \exp\left\{-\frac{g^2}{2}\int~dt~dt' ~Q_x~q_{\tau_0}(|t-t'|) \hat{y}(t)\hat{y}(t') \right\}\ .
    \end{aligned}
\end{equation}
By construction we have the normalization for $Z$, namely, $Z[\psi=0,\phi=0, \theta, \chi]= 1$. Therefore, 
\begin{equation}
 \langle \hat{x}(t) \rangle = i  \lim_{\psi\rightarrow 0, \phi\rightarrow 0} \frac{\delta Z[\psi,\phi, \theta, \chi]}{\delta \theta(t)} = \frac{\delta }{\delta \theta(t)}\left\{\lim_{\psi\rightarrow 0, \phi\rightarrow 0} \delta Z[\psi,\phi, \theta, \chi]\right\} =0 \,,\qquad  \langle \hat{y}(t) \rangle = i  \lim_{\psi\rightarrow 0, \phi\rightarrow 0} \frac{\delta Z[\psi,\phi, \theta, \chi]}{\delta \chi(t)} =0 \ .
\end{equation}
Therefore, we obtain
\begin{equation}
 \langle \hat{x}(t) \hat{y}(t') \rangle = i^2  \lim_{\psi\rightarrow 0, \phi\rightarrow 0} \frac{\delta^2 Z[\psi,\phi, \theta, \chi]}{\delta \theta(t) \delta \chi(t')} =0 \ .
\end{equation}

\section{Derivation of Entropy production rate~Eq.~(5)}
\label{app:EPR}
In this section, we compute the entropy production for the original dynamics~\eqref{eq:eqn1} and compare it with that of the effective dynamics~\eqref{effective}. In the original dynamics given by Eq.~\eqref{eq:eqn1}, $\zeta_i(t)$ is the Gaussian thermal white noise coming from the environment. The strength of this noise $\sigma^2$
is related to the environment's temperature $T$ by the Einstein fluctuation-dissipation relation as $\sigma^2 = 2k_BT$, where $k_B$ is Boltzmann's constant. (Notice that we have set the dissipation constant $\gamma=1$.) 
Since the elements $x_i$ in Eq.~\eqref{eq:eqn1}, are
coupled through a fully-connected network $\mathbf{J}$ of {\it nonreciprocal} interactions, i.e., $J_{ij}\neq J_{ji}$ for every pair $(i,j)$, the system breaks down the detailed balance condition. 
Therefore, the system constantly produces entropy, even in  non-equilibrium  steady states (NESS).

Following the decomposition of the total entropy production rate along a single stochastic trajectory~\cite{Seifert_2012}, we have
\begin{eqnarray}
   \dot{S}_{\rm tot} = \dot{S}_{\rm res}+\dot{S}_{\rm sys}\ ,
\end{eqnarray}
where $\dot{S}_{\rm res}(t)$ is the rate at which entropy is dissipated into the environment and $\dot{S}_{\rm sys}(t)$ is the change in the system's entropy, respectively, along a single stochastic trajectory, and these are given by 
\begin{align}
     \dot{S}_{\rm res}&=\displaystyle \frac{2}{\sigma^2}\sum_{i} \bigg(F\big[\sum_j J_{i,j}(t)x_j(t)\big] -x_i(t) \bigg)\circ \frac{\rmd x_i}{\rmd t}\ ,\label{entropy_reservoir}\\
     \dot{S}_{\rm sys}&=\displaystyle -k_B\sum_i\left(\frac{\partial}{\partial x_i} \ln p(\mathbf{x},t)\right)\circ \frac{\rmd x_i}{\rmd t}-k_B\frac{\partial}{\partial t}\ln p(\mathbf{x},t) \ .
\end{align}
Here ``$\circ$" denotes the Stratonovich convention, and $p(\boldsymbol{x},t)$ is the  joint probability distribution of $\boldsymbol{x}$ at time $t$. 
In the non-equilibrium steady state, the average rate of system entropy production vanishes, i.e., $\langle \dot{S}_{\rm sys}^{\rm (ss)} \rangle = 0$, and hence the average rate of total entropy production becomes equal to the average rate of environmental entropy production, i.e.,  
$\big\langle \dot{S}_{\rm tot}^{\rm (ss)}\big\rangle =\big\langle \dot{S}_{\rm res}^{\rm (ss)}\big\rangle$, where the $\langle\cdot\rangle$ represent the average taken over noise realizations.

Since the original dynamics in Eq.~\eqref{eq:eqn1} can be represented by a one-dimensional effective process~\eqref{si:eff-eqn}
the average of the rate of environmental entropy production in Eq.~\eqref{entropy_reservoir} can be  shown to be equivalent to that of 
the effective process~\eqref{si:eff-eqn}. Specifically, denoting $\dot{s}_{\rm res}(t):= \dot{S}_{\rm res}(t)/N$ as the rate of the environmental entropy production per particle for the original dynamics~\eqref{eq:eqn1}, we will prove 
that
\begin{equation}
\langle \dot{s}_{\rm res}(t) \rangle =\langle \dot{s}_{\rm res}(t) \rangle^{\rm (MF)}\ ,
\label{numerical_simulation_DMFT_entropy}
\end{equation}
where the right-hand side is the average rate of environmental entropy production of the mean field (effective) dynamics~\eqref{si:eff-eqn}. Notice that the angular brackets on the right-hand side indicate the average with respect to the measure associated with the effective dynamics~\eqref{si:eff-eqn}. Furthermore, similar to the rate of environmental entropy production~\eqref{entropy_reservoir} for the original dynamics,  
one can show that the stochastic rate of  environmental entropy production of the effective dynamics is
\begin{equation}
  \dot{s}_{\rm res}(t)  \equiv\frac{2}{\sigma^2}\big(F[y(t)] - x(t)\big)\circ \frac{\rmd x(t)}{\rmd t}\ ,\label{env-eff}
\end{equation}
 In the following, our aim is to show how to obtain Eq.~\eqref{env-eff}. 
To this end, we write the moment generating function for the rate of entropy production
\begin{align}
    \mathcal{Z}[\Lambda'] = \overline{\bigg\langle e^{i \sum_k \int~dt~\Lambda_k'(t)\dot{S}_k}\bigg\rangle}_{\rm paths}\ ,
\end{align}
where again the angular brackets are the average over the trajectories of the system, and the overline denotes the average over the disorder present in the dynamics. 
\begin{align}
\mathcal{Z}[\Lambda']&\equiv \overline{\langle (\cdots) \rangle_{\rm paths}}\nonumber\\
&= \int \mathcal{D}[x,\hat{x},y,\hat{y}]~\underbrace{\exp\bigg[i\sum_k \int~dt~\hat{x}_k(t)\bigg(\dot x_{k}(t) +x_k(t) -  F[y_k(t)] -\theta_k(t) + iD\hat{x}_k(t)\bigg) \bigg]}_{A_2[x,\hat{x},y,\hat{y}]} \nonumber\\
&\times\underbrace{\exp\bigg[i\sum_{k'} \int~dt~\hat{y}_{k'}(t)\bigg(y_{k'} - \chi_{k'}(t)\bigg)\bigg]}_{B_2[y,\hat{y}]}\underbrace{\overline{\exp\bigg[-i\sum_{k',j} \int~dt~\hat{y}_{k'}(t) J_{k',j}(t)~x_j(t)\bigg]}}_{\Delta[x,\hat{x},y,\hat{y}]}\nonumber\\
&\times e^{i\sum_k \int~dt~ \Lambda_k(t)  \dot{x}_k(t)\circ(F[y_k] - x_k)}\ ,
\label{decompo-2}
\end{align}
where we have substituted the value of the rate of environmental entropy production $\dot{S}_k$ for each $k$ [see Eq.~\eqref{entropy_reservoir}], and we define $\Lambda_k\equiv \Lambda'_k/D$ with $\Lambda'_k$ as the conjugate field of $\dot{x}_k\big(F[y_k] - x_k\big)$ in the Onsager-Machlup action representation of the dynamics~\eqref{eq:eqn1}. 
Note that in writing Eq.~\eqref{decompo-2}, we do not consider the change of the Shannon entropy along the evolution of the system, but only the rate of environmental entropy production $\dot{S}_{\rm res}$. This is because in the non-equilibrium steady state, the average change in the system entropy vanishes.  Therefore, we shall focus on calculating the entropy production in the reservoir $\dot{S}_{\rm res}$ only.  
Therefore, we simply rewrite the above equation as follows 
\begin{equation}
    \mathcal{Z}[\Lambda]= \int \mathcal{D}[x,\hat{x},y,\hat{y}]A_2[x,\hat{x},y,\hat{y}] B_2[y,\hat{y}] \Delta[x,\hat{x},y,\hat{y}]~e^{i\sum_k \int~dt~ \Lambda_k(t)  \dot{x}_k(t)\circ(F[y_k] - x_k)}\ .
\label{generating_function_for_the_entropy}
\end{equation} 
After repeating all the steps for computing $\overline{\langle (\cdots) \rangle_{\rm paths}}$ as in Sec.~\ref{app-eff-eqn},
in terms of the macroscopic quantities,
\begin{subequations}
\label{allequations7}
 \begin{eqnarray}
 \rho_x(t) & =&  \frac{1}{N}\sum_j x_j (t)  \,,\qquad \qquad\qquad \lambda_y(t)= \frac{1}{N}\sum_k \hat{y}_k (t)   \label{auxiliary3} 
  \\ Q_x(t,t') &= &  \frac{1}{N}\sum_{k} x_k(t) x_k(t') \label{auxiliary4}
\,,\qquad  L_y(t,t') =  \frac{1}{N}\sum_{k} \hat{y}_k(t) \hat{y}_k(t')   \label{auxiliary5}
\end{eqnarray}
\end{subequations}
the ensemble-averaged moment-generating functional reads as follows
\begin{equation} 
   \mathcal{Z}[\Lambda']= \int D\big[\Pi, \hat{\Pi}\big] ~e^{N[\Omega+ \Phi_{\rm tot}]}\label{generating_function2}\ ,
\end{equation}
where, as in the previous Section, we define $\Pi = (\rho_x,\lambda_y,Q_x,L_y)$ and $\hat{\Pi} = (\hat{\rho}_x,\hat{\lambda}_y,\hat{Q}_x,\hat{L}_y)$. In addition, we have
\begin{equation}
    \Phi_{\rm tot} \equiv -i \bar{J} \int dt\, \rho_x\lambda_y  - \frac{g^2}{2}\int dt \, dt' q_{\tau_0}(|t-t'|)Q_xL_y  +i \int dt\,\big[\hat{\rho}_x\rho_x + \hat{\lambda}_y\lambda_y \big]   +i\int dt\, dt' \big[\hat{Q}_xQ_x + \hat{L}_yL_y \big]\ ,
\label{generating_function3}
\end{equation}
and 
\begin{equation}
\Omega \equiv \dfrac{1}{N}\sum_k \ln I_k\ ,
\end{equation}
for
\begin{equation}
\begin{aligned}
I_k &\equiv \int~\mathcal{D}[\dots] \exp\bigg\{i \int~dt~S_k(t) -i \int\int~dt~dt'\Big[ \hat{Q}_x(t,t')x_k(t)x_k(t') +  \hat{L}_y \hat{y}_k(t) \hat{y}_k(t')\Big]\bigg\}\ ,\\
S_k&\equiv -\hat{\rho}_x x_k -\hat{\lambda}_y \hat{y}_k  +\hat{y}_ky_k +\hat{x}_k\Big(\dot x_{k} + x_{k} -F(y_k)\Big)+  iD\hat{x}^2_k(t)+\Lambda_k(t)  \dot{x}_k(t)\circ \Big(F(y_k) - x_k\Big) \ .
\label{generating_function4}
\end{aligned}
\end{equation}
Substituting the saddle-point values [similar to what we did in the previous section in Eqs.~\eqref{ex-1} and \eqref{ex-2}] of these auxiliary fields  $\Pi^*, \hat{\Pi}^*$, and noting that the $N$ identical integrals are uncoupled, we can simplify Eq. \eqref{generating_function3}-\eqref{generating_function4} to
\begin{equation}
\begin{aligned}
\Phi_{\rm tot} &= 0\ ,
\,  \\ 
I&=\int~\mathcal{D}[\dots] \exp\bigg\{i \int~dt~S(t) -\frac{g^2}{2} \int\int~dt~dt'~q_{\tau_0}(|t-t'|)Q_x \hat{y}(t) \hat{y}(t')\bigg\}\ ,\\ 
S&=-\bar{J}\rho_x\hat{y} +\hat{y}y +\hat{x}\Big(\dot x + x_{k} -F(y)\Big)+  i\frac{\sigma^2}{2}\, \hat{x}^2+\Lambda(t)  \dot{x}(t)\circ \Big(F(y) - x\Big)\ .
\label{generating_function5}
\end{aligned}
\end{equation}
Taking into account the above expressions leads to the final result
 \begin{equation}
 \begin{aligned}
     \mathcal{Z}[\Lambda']&=\left( \int D\big[\dots \big] e^\Theta\right)^N\ , \label{generating_function6}
     \end{aligned}
 \end{equation}
 where, for convenience, we define
 \begin{align}
     \Theta \equiv i\int~dt~\Big[\big(y-\bar{J}\rho_x \big)\hat{y} +\hat{x}\Big(\dot x + x -F(y)\Big)+  i\frac{\sigma^2}{2}\, \hat{x}^2+\beta\Lambda'  \dot{x}\circ \Big(F(y) - x\Big)\Big] -\frac{g^2}{2} \int\int ~dt~dt'~q_{\tau_0}(|t-t'|)  Q_x \hat{y}(t) \hat{y}(t')\ .
 \end{align}

 $\langle \dot{S}_{\rm res}(t)\rangle^{\rm (MF)}$ can be easily obtained from the above moment generating function:
\begin{align}
\left\langle\dot{S}_{\textrm{res}}(t)\right\rangle^{\rm (MF)} & =\beta~\left.\frac{1}{i}\frac{\delta \mathcal{Z}}{\delta \Lambda(t)}\right|_{\Lambda=0} =N \left\langle\dot{x}(t)\circ \big[F(y(t)) - x(t)\big]\right\rangle\ , \\
\left\langle\dot{s}_{\textrm{res}}(t)\right\rangle^{\rm (MF)} &= N^{-1}\left\langle\dot{S}_{\textrm{res}}(t)\right\rangle^{\rm (MF)}= \beta \Big[\big\langle x^2(t)\big\rangle + \big\langle F^2(y(t))\big\rangle -2 \big\langle x(t)\circ F(y(t))\big\rangle  + \big\langle \zeta(t)\circ F(y(t))\big\rangle   -  \big\langle \zeta(t)\circ x(t)\big\rangle  \Big
]\ ,
\label{entropy4}
\end{align}
where $\beta = 1/(k_{\rm B} T)$  and in the last line we have substituted the effective dynamics~\eqref{si:eff-eqn}.
The right-most term $\left\langle  x(t)\circ \zeta(t)\right\rangle$ has been obtained in Eq.~\eqref{x_linear_response}. However, the third term vanishes due to the fact that the external and effective noises are not correlated, as proved in Sec.~\ref{no_correlation_between_noises}. Together with all this, Eq.~\eqref{entropy4} reduces to Eq.~\eqref{EPR}.

\section{Time-dependent comparison}
\label{TDC}
\begin{figure}
    \centering
    \includegraphics[width=\textwidth]{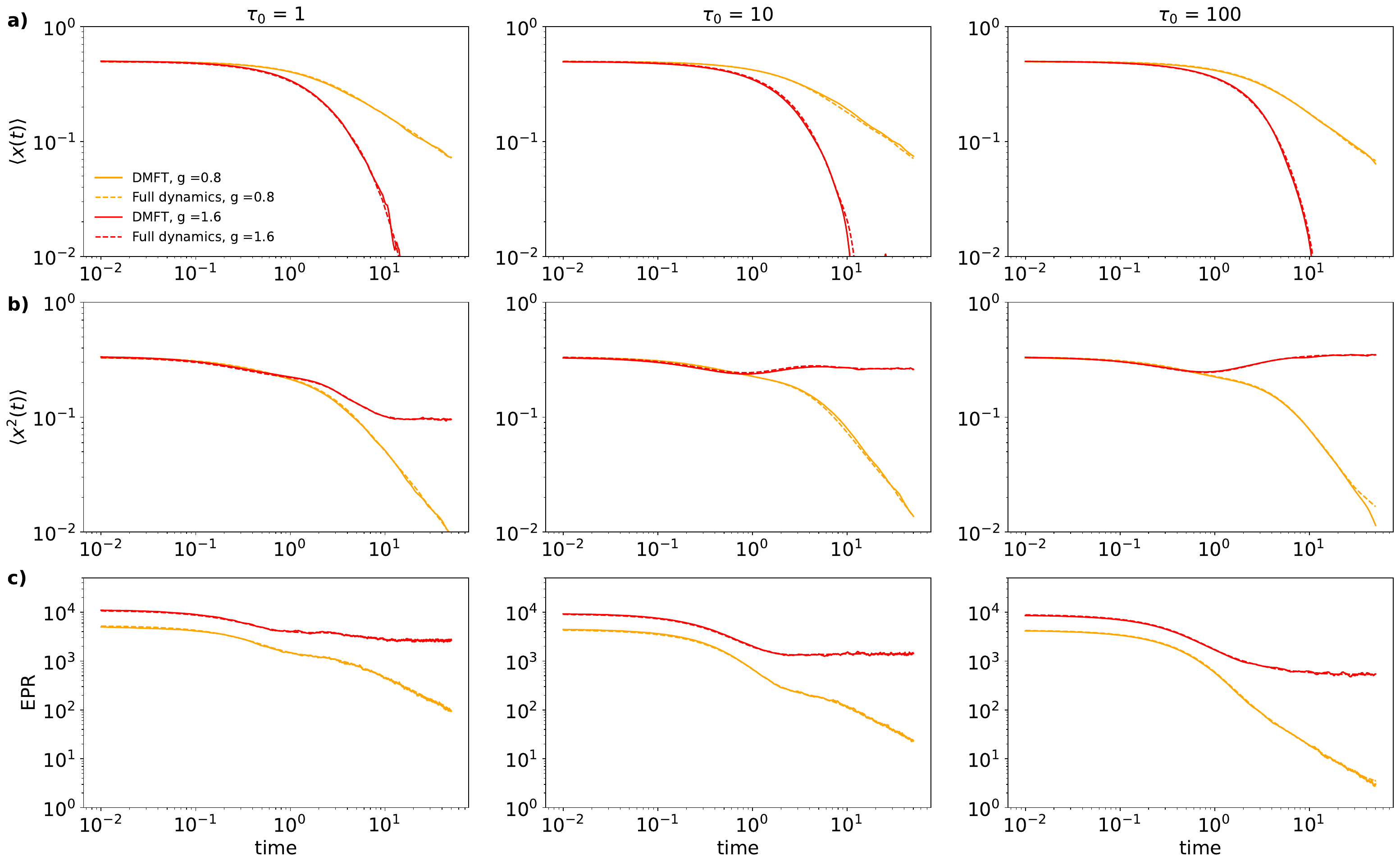}
    \caption{Comparison of average (a), second moment (b), and entropy production rate (EPR) (c) obtained using full dynamics~\eqref{eq:eqn1} and effective dynamics (DMFT equation)~\eqref{effective}, each as a function of time for three different correlation times $\tau_0=1,10,100$. For full dynamics: we show the scaled average (scaled by system's size $N$).
    Solid: DMFT. Dashed: Full dynamics. Other parameters: Number of particles $N = 2000$, discretization time $dt = 0.01$, temperature $T = 5\times 10^{-5}$, coupling parameter $\mu = 1$, number of realizations for full dynamics: 100, and number of iterations for DMFT: 1000.}
    \label{TD-comp-FD-and-DMFT}
\end{figure}

Figure~\ref{TD-comp-FD-and-DMFT} shows the comparison of $\langle x(t)\rangle$, the second moment $\langle x^2(t)\rangle$, and the entropy production rate obtained using full dynamics~\eqref{eq:eqn1} and effective dynamics (DMFT equation)~\eqref{effective}, each as a function of time for three different correlation times $\tau_0=1,10,100$.
 
\section{Comparison of the long-time behavior obtained by solving Eq.~(1) and Eq.~(3)}
\begin{figure}[!b]
    \centering
    \includegraphics[width=\textwidth]{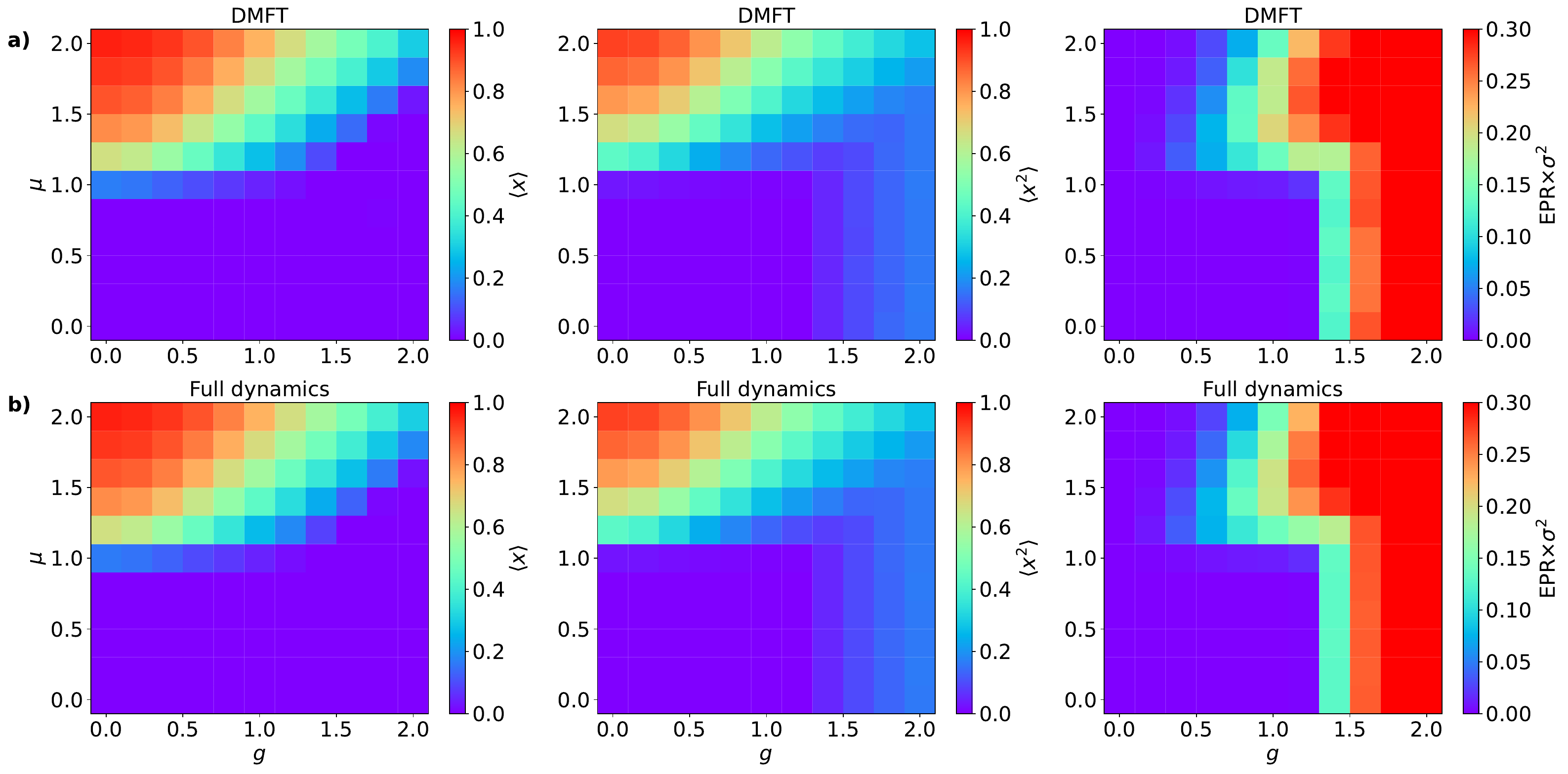}
    \caption{Comparison of scaled average, scaled second moment, and entropy production rate (EPR) obtained using full dynamics~\eqref{eq:eqn1} (panel a) and effective dynamics (DMFT equation)~\eqref{effective} (panel b) in $g,\mu$ plane at time $t=50$ and $\tau_0=1$. All other parameters are same as in Fig.~\ref{TD-comp-FD-and-DMFT}.}
    \label{fig:DMFT-SP-tau1}
\end{figure}

\begin{figure}
    \centering
    \includegraphics[width=\textwidth]{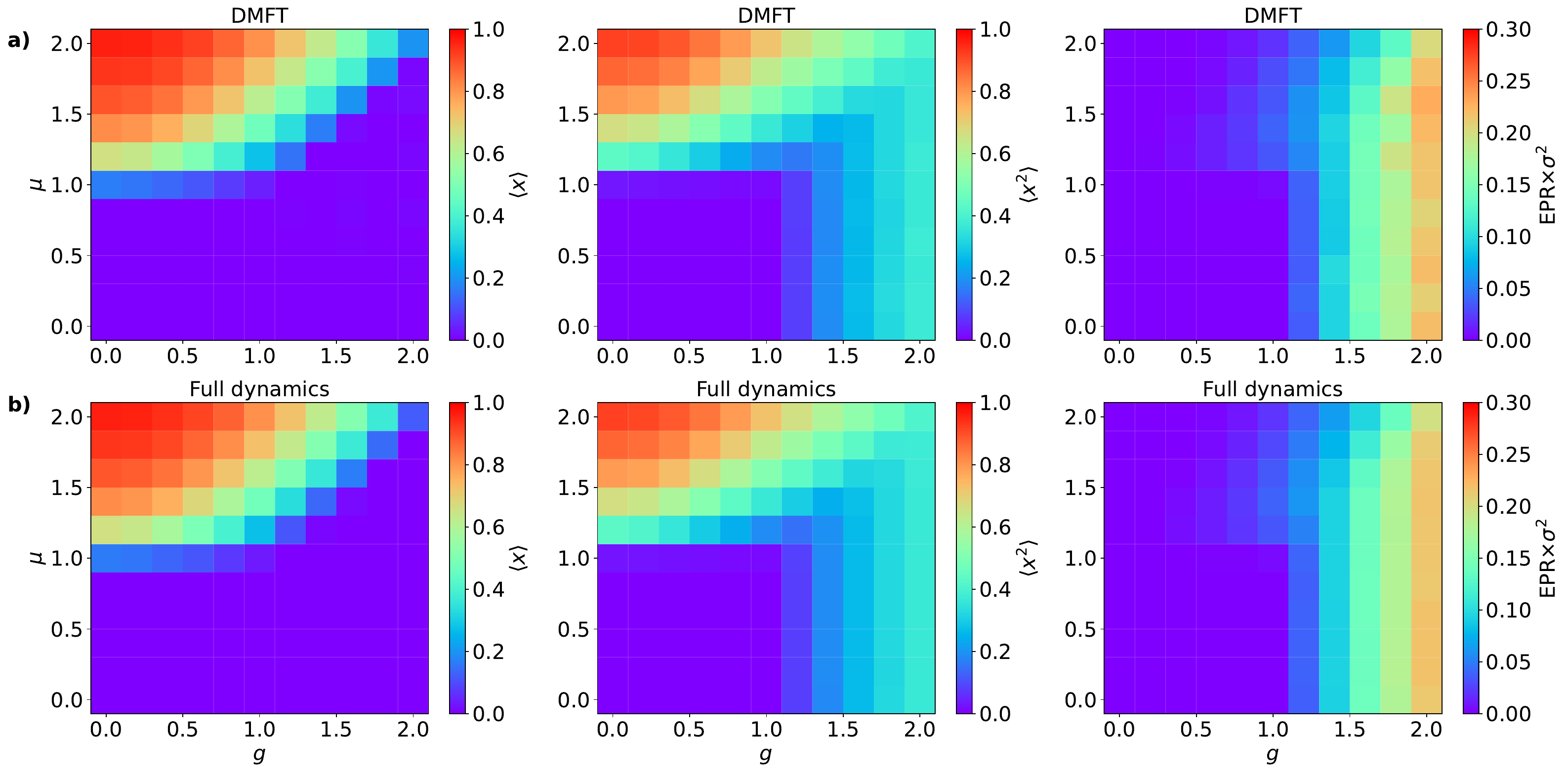}
    \caption{Comparison of scaled average, scaled second moment, and entropy production rate (EPR) obtained using full dynamics~\eqref{eq:eqn1} (panel a) and effective dynamics (DMFT equation)~\eqref{effective} (panel b) in $g,\mu$ plane at time $t=50$ and $\tau_0=10$. All other parameters are same as in Fig.~\ref{TD-comp-FD-and-DMFT}.}
    \label{fig:DMFT-SP-tau10}
\end{figure}

\begin{figure}[!b]
    \centering
    \includegraphics[width=\textwidth]{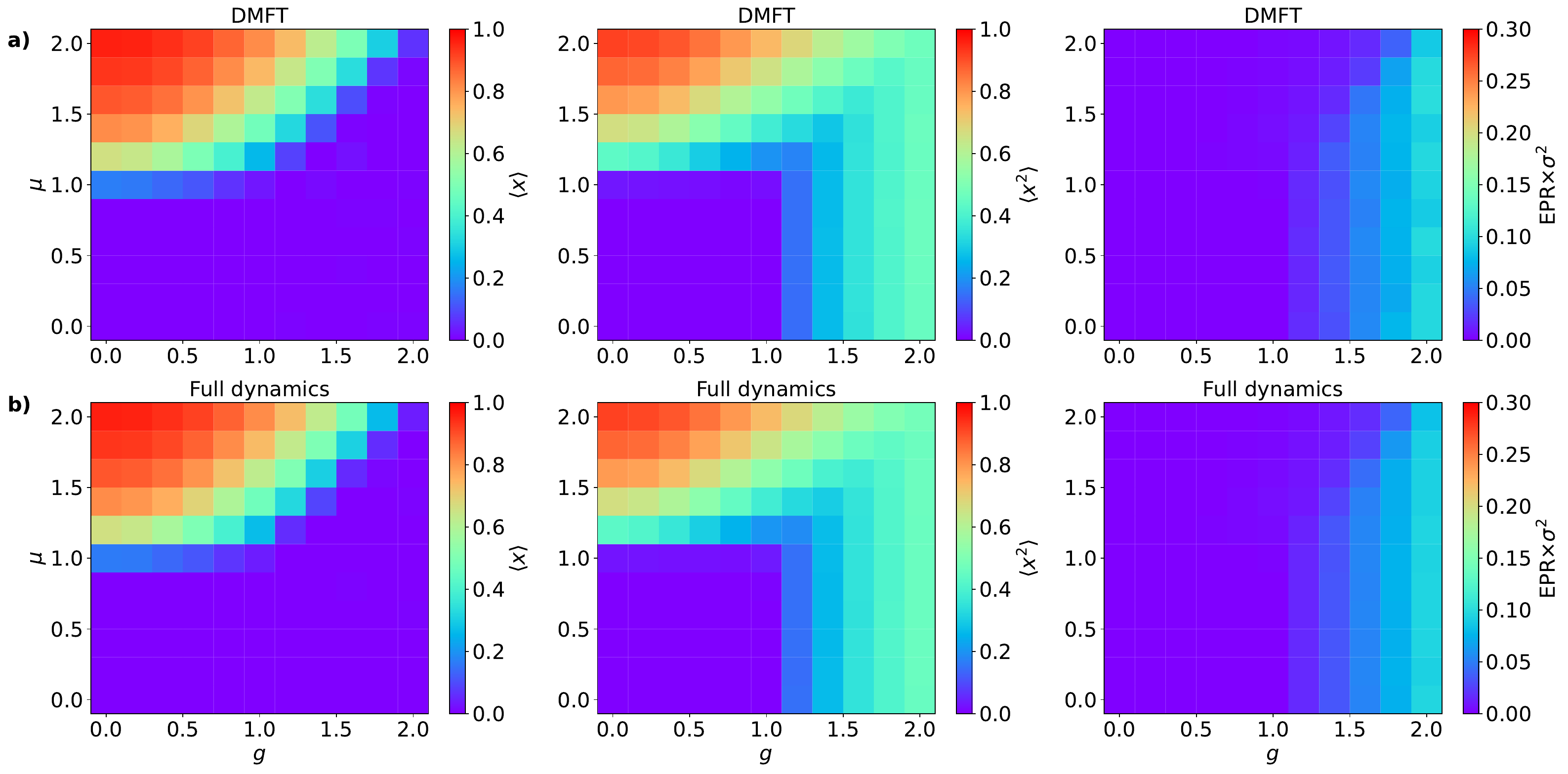}
    \caption{Comparison of scaled average, scaled second moment, and entropy production rate (EPR) obtained using full dynamics~\eqref{eq:eqn1} (panel a) and effective dynamics (DMFT equation)~\eqref{effective} (panel b) in $g,\mu$ plane at time $t=50$ and $\tau_0=100$. All other parameters are same as in Fig.~\ref{TD-comp-FD-and-DMFT}.}
    \label{fig:DMFT-SP-tau100}
\end{figure}

The long-time behavior of the scaled mean, the scaled second moment, and the entropy production rate obtained from the full dynamics~\eqref{eq:eqn1} and the effective dynamics~\eqref{effective}, as a function of $g$ and $\mu$, for different $\tau_0$ is shown in Figs.~\ref{fig:DMFT-SP-tau1}, \ref{fig:DMFT-SP-tau10}, and \ref{fig:DMFT-SP-tau100}.

\section{Example of the dependence on $\tau_0$}
\label{tau0_dependence}
\begin{figure}[!h]
    \centering
    \includegraphics[width=\textwidth]{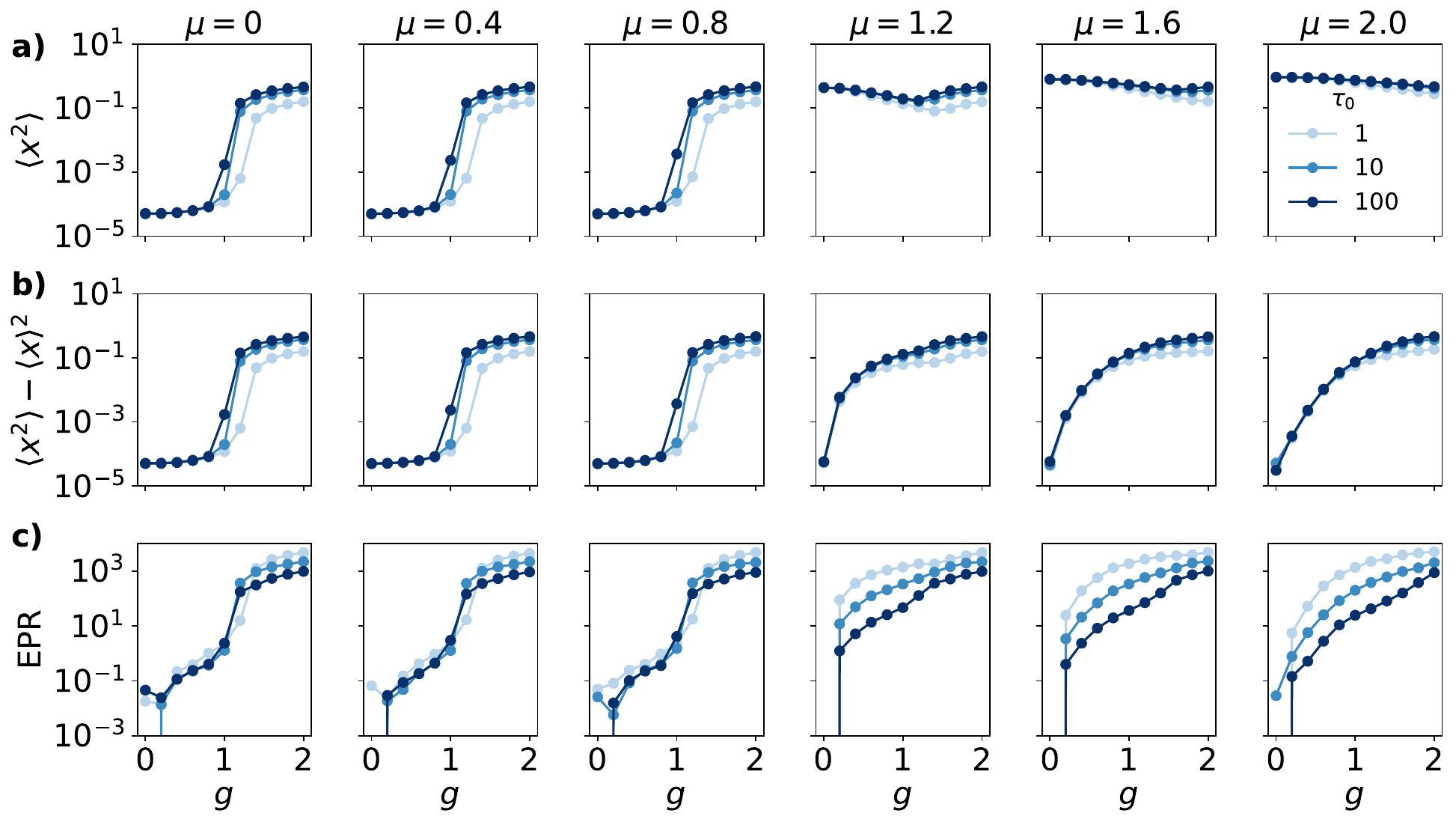}
    \caption{Long-time second moment $\langle xx \rangle$ (a), variance $\langle x x\rangle - \langle x\rangle^2$ (b), and entropy production rate (EPR) (c) as a function of $g$ for different $\mu$. Connecting lines are guide to the eye. The color intensity increases with increasing the value of $\tau_0.$}
\label{fig:compare1}
\end{figure}

\begin{figure}[!b]
    \centering
    \includegraphics[width=\textwidth]{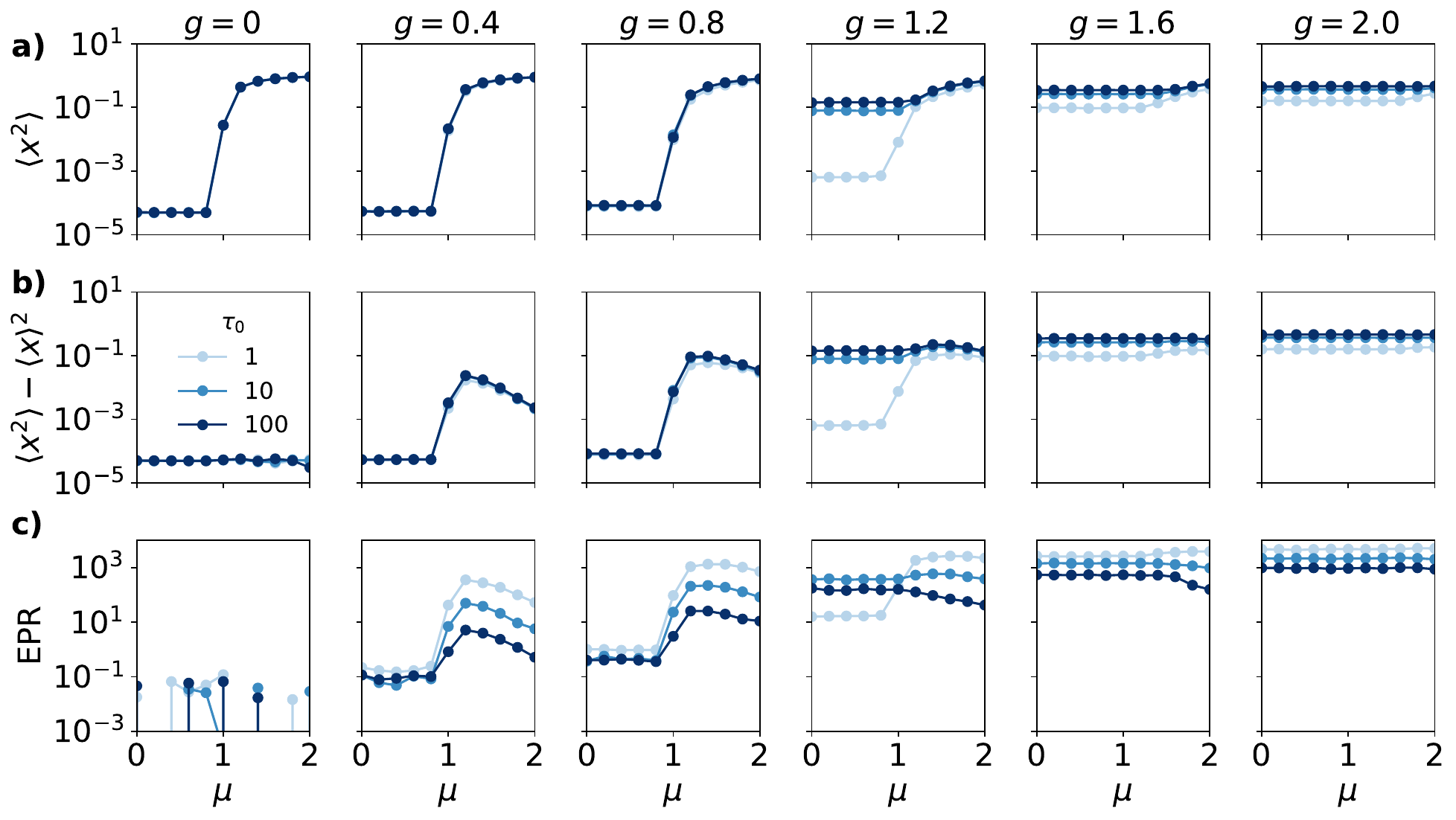}
    \caption{Long-time second moment $\langle xx \rangle$ (a), variance $\langle x x\rangle - \langle x\rangle^2$ (b), and entropy production rate (EPR) (c) as a function of $\mu$ for different $g$. Connecting lines are guide to the eye. The color intensity increases with increasing the value of $\tau_0.$}
\label{fig:compare2}
\end{figure}

Fig.~\ref{fig:compare1} shows the second moment, the variance, and the entropy production rate for three different $\tau_0$ by solving the effective equation~\eqref{effective}, as a function of $g$ for different $\mu$. Decreasing the correlation time $\tau_0$ increases the critical value of $g$, i.e., $g_c$, after which the second moment (or variance) becomes non-zero. Similarly, Fig.~\ref{fig:compare2} shows the behavior of these quantities for three different $\tau_0$ by solving the effective equation~\eqref{effective}, as a function of $\mu$ for different $g$, where for $g=0.4, 0.8, 1.2$ a non-monotonic dependence on $\mu$ is observed for the variance and the entropy production rate.

\section{Equation of motion  for NESS autocorrelation for arbitrary $\tau_0$}
\label{EOM_for_Cx}
 Let us derive equations for the evolution of the mean and the autocorrelation function. Specifically, for the mean, by averaging both sides of Eq.~\eqref{effective}, we get
\begin{equation}
       \langle  \dot{x}(t)\rangle^{\rm (MF)} = -\langle x(t)\rangle^{\rm (MF)} +\langle F[\mu \langle x \rangle^{\rm (MF)}(t) + g\eta(t)]\rangle^{\rm (MF)} 
       \label{EOM_for_M}\ .
\end{equation}
Hence, the stationary solution for the mean value satisfies
the nonlinear self-consistency equation
\begin{equation}
M\equiv \lim_{t\rightarrow \infty}\langle x(t)\rangle^{\rm (MF)} =\lim_{t\rightarrow \infty}\langle  F[\mu \langle x(t) \rangle^{\rm (MF)} + g\eta(t)]\rangle^{\rm (MF)} \ .
\label{mean}
\end{equation}

Defining $\delta x(t) = x(t) - \langle x(t)\rangle^{\rm (MF)}$ the deviation from the mean, we get a differential equation for the stationary
covariance function $C_x(t,t') = \langle \delta x(t)  \delta x(t')\rangle^{\rm (MF)}$.
The equation for $C_x$ is obtained by simple manipulations
\begin{equation}
 (\partial_t +1 )\delta x(t)= \underbrace{-\langle x(t)\rangle^{\rm (MF)} + F[\mu \langle x(t) \rangle^{\rm (MF)} + g\eta(t)]}_{\displaystyle \delta F[\cdot]\equiv - \langle F[\cdot]\rangle^{\rm (MF)} + F[\cdot]} + \zeta(t) \ ,
     \label{EOM_deviation}
\end{equation}
where, for convenience, we define $\partial_t\equiv d/dt$. 
Next, by multiplying Eq.~\eqref{EOM_deviation} with the same equation at different times $t'\neq t$ and taking the average, we get the following 
\begin{equation}
(\partial_t +1)(\partial_{t'} +1)\bar{C}_x(t,t')  = \bar{C}_F(t,t') +\sigma^2 \delta(t-t')\ ,
\label{two_time}
\end{equation}
where $\bar{C}_x(t,t') \equiv\langle \delta x(t) \delta x(t')\rangle$, $\bar{C}_F(t,t') \equiv   \left \langle \delta F[u(t)]\delta F[u(t')]\right\rangle$
for $u(t) \equiv \mu \langle x(t) \rangle^{\rm (MF)} + g\eta(t)$. Furthermore, to obtain Eq.~\eqref{two_time}, we have used the following relation  proved in Eq.~\eqref{most_important_identity}: $\langle \eta(t)\zeta(t')\rangle =0$.

In NESS (with time-translational invariance), for $\tau=t-t'\geq 0$, Eq. \eqref{two_time} becomes
\begin{equation}
\partial_\tau^2\bar{C}_x(\tau) = \bar{C}_x(\tau) - \bar{C}_F(\tau) -\sigma^2\delta(\tau)\ .
\label{EOM_for_variance}
\end{equation}
Furthermore, multiplying both sides of Eq.~\eqref{EOM_deviation} by $\delta x(t') = x(t') - \langle x(t') \rangle^{\rm (MF)}$ and then averaging, we get the following equation
\begin{equation}
    (\partial_t +1)\langle \delta x(t) \delta x(t')\rangle^{\rm (MF)} = \langle \delta F[u(t)] \delta x(t')\rangle^{\rm (MF)} + \langle \zeta(t) \delta x(t')\rangle^{\rm (MF)}\ ,
\end{equation}
where $\langle \zeta(t) \delta x(t')\rangle^{\rm (MF)} = 2D G_\theta(t',t) = 0$ is due to causality ($t'<t$), and the first equality is based on Eq.~\eqref{x_linear_response_unequal_time}. Then, in NESS with $t-t'=\tau>0$, this equation becomes
\begin{equation}
       \partial_\tau \bar{C}_x(\tau) = - \bar{C}_x(\tau) +  \bar{C}_{xF}(\tau)\ .\label{cbar-2eqn}
\end{equation}

\subsection{Derivation of EPR at NESS Eq.~(9)}
We next do a simple algebra to show how Eq.~\eqref{EPR} in the long-time limit ($t\to \infty$) from Eq.~\eqref{entropy4} can be equivalently written in terms of connected correlations $\bar{C}_x(t,t) \equiv C_x(t,t) - \langle x(t) \rangle^2$; $\bar{C}_{xF}(t,t) \equiv  \big\langle \delta x(t)\delta F(t)]\big\rangle$ and $\bar{C}_F(t,t) \equiv C_F(t,t) - \langle F(t) \rangle^2$. In fact, for $t\rightarrow \infty$
 \begin{align} \nonumber
 \lim_{t\rightarrow \infty}\bigg[C_x(t,t) + C_{F}(t,t) - 2C_{xF}(t,t)\bigg] &= \lim_{t\rightarrow \infty}\bigg[\bar{C}_x(t,t) +   \langle x(t) \rangle^2 +   \bar{C}_F(t,t)  + \langle F(t) \rangle^2 - 2  \bigg\langle \big[\langle x(t) \rangle + \delta x(t)\big] \big[\langle F(t) \rangle +\delta F(t)\big]\bigg\rangle \bigg] \\ &= \lim_{t\rightarrow\infty} \bigg[\bar{C}_x(t,t) +     \bar{C}_F(t,t)  - 2 \bar{C}_{xF}(t,t) + \underbrace{\langle x(t) \rangle^2 +  \langle F(t) \rangle^2 - 2\langle x(t) \rangle \langle F(t) \rangle}_{=0 \,{\rm due\,to\,Eq.~\eqref{mean}}} \bigg] \ .
 \end{align}
Then, using Eqs.~\eqref{EOM_for_variance} and \eqref{cbar-2eqn}, we can show EPR~\eqref{NESS_entropy} in NESS, as discussed in the main text.

\section{Stability analysis and EPR for the linear model}
\label{app:linear}
In the following section, we specialize to the case of linear dynamics $F(z)=z$. Here, we have the following 
\begin{align}
    F[u(t)] -\langle F[u(t)]\rangle^{\rm (MF)} = u(t) - \langle u(t)\rangle^{\rm (MF)} = \mu \langle x (t)\rangle^{\rm (MF)} + g\eta(t) - \mu \langle x \rangle^{\rm (MF)}(t) =g\eta(t)
\end{align} as $\langle\eta(t)\rangle^{\rm (MF)} = 0$. 
Therefore, $\bar{C}_F(t,t') = g^2q_{\tau_0}(|t-t'|) C_x(t,t') = g^2q_{\tau_0}(|t-t'|) \bar{C}_x + g^2q_{\tau_0}(|t-t'|) M(t)M(t')$ 
subsequently,
\begin{equation}
\partial_\tau^2\bar{C}_x(\tau) =\big[1 - g^2q_{\tau_0}(|\tau|)\big] \bar{C}_x(\tau) -g^2M^2q_{\tau_0}(|\tau|) -\sigma^2\delta(\tau)\ ,
\label{full_equation}
\end{equation}
where we write the above equation in stationarity for which $M$ becomes time-independent.
In the following, we discuss the case for $M=0$ in the stationary state for the linear system. In general, for a linear system, the DMFT equation~\eqref{effective} reduces to
\begin{align}
    \dot x(t) = -x(t) + \mu \langle x\rangle(t) + g \eta(t) + \zeta(t)\ , \label{eff-sieqn}
\end{align}
where ${M}(t) = \langle x\rangle(t)$. Averaging the above equation~\eqref{eff-sieqn}, we obtain
\begin{align}
    \dot{M}(t) = - (1 - \mu){M}(t)\ .
\end{align}
Its solution is 
\begin{align}
    M(t) = M_0 e^{-(1-\mu)t}\ .
    \label{Mzero_for_linear}
\end{align}
Therefore, for $\mu < 1$, $M(t)$ reaches a stationary state and then $M(t\to \infty) \to 0$, which is consistent with~\cite{Ferraro_2025}. In the following, we shall only be interested in this stationary case with $\mu<1$.


To understand the transition from the fixed point to chaos for the model $F(z)=z$, we consider Eq.~\eqref{main:EOM_for_Cx} by substituting $F(z)=z$. This yields, 
\begin{equation}  \partial_\tau^2\bar{C}_x(\tau) = [1-g^2q_{\tau_0}(\tau)]\bar{C}_x(\tau)  -\sigma^2\delta(\tau)\ .
\label{formal_linear}
\end{equation}
Changing variable $u= 2\tau_0 g\sqrt{q_{\tau_0}(0)} e^{-\tau/(2\tau_0)}$ for $\tau>0$ translates Eq.~\eqref{formal_linear} into the  following Bessel equation:
\begin{equation}
    u^2\partial_u^2\bar{C}_x(u) +u\partial_u \bar{C}_x(u) +(u^2-\nu^2)\bar{C}_x(u) = 0\ ,
\end{equation}
where we defined $\nu\equiv 2\tau_0$. 
 Thus, for any $\tau$, we obtain the  solution with $\dot{\bar{C}}_x(0^+) = -\sigma^2/2$ in terms of the Bessel function $J_\nu$ as follows:
\begin{equation}
   \bar{C}_x(\tau) = \frac{\sigma^2}{2 g \sqrt{q_{\tau_0}(0)}}~\frac{J_{\nu}\Big(2\tau_0 g\sqrt{q_{\tau_0}(0)} e^{-|\tau|/(2\tau_0)}\Big)}{J_{\nu}'\Big(2\tau_0 g\sqrt{q_{\tau_0}(0)}\Big)}\ .
   \label{Bessel}
\end{equation}
\textcolor{black}{For stochastic differential equations with additive noise, such as our model~\eqref{eq:eqn1}, linear stability analysis corresponds to requiring fluctuations around a given fixed point $x_*$ of the drift part to remain bounded, i.e., $\bar{C}_x(0) < \infty$. From Eq.~\eqref{Bessel} it is trivial to see that $\bar{C}_x(0)$ would diverge if $J_{\nu}'\Big(2\tau_0 g\sqrt{q_{\tau_0}(0)}\Big) =0$, and this condition, which is also in agreement with Ref.~\cite{Ferraro_2025}, is displayed in Fig.~\ref{fig:region-plot}. Thus, combining the conditions $\mu=1$ and $J_{\nu}'\Big(2\tau_0 g\sqrt{q_{\tau_0}(0)}\Big) =0$ provides the phase diagram for the linear model $F(z)=z$.} 

\begin{figure}
    \centering
\includegraphics[width=0.6\columnwidth]{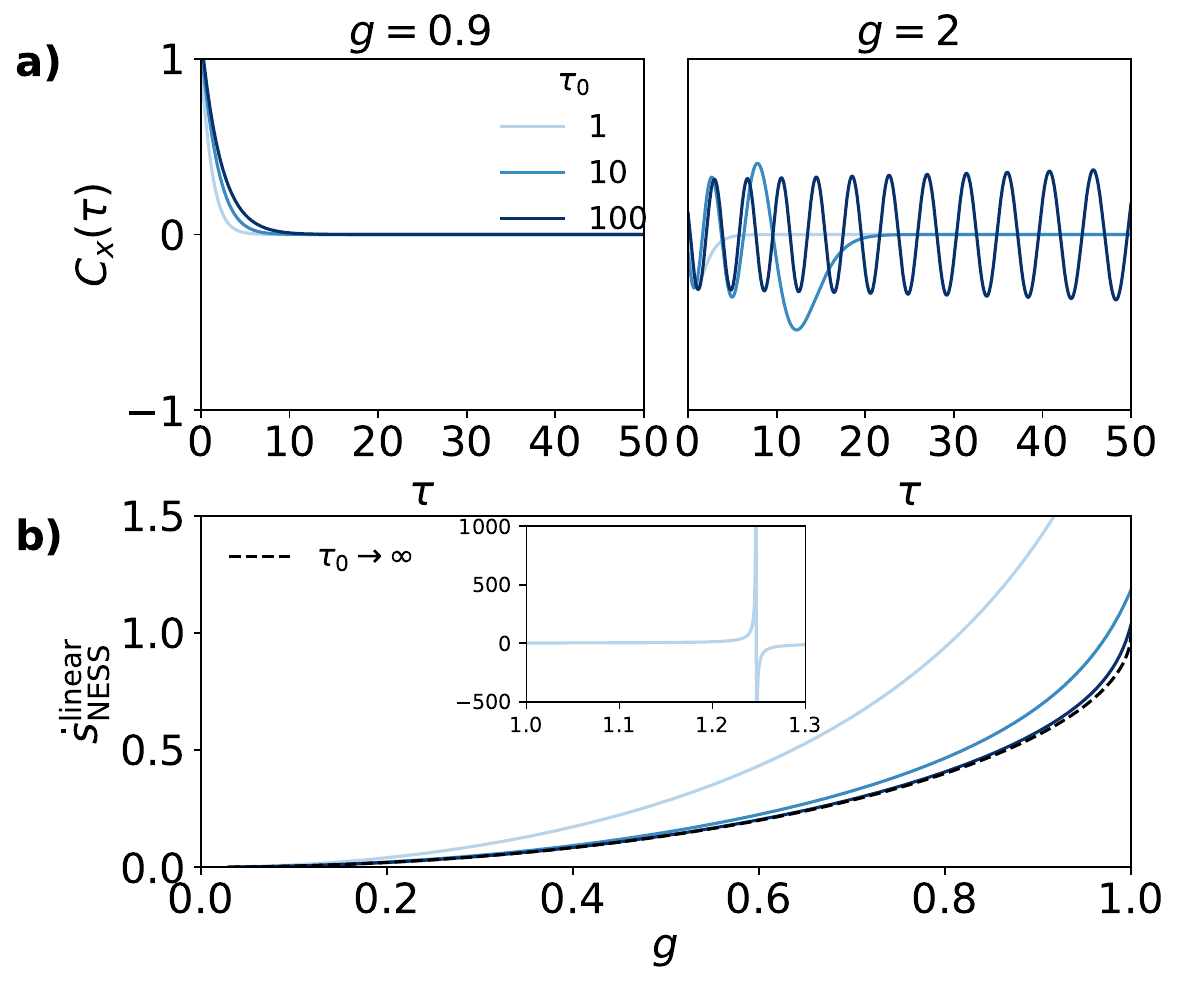}
    \caption{a) Correlation function $C_x(\tau)$~\eqref{Bessel} as a function of time $\tau$ for two different values of $g$ and $\sigma=1$. b) EPR~\eqref{final} as function of $g$. The black dashed line is for the quenched case $\tau_0\to \infty$. Inset: Data for $\tau_0=1$ for $g\geq 1$.  In both panels, the blue color intensity increases with $\tau_0=1,10,100$.}
    \label{fig:Bessel_solution}
\end{figure}

Figure~\ref{fig:Bessel_solution} shows the exponentially decaying solution~\eqref{Bessel} of Eq.~\eqref{formal_linear}, which corresponds to a physically acceptable solution of the original dynamics at $g=0.9$. In contrast, at $g=2$, the solution remains mathematically valid for Eq.~\eqref{formal_linear} but is not physically admissible. In this case, $\bar{C}_x(\tau=0)$ becomes negative for $\tau_0=1,10$, while for $\tau_0=100$ it grows unboundedly as $|\tau|\rightarrow\infty$. 
In fact, although Eq.~\eqref{Bessel} is a formal solution of Eq.~\eqref{formal_linear}, it does not necessarily correspond to a physical solution of the original dynamics~\eqref{eq:eqn1} for $F[z]=z$, since there is no finite stationary correlation beyond the stability line~(see following subsections for our self-contained presentation).

\subsection{Entropy production rate of a linear model $F(z)=z$}
\label{sec:linear-entropy}

Given the correlation function~\eqref{Bessel}
using Eq.~\eqref{NESS_entropy}, we obtain the exact expression for steady-state EPR:  
\begin{align}
   \dot{s}_{\rm NESS}^{\rm (linear)} = 1 +\frac{g^2q_{\tau_0}(0)-1}{g \sqrt{q_{\tau_0}(0)}}~\frac{J_{\nu}\Big(2\tau_0 g\sqrt{q_{\tau_0}(0)}\Big) }{J_{\nu}'\Big(2\tau_0 g\sqrt{q_{\tau_0}(0)}\Big)}\ ,
   \label{final}
\end{align}
where the right-hand side can also be written as $1 + [g^2q_{\tau_0}(0)-1]~\bar{C}_x(0)/T$, showing a linear dependence of EPR on the variance of $x$ as seen previously in Fig.~\ref{fig:epr-var-para}.
We plot this expression in Fig.~\ref{fig:Bessel_solution}b for the case in which $\bar{C}_x(\tau)$ decays exponentially, i.e., the original system relaxes to a stable quasi fixed-point. However, this expression is no longer valid if the solution given in Eq~\eqref{Bessel} becomes unphysical. We demonstrate the singular behavior of the EPR upon crossing this stability line in Fig.~\ref{fig:Bessel_solution}b(inset)  for $\tau_0=1$. Similar behavior is also observed for other $\tau_0=10, 100$.

Finally, we perform asymptotic analyses of EPR~\eqref{final} to approximate the behavior of EPR in the limits: 1) $g\to 0$, 2) $\tau_0\to 0$, and 3) $\tau_0\to\infty$. In the limit $g \to 0$, we  show that $J_\nu(z)/J'_\nu(z) \sim z/\nu$ for $z\equiv 2\tau_0 g\sqrt{q_{\tau_0}(0)}\to 0$. Substituting this on the right-hand side of Eq~\eqref{final}, we find $  \dot{s}_{\rm NESS}^{\rm (linear)}\sim g^2 q_{\tau_0}(0)$. For $\tau_0\to 0$, considering $z = 2\tau_0 g\sqrt{q_{\tau_0}(0)}$ as a small parameter for $\nu = 2\tau_0$, again gives $\dot{s}_{\rm NESS}^{\rm (linear)}\sim g^2 q_{\tau_0}(0)$, which diverges as $\tau_0\to 0$. For $\tau_0\to \infty$, we have $z\to \infty$ and $\nu\to \infty$, but $z/\nu = g\sqrt{q_{\tau_0}(0)}\to g$ is a fixed number. Now, using the asymptotic expression in~\cite{gradshteyn2014table} for $g<1$, we obtain $\lim_{\nu\to\infty}[J_\nu(z)/J'_\nu(z)] = g/\sqrt{1-g^2}$. Then, in this limit $\dot{s}_{\rm NESS}^{\rm (linear)}\sim 1- \sqrt{1-g^2}$~[Fig.~\ref{fig:Bessel_solution}b]. Note that the above derivations are consistent with the alternative derivations of the behavior of EPR in $\tau_0\to 0$ and $\tau_0\to \infty$ shown below in Secs.~\ref{app-tau0infty} and \ref{app-tau00}.

\subsubsection{Quenched limit: $\tau_0\to \infty$}
\label{app-tau0infty}
It is instructive to check  whether we can recover the known result for the quenched disorder such as $\tau_0\rightarrow \infty$, for which we have $q_{\tau_0}(|\tau|)=1$. Then, from Eq.~\eqref{formal_linear}, we have
\begin{equation}
\partial_\tau^2\bar{C}_x(\tau) =(1 - g^2) \bar{C}_x(\tau)  -\sigma^2\delta(\tau)\ ,\label{eq:lin-corr}
\end{equation}
with the initial condition $\dot{C}_x(\tau\rightarrow 0^+) = -\sigma^2/2$. We Fourier transform the above equation~\eqref{eq:lin-corr}, which gives 
\begin{align}
     \hat{\bar{C}}_x(\omega)  =  \dfrac{\sigma^2}{\omega^2 -(1-g^2)} \ . \label{eq:FT-corr}
\end{align}
The inverse Fourier transform of Eq.~\eqref{eq:FT-corr} depends on whether $g^2>1$ or $g^2<1$. Thus, we get
\begin{align}
    \bar{C}(\tau) = \begin{cases}
        \dfrac{\sigma^2}{2}~\dfrac{1}{\displaystyle \sqrt{1-g^2}}~\exp\left\{-|\tau| \sqrt{1-g^2} \right\} &\qquad g^2 < 1\\
        \dfrac{\sigma^2}{2}~\dfrac{1}{\displaystyle \sqrt{g^2-1} }~{\rm sin}\left\{|\tau| \sqrt{g^2-1} \right\}&\qquad g^2 > 1
    \end{cases}\ .\label{eq:I2}
\end{align}
We can simply check that both of these solutions satisfy $\dot{C}_x(\tau\rightarrow 0^+) = -\sigma^2/2$. 
However, for case $g^2>1$, the correlation function is oscillatory, which is not a physical solution of Eq.~\eqref{eq:lin-corr} as this equation is valid for the stationary state.  
Then, substituting the expression of $\bar{C}(\tau)$ into Eq.~\eqref{NESS_entropy} for $g^2<1$, we arrive at the following expression for the entropy production rate:
\begin{align}
     \langle \dot{s}_{\rm res}(t) \rangle^{\rm (MF)}_{\rm NESS}=1-\sqrt{1 - g^2}\ .
\end{align}
This agrees with the asymptotic analysis of the entropy production rate for the linear model in Sec.~\ref{sec:linear-entropy}.

\subsection{$\tau_0\to 0$ limit}
\label{app-tau00}
In the limit $\tau_0\to 0$, we have $q_{\tau_0}(|\tau|) = \delta(\tau)$. From Eq.~\eqref{formal_linear}, the evolution of the correlation function is given by
\begin{equation}
\partial_\tau^2\bar{C}_x(\tau) =\big[1 - g^2 \delta(\tau)\big] \bar{C}_x(\tau)  -\sigma^2\delta(\tau)\ ,
\label{full_equation}
\end{equation}
The Fourier transform of above equation~\eqref{full_equation} gives
\begin{align}
\hat{\bar{C}}_x(\omega) = \dfrac{g^2   \bar{C}_x(0)  + \sigma^2}{1+\omega^2}\ ,
\end{align}
and then inverting the Fourier transform, we obtain
\begin{align}
\bar{C}_x(\tau) = [g^2 \bar{C}_x(0) + \sigma^2]\dfrac{e^{-|\tau|}}{2}\ .
\end{align}
Substituting $\tau=0$ on both sides, we obtain the variance as follows
\begin{equation}
    Q\equiv\bar{C}_x(0) = \frac{  \sigma^2}{2-g^2}\ .
    \label{NESS_variance_singular}
\end{equation}
The above expression~\eqref{NESS_variance_singular} is in agreement with that observed in~\cite{Ferraro_2025}, where the latter shows that the variance vanishes in the stationary state in the absence of external thermal noise. Furthermore, the positivity of $\bar{C}_x(0)$ implies that $g^2 <2$.
so that the DMFT $x$-dynamics behaves like an  Ornstein–Uhlenbeck process with
\begin{align}   
\bar{C}_x(\tau)=\frac{\sigma ^2}{2-g^2}e^{-|\tau|}\ ,
\end{align}
while for $g^2>2$, our assumption that time-translationally invariance that $C_x(\tau)$ depends only on $\tau=t-t'$ and remains bounded, is no longer valid. We  note by passing that in the $\sigma\rightarrow 0$ limit, the stationary distribution becomes  a delta function.

 Then, given the formula of the EPR~\eqref{NESS_entropy},  we find
\begin{align}
    \langle \dot{s}_{\rm res}(t) \rangle^{\rm (MF)}_{\rm NESS} 
    =1-\frac{\sigma ^2}{\left(2-g^2\right)T}
   =\dfrac{g^2}{g^2-2}
\end{align}
where we used $\sigma^2 = 2T$. The above expression shows that the EPR is positive only when $g^2 > 2$. This seems to contradict the condition of $g^2<2$ for having a well-defined variance $\bar{C}_x(0)$ in Eq.~\eqref{NESS_variance_singular}, hinting at a non-trivial limiting behavior of EPR in this case.  It turns out that in this case the EPR diverges, and therefore, the formula~\eqref{NESS_entropy} fails to predict the entropy production.

To analyze this divergence carefully, we begin by using Eq.~\eqref{entropy4} for the linear case. It reads for $y(t) \equiv \mu M(t) + g  \eta(t)$:
\begin{equation}
    \langle \dot{s}_{\rm res} \rangle^{\rm (MF)} = \beta \Big[\big\langle x^2(t)\big\rangle + \big\langle y(t)^2\big\rangle -2 \big\langle x(t)\circ y(t)\big\rangle  + \big\langle \zeta(t)\circ y(t)\big\rangle   -  \big\langle \zeta(t)\circ x(t)\big\rangle  \Big
]\ ,
\label{entropy_linear}
\end{equation}
 To evaluate the right-hand side of the above equation~\eqref{entropy_linear}, we do the regularization to take care of the double limit $\tau_0\rightarrow 0$ and $\tau\rightarrow 0^+$. To this end, we remind ourself that $y(t)$ is actually our OU noise that represents the original term $\sum_{j}J_{ij}(t)x_j$ in Eq.~\eqref{eq:eqn1} in the limit of $N\rightarrow \infty$. Since the $\tau_0\rightarrow 0$ will result in a singular $\delta$-function term when we take $t\rightarrow \infty$ for the equal-time correlators in Eq.~\eqref{entropy_linear}, we first need to replace $x(t)$ by $x_{\tau_0} (t)$ and similarly, $y(t)$ by $y_{\tau_0}(t)$, in computing EPR using Eq.~\eqref{entropy_linear}
then take $\tau_0\rightarrow 0$ afterwards. So our DMFT system in NESS Eq.~\eqref{eff-sieqn} becomes 
\begin{equation} 
\left \{ \begin{aligned} \displaystyle
    \dot{x}_{\tau_0} &= - x_{\tau_0}(t) + y_{\tau_0}(t)
+ \zeta(t)
\\  \displaystyle  \dot{y}_{\tau_0} & = -\frac{y_{\tau_0}}{\tau_0} + \frac{1}{\tau_0}~g\sqrt{Q}\xi(t)\end{aligned} \right.\  , 
\label{regularised_ODEs}
\end{equation}
where we have removed the time-dependence of $Q(t)$ by its stationary time-independent value $Q$ given in Eq.~\eqref{NESS_variance_singular}.
In this formulation, the NESS restriction of Eq.~\eqref{entropy_linear} requires one to compute three correlators:
\begin{equation}
   Q_{\tau_0}\equiv \lim_{t\rightarrow \infty}  \langle x^2_{\tau_0}(t)\rangle\,,\qquad Y_{\tau_0}\equiv \lim_{t\rightarrow \infty}  \langle y^2_{\tau_0}(t)\rangle\,,\qquad R_{\tau_0}\equiv \lim_{t\rightarrow \infty}  \langle x_{\tau_0}(t)\circ y_{\tau_0}(t)\rangle
\end{equation}
We have straightforwardly  that $\lim_{\tau_0 \rightarrow 0}Q_{\tau_0 } = Q$  from Eq.~\eqref{NESS_variance_singular}, while from the above definition and the  OU process for $y_{\tau_0}$~\eqref{regularised_ODEs}, we know already 
\begin{equation}
   Y_{\tau_0} = \frac{g^2Q}{2\tau_0} \ .
\end{equation}
Since $y_{\tau_0}$ has  a finite correlation time $\tau_0>0$, we have
\begin{equation}
   \big\langle x_{\tau_0}(t)\circ y_{\tau_0}(t)\big\rangle = \lim_{\tau\rightarrow 0^+}  \big\langle x_{\tau_0}(t) y_{\tau_0}(t+\tau)\big\rangle=  \lim_{\tau\rightarrow 0^+} \big\langle x_{\tau_0}(t+\tau) y_{\tau_0}(t)\big\rangle  = \big\langle x_{\tau_0}(t) y_{\tau_0}(t)\big\rangle\ .
\end{equation}
From the regularized stochastic differential equations~\eqref{regularised_ODEs}
\begin{equation}  
\dot{x}_{\tau_0} = - x_{\tau_0}(t) + y_{\tau_0}(t)
+ \zeta(t) \Rightarrow x_{\tau_0}(t) = \int_{0}^\infty ds~ e^{-s}\Big[y_{\tau_0}(t-s)
+ \zeta(t-s) \Big]\ .
\end{equation}
Therefore, [again $\langle y_{\tau_0}(t) \zeta(t')\rangle =0$ -- this term vanishes in virtue of our central identity Eq.~\eqref{most_important_identity}]
\begin{equation}
    \langle x_{\tau_0}(t) y_{\tau_0}(t)\big\rangle =  \int_{0}^\infty ds~e^{-s}\Big\langle y_{\tau_0}(t-s) y_{\tau_0}(t)
 \Big\rangle = \int_{0}^\infty ds~e^{-s} \frac{g^2Q}{2\tau_0}~e^{-|s|/\tau_0} = \frac{\tau_0}{\tau_0+1}~\frac{g^2Q}{2\tau_0}=\frac{g^2Q}{2(\tau_0+1)}
\end{equation}
Putting all that we get  into the  NESS restriction of Eq.~\eqref{entropy_linear} and taking the $\tau_0 \rightarrow 0$ limit afterwards  yields 
\begin{equation}
\begin{aligned}
    \langle \dot{s}_{\rm res} \rangle^{\rm (MF)}_{\rm NESS}&= \beta \lim_{\tau_0 \rightarrow 0} \left\{\lim_{t\rightarrow\infty} \Big[\big\langle x_{\tau_0}^2(t)\big\rangle + \big\langle y_{\tau_0}^2\big\rangle -2 \big\langle x_{\tau_0}(t)\circ y_{\tau_0}(t)\big\rangle  + \underbrace{\big\langle \zeta(t)\circ y_{\tau_0}(t)\big\rangle}_{=0}   -  \underbrace{\big\langle \zeta(t)\circ x_{\tau_0}(t)\big\rangle}_{=T}  \Big]\right\}\\ &= \beta  \Big[\underbrace{Q}_{\langle x^2\rangle} + \lim_{\tau_0 \rightarrow 0} \left(\underbrace{\frac{g^2 Q}{2\tau_0}}_{\langle y^2\rangle} -2\underbrace{\frac{g^2Q}{2(\tau_0+1)}}_{\langle xy\rangle} \right)  -  T \Big] = \dfrac{1}{T}\lim_{\tau_0 \rightarrow 0} \underbrace{\frac{g^2 Q}{2\tau_0}}_{\langle y^2\rangle} - \frac{\sigma^2g^2}{2T(2-g^2)} \rightarrow \infty \ .
\end{aligned}
\end{equation}
Alternatively to the regularization approach of coupled stochastic differential equations~\eqref{regularised_ODEs}, by  defining entropy production with respect to the thermal bath only, we can use Eq.~\eqref{env-eff} to compute EPR. In this case, it is even much more straightforward to show the same thing
\begin{equation}
    \langle \dot{s}_{\rm res} \rangle^{\rm (MF)}_{\rm NESS} = T^{-1}\langle (y - x)\circ \dot{x}\rangle = T^{-1}\langle y\circ \dot{x}\rangle= T^{-1}\underbrace{\langle y^2\rangle}_{g^2Q \delta(0)} - T^{-1}\langle xy\rangle \rightarrow \infty\ , \label{EPRinfty}
\end{equation}
where the second term vanishes because $\langle x\circ \dot{x}\rangle = d \langle x^2\rangle/(2dt) = 0$ in NESS.
In both derivations, we find that the EPR diverges due to the divergence of $\langle y_{\tau_0}^2\rangle$ when $\tau_0\rightarrow 0$. Physically, this means that the housekeeping cost of maintaining an infinitely fast stochastic protocol diverges, despite a well-defined long-term dynamics of 
$x$ that converge to finite variance.

\begin{figure}
    \centering
    \includegraphics[width=0.6\textwidth]{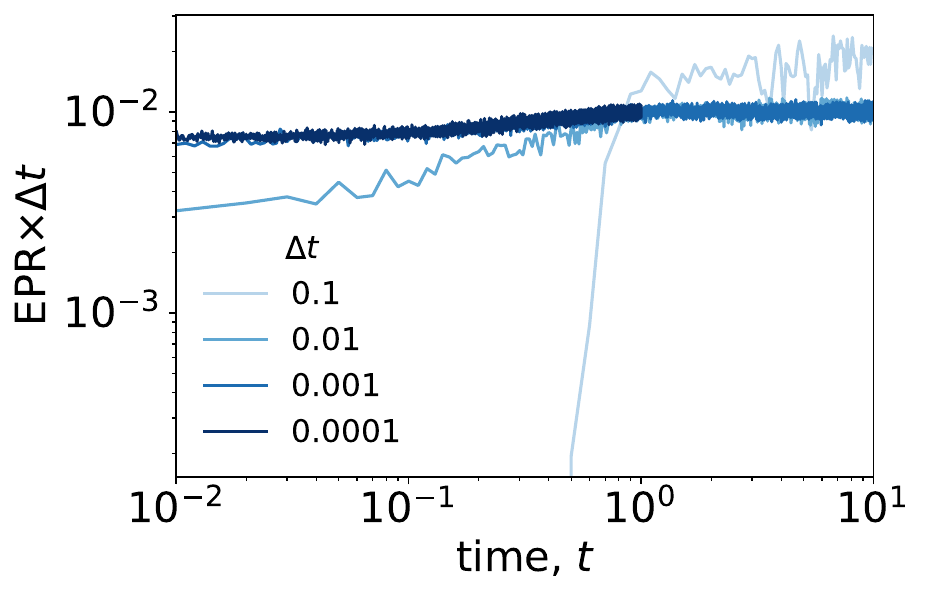}
    \caption{Case: $\tau_0\to 0$. Entropy production rate (EPR) times discretization time ($\Delta t$) as a function of observation time $t$. We simulate the dynamics~\eqref{eff-sieqn} to compute the entropy production rate evaluated using Eq.~\eqref{EPR}. The color intensity increases with decreasing $\Delta t$. Here, we fixed $g=\mu=0.1$. The other parameters for these numerical simulations are same as in the main text. (For $\Delta t= 10^{-4}$, EPR is evaluated up to time $t=1$; the observation time higher than this is computationally expensive.)}
    \label{fig:EPRdt}
\end{figure}

Figure~\ref{fig:EPRdt} shows the behavior of the EPR for different discretization times $\Delta t$, as a function of the observation time $t$. This confirms the divergence of the entropy production rate as predicted by Eq.~\eqref{EPRinfty}. (Notice that the negative EPR values at early times for $\Delta t = 0.1$ are numerical artifacts arising from the breakdown of the Euler discretization scheme at large time steps.) This is again in agreement with the asymptotic analysis discussed in Sec.~\ref{sec:linear-entropy} for EPR. 


{\color{black}
\section{Transitions from fixed point to chaos and from symmetric to symmetry-broken solutions in the presence of noise $\sigma>0$}
\label{lyapunov_analysis}

In the following Subsecs.~\ref{sec:fixed-point-m-0} and \ref{lyapunov_analysis-main}, we discuss two different types of instability, namely:
\begin{enumerate}
        \renewcommand{\labelenumi}{(\roman{enumi})}
        \item Destabilization of the  symmetric  fixed-point $M_*=0$. Strictly speaking, under noise $\sigma>0$, the dynamics do not relax to a fixed point but to an ensemble of stochastic trajectories, whose mean is non-vanishing $M(t\rightarrow \infty)>0$.
        \item Transverse instability results in a transition from fixed-point to chaotic behavior.
\end{enumerate}

The joint action of these instabilities gives rise to two different dynamical phases that are $\tau$-dependent solutions of Eq.~\eqref{main:EOM_for_Cx}, namely, $(M=0, C_x(0)>0, C_x(\infty) \rightarrow 0)$ -- the so-called \textit{asynchronous} chaos, and $(M>0, C_x(0)>0, C_x(\infty) > M^2)$ -- the so-called \textit{synchronous} chaos [note the difference between synchronous chaos and the $\tau$-independent ferromagnetic solution with $C_x(0) = C_x(\tau) =C_x(\infty) = Q, \forall~\tau$].

Finally, in Subsec.~\ref{stability}, we discuss the linear stability analysis for our system in the limit $\tau\to \infty$ for fixed $\sigma=0$, where we show that the linear stability analysis provides a lower bound to the boundary obtained in the above point (ii).

\subsection{Boundary from condition (i)}
\label{sec:fixed-point-m-0}

For (i), we consider a time-dependent perturbation around  $M_*$ as $M(t) = M_* + \epsilon~\delta M(t)$ for $\epsilon\ll 1$. To understand the instability of $M_*$, we require this perturbation $\delta M(t)$ not to die out as $t\rightarrow \infty$. We proceed by linearizing Eq.~\eqref{EOM_for_M}, for $F()=\tanh()$, to an order $\epsilon$.
\begin{align}
     \frac{d}{dt}\delta M(t) =  - \delta M(t) + \mu \delta M(t) \langle 1 - \tanh^2[\mu M_* + g \eta(t)]\rangle\ ,\label{eq:fxdpnt}
\end{align}
where we substituted the equation for the fixed point $M_* = \langle \tanh[\mu M_* + g \eta(t)]\rangle.$ At this point, we remind ourselves that the two-time correlation function for $\eta(t)$, at the stationary state, is
\begin{align}
    \langle \eta(t)\eta(t') \rangle = q_{\tau_0}(|t-t'|) C_x(t-t')\ .\label{twopoint-corr}
\end{align}
Now, we evaluate the right-most term of Eq.~\eqref{eq:fxdpnt}
\begin{align}
 \langle \tanh^2(\mu M_* + g \eta(t))\rangle  = \langle [\mu M_* + g \eta(t)]^2 
 -\frac{2}{3}[\mu M_* + g \eta(t)]^4 +\frac{17}{45}[\mu M_* + g \eta(t)]^6 + \dots] \rangle  
\end{align}
It turns out that the average of each term can be calculated by employing Wick's theorem and Eq.~\eqref{twopoint-corr}
\begin{align}
    \langle [\mu M_*+ g \eta(t)]^n\rangle 
    &=  \langle [\mu M_*+ g z \sqrt{q_{\tau_0}(0)C_x(0)}]^n\rangle \ \qquad n\geq 2\ , 
\end{align}
 where $z$ is a Gaussian random variable with zero mean and unit variance.
Therefore, Eq.~\eqref{eq:fxdpnt} becomes
\begin{align}
    \dfrac{d}{dt}\delta M(t) = -\delta M(t) + \mu \delta M(t) \big\langle 1- \tanh^2[\mu M_* + z g\sqrt{q_{\tau_0}(0)~C_x(0)}]\big\rangle\ . \label{fineqndm}
\end{align}

We are interested in the stability of the fixed point $M_*=0$. Therefore, the above Eq.~\eqref{fineqndm} gives us the condition for the instability of the fixed point $M_*$:
\begin{equation}
    \frac{1}{\mu} = \bigg\langle 1 - \tanh^2[z~g\sqrt{q_{\tau_0}(0)~C_x(0)}] \bigg\rangle\ ,
    \label{eq:m-stab}
\end{equation}
where $C_x(0)$ is the solution of Eq.~\eqref{EOM_for_variance} with the initial condition $\dot{C}_x(\tau\rightarrow 0^+) = -\sigma^2/2$. Fig.~\ref{fig:stability-plot} displays the stability contour dividing the stability of the fixed point $M_*=0$ using Eq.~\eqref{eq:m-stab}.

\subsection{Boundary from condition (ii)}
\label{lyapunov_analysis-main}
For (ii), we  generalize the calculation of the  Lyapunov exponent, first pioneered by Sompolinsky et al \cite{Sompolinsky} and later detailed in~\cite{Schuecker2018}. Specifically, we need to discuss the temporal evolution of the distance $d(t)$ between two stochastic trajectories, referred to as two replicas. This distance can be expressed in terms of  the cross-correlation function between the two replicas, $\bar{C}^{12}_x(t,s)$, and their respective autocorrelation functions $\bar{C}^{11}_x(t,s) = \bar{C}^{22}_x(t,s)= \bar{C}_x(t-s)$, where $\bar{C}_x(t-s)$ denotes the background stationary solution of Eq.~\eqref{EOM_for_variance}. A small perturbation around this solution of the form $\bar{C}^{12}_x(t,s) = \bar{C}_x(t-s) + \epsilon k^{(1)}(t,s) $ allows us to study the non-stationary behavior as well as compute $d(t) = -2\epsilon k^{(1)}(t,t)$. The latter quantifies the growth rate of the distance between the two replicas. The non-stationary correlation functions of these replicas, subjected to exactly the same realization of the noises $\{\boldsymbol{\xi}, \boldsymbol{\zeta}\}$, satisfy the following equations similar to Eq.~\eqref{two_time} with $\alpha, \beta\in \{1,2\}$:
\begin{equation}
    (\partial_t +1)(\partial_{t'} +1)\bar{C}^{\alpha \beta}_x(t,t')  = \bar{C}^{\alpha \beta}_F(t,t') +\sigma^2 \delta(t-t')\ ,
\label{two_replica}
\end{equation}
where  $\bar{C}^{\alpha\beta}_F(t,t') \equiv   \left \langle \delta F[u_\alpha(t)]\delta F[u_\beta(t')]\right\rangle$
for $u_\alpha(t) \equiv \mu \langle x_\alpha(t) \rangle^{\rm (MF)} + g\eta_\alpha(t)$ with
\begin{equation}
  \big\langle \eta_\alpha(t)\eta_\beta(t')\big\rangle = q_{\tau_0}(|t-t'|) \big\langle x_\alpha(t)x_\beta(t')\big\rangle\ .
\end{equation}
Expanding  up to linear order in $\epsilon$:
\begin{equation}
\bar{C}^{\alpha\beta}_F(t,s) \simeq \bar{C}_F(t-s) +~\epsilon k^{(1)}(t,s)~\dfrac{\partial \bar{C}^{\alpha\beta}_F(t,s)} {\partial \bar{C}^{\alpha\beta}_x(t,s)}\bigg|_{\displaystyle\bar{C}^{\alpha\beta}_x(t,s) = \bar{C}_x(t-s)}\ .
\end{equation}
Then, using Price's theorem, namely:
\begin{equation}
    \frac{\partial} {\partial \bar{C}^{\alpha\beta}_x(t-t')}\Big\langle \delta F[u_\alpha(t-t')]\delta F[u_\beta(0)]\Big\rangle_{\eta_\alpha, \eta_\beta} = \Big\langle \delta F'[u_\alpha(t-t')]\delta F'[u_\beta(0)]\Big\rangle_{\eta_\alpha, \eta_\beta}\ ,
\end{equation}
where $F'(z)\equiv  dF/dz$. Inserting this expression into Eq.~\eqref{two_replica} and comparing the terms of order $\epsilon$ leads to 
\begin{equation}
    (\partial_t +1)(\partial_{t'} +1)k^{(1)}(t,t')  = \Big\langle \delta F'[u_\alpha(t-t')]\delta F'[u_\beta(0)]\Big\rangle k^{(1)}(t,t')
\end{equation}
Under the standard separation ansatz used in quantum mechanics, namely 
\begin{equation}
    k^{(1)}(T,\tau)= e^{\lambda T}\psi(\tau)\ ,
    \label{k_ansatz}
\end{equation}
for $T=t + t'$ and $\tau = t- t'$ we arrive at the following Schrödinger equation for $\psi(\tau)$, that generalized the one of \cite{Schuecker2018} for annealed case: 
\begin{equation}
    \Big[-\partial_\tau^2 +1 - g^2q_{\tau_0}(\tau)K(\tau)\Big]\psi(\tau) = \underbrace{\left[1- \big(\lambda +1\big)^2\right]}_{\displaystyle \equiv E}\psi(\tau)\ ,
    \label{Schrodinger}
\end{equation}
where $\tau$ now plays the role of a spatial coordinate. 
In Eq.~\eqref{Schrodinger}, we defined
\begin{equation}
\begin{aligned}
    K(\tau)&\equiv \int\int~\frac{d\eta(\tau) d\eta(0)}{2 \pi  \sqrt{A^2-B^2}}\exp \left\{-\displaystyle\frac{A\big(\eta^2(\tau)+\eta^2(0)\big)-2 B \eta(\tau) \eta(0) (\tau) }{2 (A^2-B^2)}\right\} \delta F'(\mu M + g \eta(\tau))\delta  F'(\mu M + g \eta(0))\\ A :=& q_{\tau_0}(0)\bar{C}_x(0)\,,\qquad B:=q_{\tau_0}(\tau)\bar{C}_x(\tau)
    \end{aligned}
\end{equation}
By introducing $\eta(0) = z_2\sqrt{A}$ and $\eta(\tau) = \rho_\tau\eta(0) + z_1\sqrt{A(1-\rho_\tau^2)}$, with $\rho_\tau= B/A$ and $z_1, z_2$ as independent $\mathcal{N}(0,1)$ variables, this form is equivalent to the following, according to Fubini’s theorem~\cite{martorell2023dynamically}.
\begin{equation}
    K(\tau):= \int\int Dz_1 Dz_2~ \delta F'\left(g  \sqrt{q_{\tau_0}(0)C(0)}\left\{z_1 \sqrt{1- \left(\dfrac{q_{\tau_0}(\tau)C(\tau)}{q_{\tau_0}(0)C(0)}\right)^2} + z_2~\dfrac{q_{\tau_0}(\tau)C(\tau)}{q_{\tau_0}(0)C(0)} \right\}+\mu M\right)\delta  F'\Big(g  \sqrt{q_{\tau_0}(0)C(0)}z_2 +\mu M \Big)\ ,
    \label{gain}
\end{equation}
where $Dz \equiv e^{-z^2/2}/\sqrt{2\pi}$. Note that the Jacobian of the transformation from $(z_1, z_2)$ to $(\eta(0), \eta(\tau))^T$ is equal to $A\sqrt{1-\rho_\tau^2}$, just canceling with the Gaussian normalization.

The ground-state energy $E_0$ of the above Schrödinger operator determines the growth rate $k^{(1)}(t,t)$ in the limit of $t\rightarrow \infty$, and hence the maximum Lyapunov exponent $\lambda_{\rm max}$ via
\begin{equation}
  \lambda_{\rm max} = - 1 + \sqrt{1-E_0} \ . 
  \label{lyapunov_exponent}
\end{equation}
If $\lambda_{\rm max}>0$, then according to Eq.~\eqref{k_ansatz}, $k(t,t)$ grows in the asymptotic limit $t\rightarrow \infty$. 
The DMFT dynamics hence become chaotic if $E_0 < 0$. The condition $E_0 = 0$ is what determines the transition from a fixed point to chaos.  The calculation of $E_0$ is straightforward, as it basically consists of 2 steps: (i) First, solving Eq.~\eqref{EOM_for_variance} to obtain $\bar{C}_x(\tau)$, from which $K(\tau)$ can be computed for any given $F()$ via the standard Gauss–Hermite quadrature; and subsequently solving Eq.~\eqref{Schrodinger} via a finite-difference discretization. Figure~\ref{fig:lambdamax-LSA} shows the boundary $\lambda_{\rm max}=0$ obtained using Eq.~\eqref{lyapunov_exponent} for $\sigma=0.01$ and $\sigma=0.1$, where  the transition toward chaos is shifted rightwards as $\tau_0$ decreases.

\subsection{Numerical recipe to solve Eq.~(S110) and finding spectrum of Eq.~(S155)}
\label{sec:algorithm}
\begin{algorithm}[H]
\caption{Solving Eq.~(S110) and finding spectrum of Eq.~(S155)}
\label{alg:solving}
\begin{minipage}{.96\columnwidth}
\begin{algorithmic}[1]
\Require Parameters $\mu,g,\tau_0,\sigma$. Choose, for example, $\alpha=0.2$  and number of {\rm iterations} = 1000.
\State Initialize $M$ and $C_x(\tau)$ 
\While{${\rm iteration}<  {\rm number~of~iterations}$}
    \State Compute $M^{\rm upd}$ using Eq.~\eqref{mean} and $\bar{C}_F(\tau)$
    \State  Do  Fourier transform for
            $\bar{C}_F(\tau) \rightarrow \hat{C}_F(\omega)$
and from Eq.~\eqref{EOM_for_variance}
            $\hat{C}_x(\omega)
              = [ \hat{C}_F(\omega) + \sigma^2]
             / [1 + \omega^2]$
  \State Inverse Fourier transform with  enforcing evenness to get $\bar{C}^{\rm upd}_x(\tau)$
    \State Update $M$ and $\bar{C}_x(\tau)$ using
    $\bar{C}_x(\tau) = (1-\alpha) \bar{C}_x(\tau) + \alpha\bar{C}^{\rm upd}_x(\tau)$ and  $M = (1-\alpha) M + \alpha M^{\rm upd}$
\EndWhile
\State \Return $(M,\bar{C}_x(\tau))$
\State Compute $K(\tau)$ using Eq.~\eqref{gain} to build the Schrödinger operator in Eq.~\eqref{Schrodinger} 
\State Solve Eq.~\eqref{Schrodinger}  on $\tau \in [0,L]$ with the
    boundary conditions:
  $\psi'(0) = 0$ and 
        $\psi(L) = 0$ 
\State Output $E_0$ 
\end{algorithmic}
\end{minipage}
\end{algorithm}

We remark that there is a numerical convergence weakness in this iterative scheme (due to critical slowing down) near the transition from $M=0$ to $M>0$. Therefore, close to this boundary, one needs to increase the number of iterations to find the root of Eq.~\eqref{mean}.}

\subsection{Linear stability analysis in the $\tau_0\to \infty$ limit and for $\sigma=0$}
\label{stability}

In the following, we discuss the fixed points and their stability in the quenched limit $\tau_0\to \infty$. To this end, we begin by writing the effective mean field equation in the absence of external noise (i.e., $\sigma = 0$)
\begin{align}
    \dot x(t) &= -x(t) +F[\mu M(t) + g \eta(t)]\ ,\label{eff-eqn-si}
\end{align}
where we recall that the correlation
\begin{align}
 C_\eta(t,t')\equiv \langle \eta(t)\eta(t')\rangle^{\rm (MF)} &= q_{\tau_0}(|t-t'|) C_x(t,t')\ ,
\end{align}
for which 
\begin{align}
    q_{\tau_0}(|t-t'|)=\dfrac{1+2\tau_0}{2\tau_0}e^{-|t-t'|/\tau_0}
\end{align}

In this case, $q_{\tau_0}(|t-t'|)\to 1$, i.e., the quenched noise limit, we have 
\begin{align}
 C_\eta(t,t') &= \langle x(t)x(t') \rangle^{\rm (MF)} \ .\label{corr-eta-x}
\end{align}

\begin{figure}[!h]
    \centering
    \includegraphics[width=\textwidth]{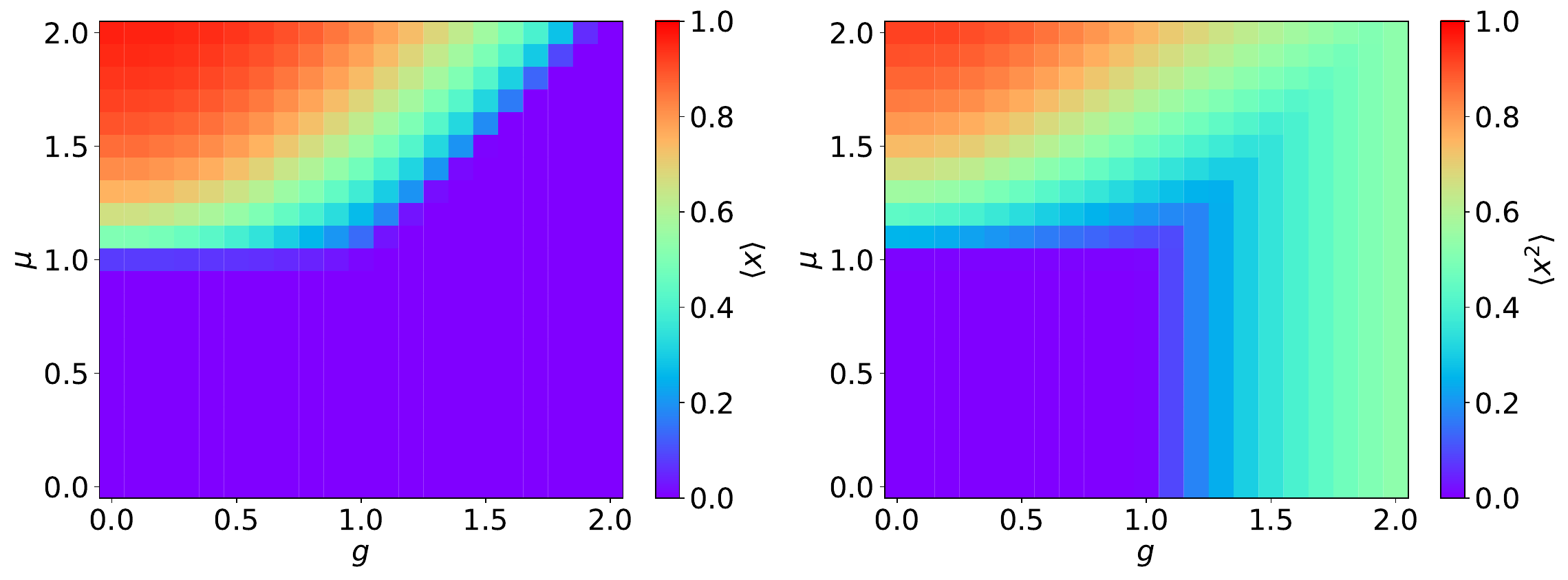}
    \caption{Phase diagram of the mean and  second moment obtained by simultaneously numerically solving Eqs.~\eqref{eq:si-mean}-\eqref{eq:si-secmom}. }
    \label{fig:quench-PD}
\end{figure}

In any of  the (multiple) stationary states $x(t\rightarrow \infty)\to x^*$, the left-hand side of Eq.~\eqref{eff-eqn-si} vanishes. Therefore,
\begin{align}
C_\eta(t,t') &= \langle x(t)x(t') \rangle^{\rm (MF)} \to \big\langle (x^*)^2\big\rangle^{\rm (MF)} = Q
\end{align}
where $Q$ is the second moment in the stationary state taken with respect to the distribution of the noise $\eta$. Due to the above self-consistency relation between $x(t)$ and $\eta(t)$, $\eta(t)$ also needs to approach a {\it random} stationary state $\eta^*$ that corresponds to $x(t\rightarrow \infty)\to x^*$ (this can be shown by considering the equality $\langle [\eta(t)-\eta(t')]^2 \rangle =0$, which implies that $\phi(t)=\phi(t')$). Therefore, we can define $\eta(t\rightarrow \infty)\to \eta^*\equiv\sqrt{Q} z$, where $z$ is a quenched Gaussian random variable with zero mean and unit variance.

Let $M_* = \langle x^*\rangle^{\rm (MF)}$ be the mean value of these stationary states. Therefore, in the stationary state, we have 
\begin{align}
    x^*(z) = F[\mu M_* + g\sqrt{Q} z]\ .
\end{align}
In what follows, we consider a specific form of $F$, i.e., $F = \tanh(\cdot)$. Therefore, the $z$-dependent stationary state  becomes
\begin{align}
    x^*(z) = \tanh[\mu M_* + g\sqrt{Q} z]\ .
    \label{random_fixed_point}
\end{align}
To investigate the stability of this stationary state, we introduce a small noisy perturbation $\epsilon\xi(t)$ as follows 
\begin{align}
  \dot x(t) = -x(t) + \tanh[\mu M(t) + g\eta(t)+ \epsilon \xi(t)] \ , \label{mfeq}  
\end{align}
and then expand around the given fixed point $(x^*,\eta^*)$ as is done in the linear stability analysis,
\begin{subequations}
\label{ptrb}
\begin{align}
    x(t) = x^* + \epsilon X(t)\ ,\\
    \eta(t) = \eta^* + \epsilon \Phi(t) \ .
\end{align}    
\end{subequations}

Therefore, $\langle X(t)\rangle^{\rm (MF)} = \langle \Phi(t)\rangle^{\rm (MF)} = 0$, while the second moment becomes
\begin{align}
    \langle x(t)x(t') \rangle^{\rm (MF)} = Q +   \epsilon M_* \langle X(t)\rangle^{\rm (MF)} +\epsilon M_* \langle X(t')\rangle^{\rm (MF)}+ \epsilon^2 \langle X(t)X(t')\rangle^{\rm (MF)}\ .
\end{align}
Similarly, we can do the same for $\eta(t)$. Since the correlations are equal~\eqref{corr-eta-x}, of the order of $\epsilon^2$, we can show that 
\begin{align}
    \langle X(t)X(t')\rangle^{\rm (MF)} = \langle \Phi(t)\Phi(t')\rangle^{\rm (MF)}\ .
\end{align}

\begin{figure}
    \centering
    \includegraphics[width=0.5\linewidth]{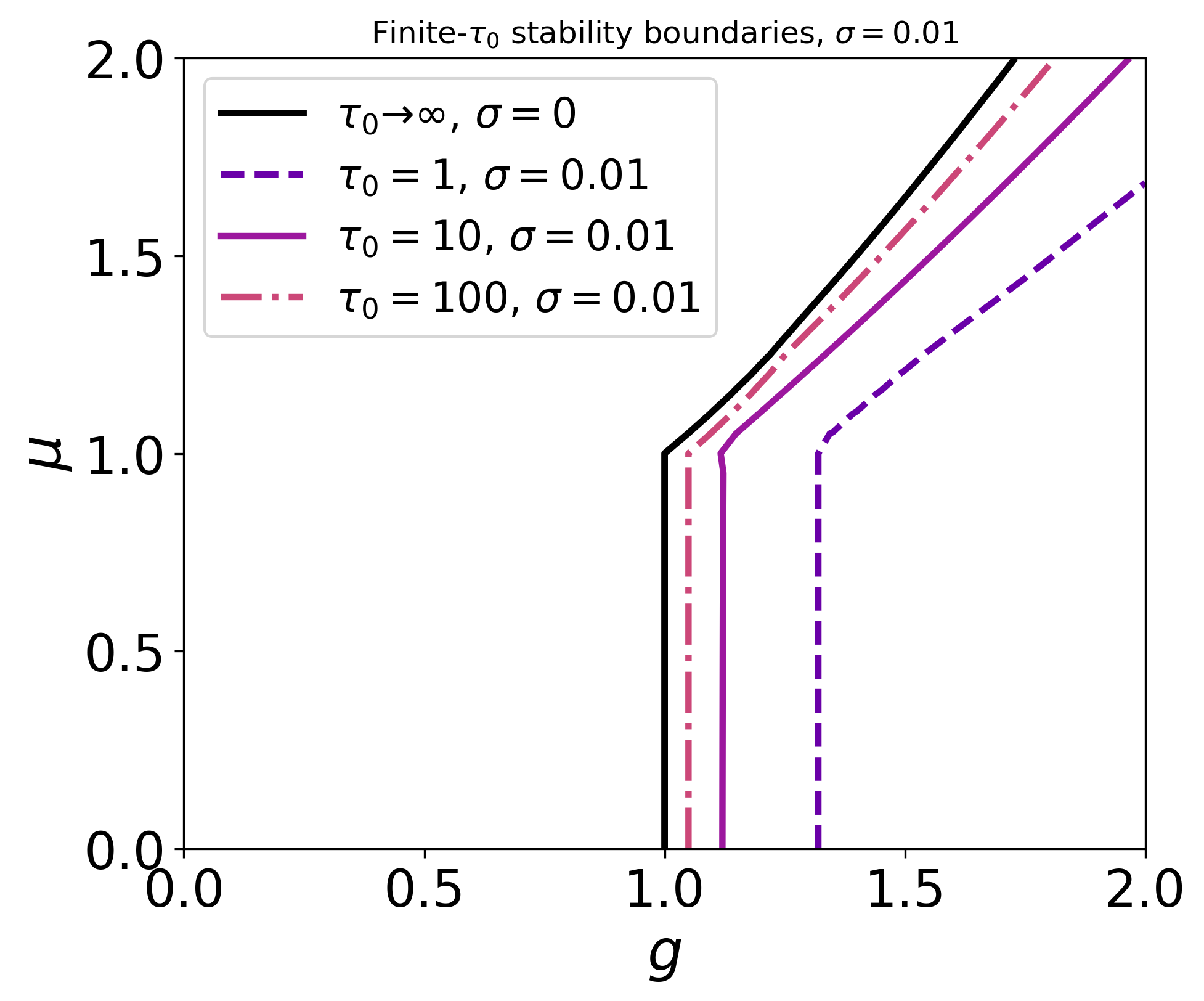}~
    \includegraphics[width=0.5\linewidth]{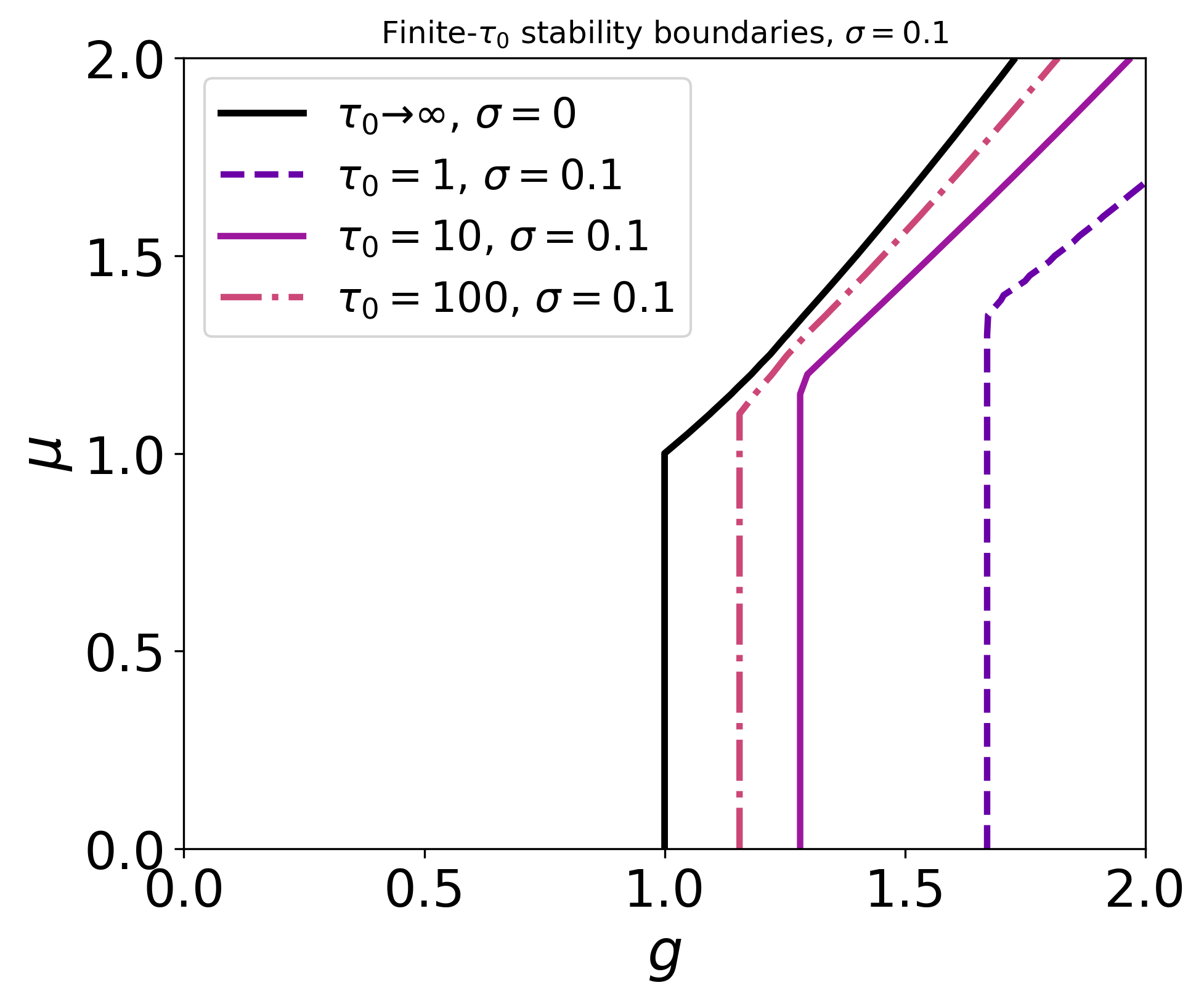}
    \caption{The boundary condition (ii) obtained using Eq.~\eqref{lyapunov_exponent} to divide the quasi-static regime with the chaotic regime. The black solid contour shows the boundary obtained using the linear stability analysis~\eqref{pd-eqn} in the quenched limit $\tau_0\to \infty$ and for $\sigma=0$}
    \label{fig:lambdamax-LSA}
\end{figure}

\begin{figure}
    \centering
    \includegraphics[width=0.33\linewidth]{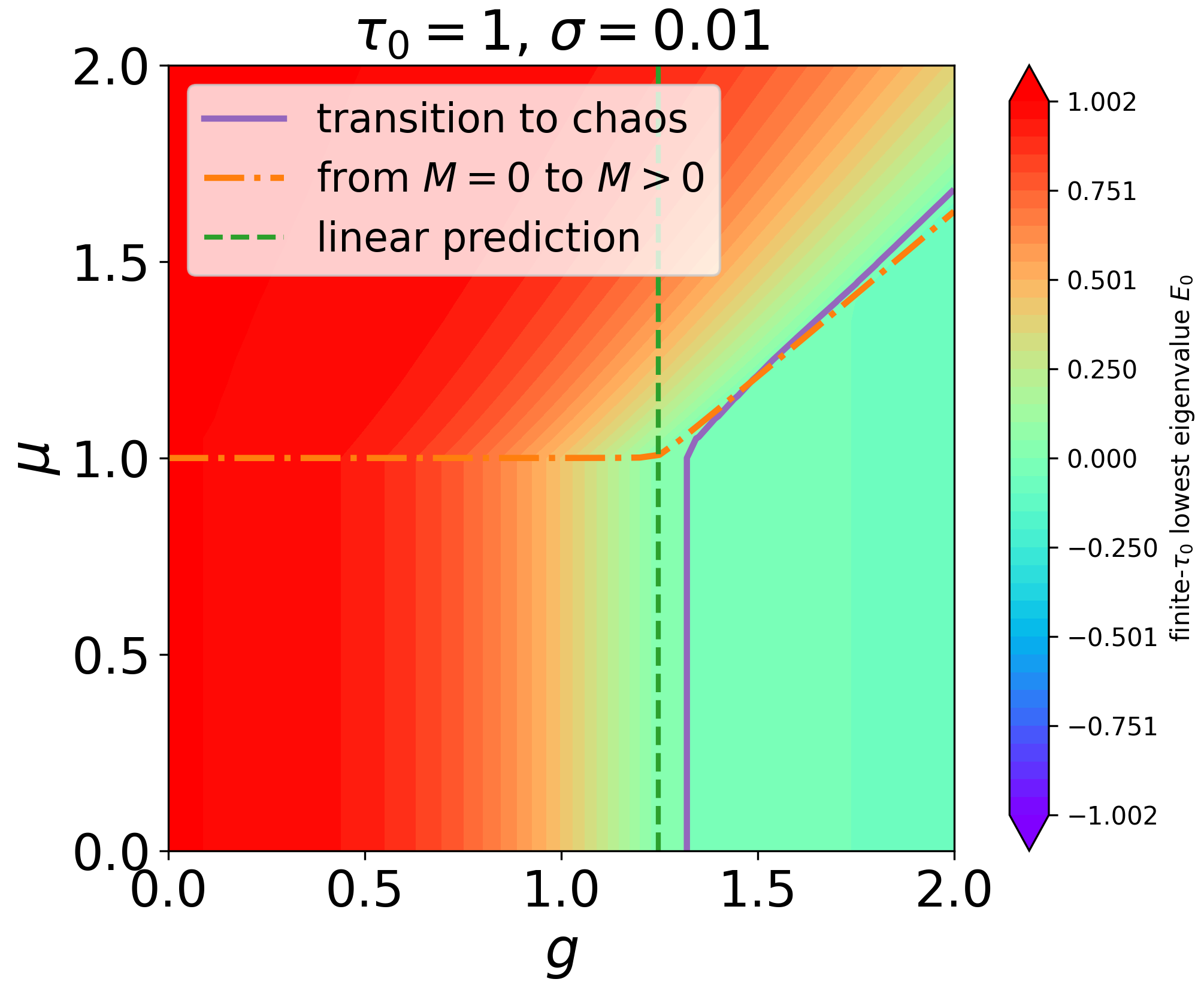}~
    \includegraphics[scale=0.33]{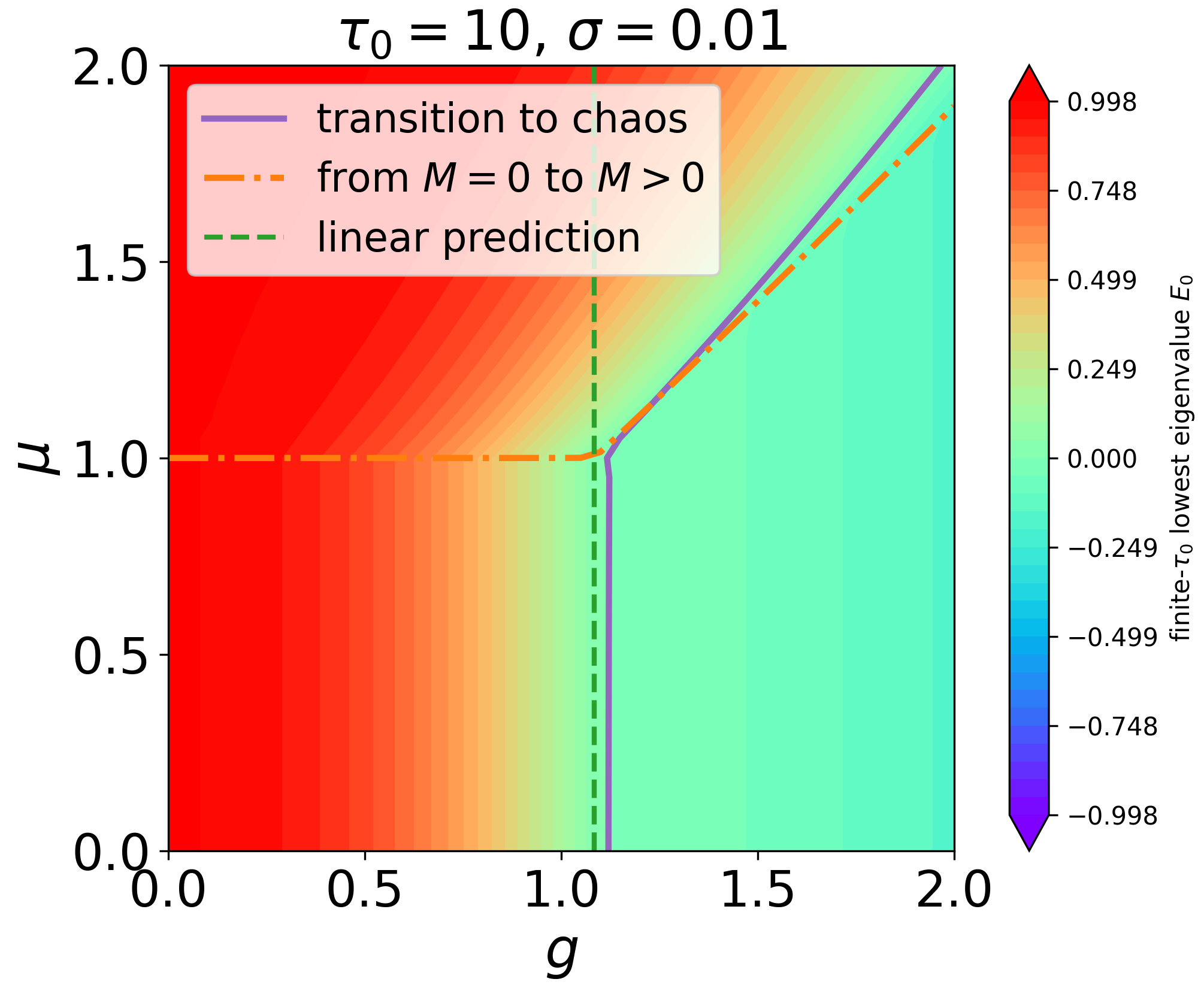}~
    \includegraphics[scale=0.33]{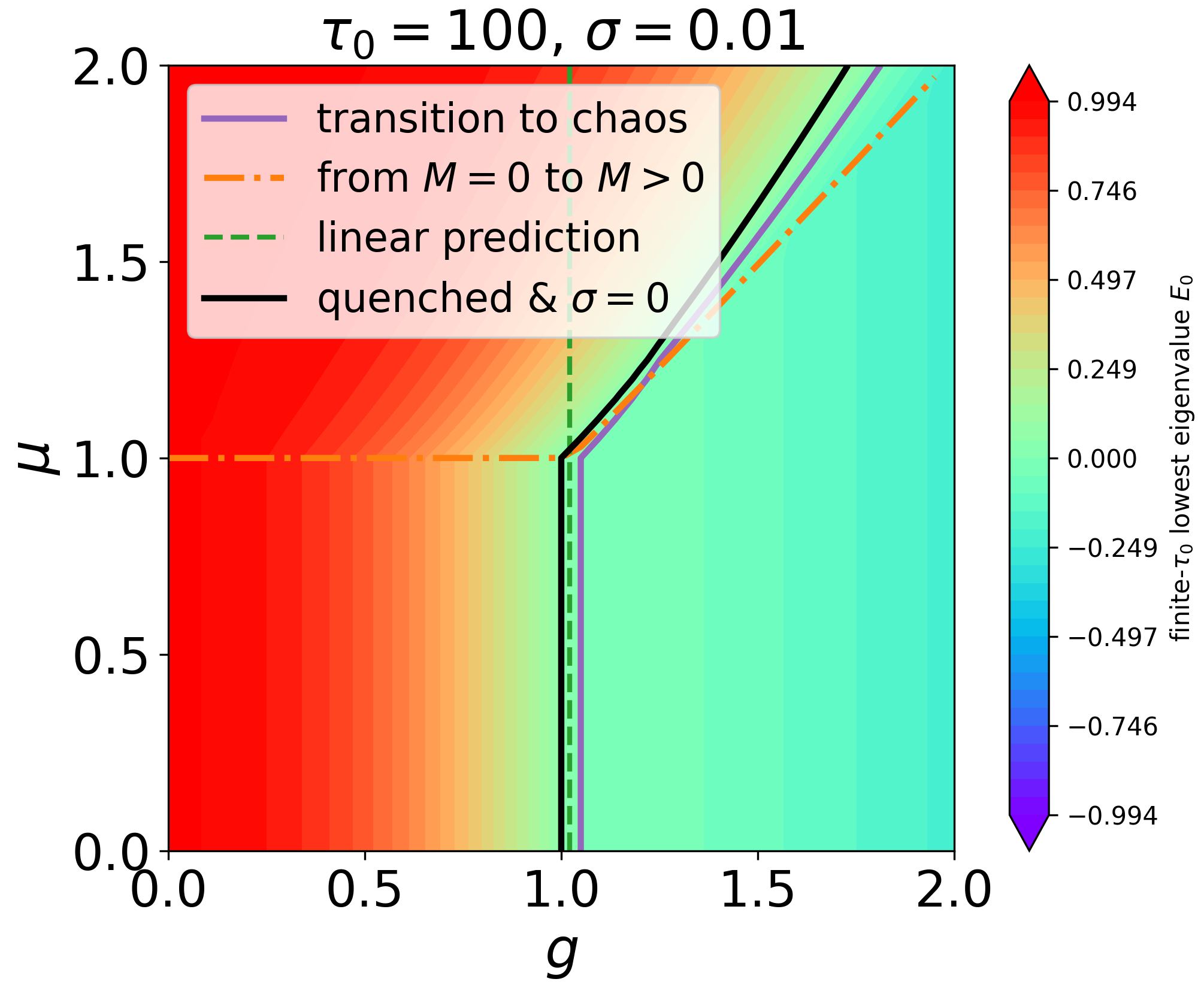}
    \caption{Compare linear stability using $J_{\nu}'\Big(2\tau_0 g\sqrt{q_{\tau_0}(0)}\Big) =0$~\eqref{Bessel} (dashed green) with the transition of the mean (orange) and the  Lyapunov stability boundary (purple) in the $(g,\mu)$ plane for at fixed $\sigma=0.01$ for  $\tau_0=1,10,100$ (from left to right). The color bar indicate the numeric value of the lowest energy $E_0$ as defined and given in Appendix~\ref{lyapunov_analysis-main}. }
    \label{fig:region-plot}
\end{figure}

Now, substituting Eq.~\eqref{ptrb} into Eq.~\eqref{mfeq}, we get
\begin{align}
  \epsilon \dot X 
  &=-x^* - \epsilon X(t) + \tanh\bigg[\underbrace{\mu M_* +  g\eta^*}_{\displaystyle Y} + \underbrace{\epsilon\Big(g\Phi(t) + \xi(t)\Big)}_{\displaystyle\Delta Y} \bigg]\label{eq:LS1}\ , \\
  &\approx \underbrace{-x^* + \tanh(Y)}_{=0} - \epsilon X(t) +\epsilon \Delta Y [1 - \tanh^2(Y)] \,\label{eq:LS2}
  \ ,
\end{align}
Notice that when coming to Eq.~\eqref{eq:LS2} from Eq.~\eqref{eq:LS1}, we drop the higher order contributions, i.e., the terms higher order than $\epsilon$. 
Therefore, we get
\begin{align}
  \dot X  =- X(t) +\big[g\Phi(t) + \xi(t)\big] [1 - \tanh^2(Y)] 
\end{align}
Performing a Fourier transform on this equation, then considering only the $\omega=0$ mode, gives us
\begin{align}
        \hat{X}(0) &= \Big[g\hat{\Phi}(0) + \hat{\xi}(0)\Big]~\Big[1 - \tanh^2(Y)\Big]\,,
\end{align}
from which, we arrive at
\begin{align}
     \langle |X(0)|^2\rangle^{-1} 
     =\bigg\langle \big[1 - \tanh^2(\mu M_* + g \sqrt{Q}z) \big]^{2}\bigg\rangle^{-1} -g^2\ .\label{pd-eqn}
\end{align}
where, with  $Dz =dz~e^{-z^2/2}/\sqrt{2\pi}$,  the mean $M_*$ and the variance $Q$ of the  stationary states $x^*$ are defined by the same form as in the  celebrated replica-symmetric solutions of the Sherrington-Kirkpatrick (SK) model~\cite{Sherrington}:
\begin{align}
    M_* &= \int~Dz~\tanh(\mu M_* + g \sqrt{Q} z)\ ,\label{eq:si-mean}\\
        Q &= \int~Dz~[\tanh(\mu M_* + g \sqrt{Q} z)]^2\ ,\label{eq:si-secmom}
\end{align}
where $M_*$ and $Q_*$ correspond to magnetization and replica overlap (i.e., the spin-glass order parameter), respectively. The pair $(M_*=0, Q=0)$ satisfying the above equations is referred to as the paramagnetic solution, and $(M_*>0,Q>0)$ as the persistent
activity (ferromagnetic) solution. {\color{black}The two self-consistent equations~\eqref{eq:si-mean}-\eqref{eq:si-secmom} can be solved simultaneously for a given $(g, \mu)$. Figure~\ref{fig:quench-PD} shows the mean~\eqref{eq:si-mean} and the second moment~\eqref{eq:si-secmom} in the $(g,\mu)$ plane.

The solution $(M_*,Q)$ in Fig.~\ref{fig:quench-PD} will be substituted into Eq.~\eqref{pd-eqn} to obtain the phase diagram numerically by imposing the condition $\langle |X(0)|^2\rangle\geq 0$. This boundary provides a transition from fixed-point to chaotic behavior (Fig.~\ref{fig:lambdamax-LSA}). Furthermore, we find that the linear-stability analysis in the absence of thermal noise $\sigma=0$ and in the quenched limit $\tau_0\to \infty$ provides a lower bound to the boundary obtained using the maximum Lyapunov exponent analysis in Sec.~\ref{lyapunov_analysis-main} in the long $\tau_0$ case. Finally, we would like to add that one can also perform the linear stability analysis by linearizing the function $F(z)=z$ and in the presence of thermal noise $\sigma\neq 0$. In this case, we obtain the vertical dashed line shown in Fig.~\eqref{fig:region-plot}, which provides the lower-bound to that of the phase boundary obtained using the maximum Lyapunov exponent (see Sec.~\ref{lyapunov_analysis-main} for more details).
}

\section{Method of Numerical simulations}
We solve the DMFT equation:
\begin{align}
    \dot x(t)= -x(t) + F[\mu M(t) + g\eta(t)] + \zeta(t)\ , \label{dmft-eq-l}
\end{align}
where we define $M(t)\equiv \langle x(t)\rangle^{\rm(MF)}$ and the noise correlation is $\langle \eta(t)\eta(t') \rangle = q_{\tau_0}(|t-t'|) \langle x(t)x(t')\rangle^{\rm (MF)} \equiv ~q_{\tau_0}(|t-t'|) C_x(t,t')$.
In the following, we discuss the method for $q_{\tau_0}=1$ (applicable for large $\tau_0$). However, the following calculations can be extended for $q_{\tau_0}\neq 1$.

\subsection{Noise $\eta(t)$ generation}
To solve the DMFT equation~\eqref{dmft-eq-l}, we have to generate Gaussian noise that has the following correlation 
\begin{align}
    \langle \eta(t)\eta(t') \rangle = C_x(t,t')\ ,
\end{align}
where the correlation $C_x(t,t')$ itself is calculated using the trajectories of $x(t)$~\eqref{dmft-eq-l}.

To obtain the noise trajectory $\eta(t)$, we start by guessing $C_x(t,t')$, then we can compute $\eta(t)$ for $t\in [1,2,3,\dots \mathcal{T}/dt]$, i.e. the correlated $\eta(t)$ over all time. 

We start by considering 
\begin{align}
    \eta = L Z\ ,
\end{align}
where $Z$ is a column vector of normal random variables with zero mean and unit variance. Then, we have 
\begin{align}
    \langle \eta \eta^\top \rangle= L \langle Z Z^\top\rangle L^\top = L L^\top\ .
\end{align}
The left hand side is recognized as the correlation matrix $C_x = \langle x x^\top \rangle$. This implies:
\begin{align}
   C_x =  L L^\top\ .
\end{align}

Therefore, we find $L = C_x^{1/2}$, and the correlation matrix $C_x^{1/2}$ can be computed using the eigenvalue decomposition:
\begin{align}
    L = S D^{1/2} S^\top \ ,
\end{align}
where $D$ is the diagonalize matrix of $C_x$ and $S$ is the similarity/orthogonal matrix. All of this gives the vector $\eta$ as
\begin{align}
    \eta =  S D^{1/2} S^\top  Z\ .
\end{align}
Since $S$ is an orthonormal matrix and the distribution of $S^\top  Z$ is also a normal distribution with zero mean and unit variance. Mathematically, this can be seen by looking at 
\begin{align}
 \langle P P^\top\rangle = S^\top  \langle Z Z^\top \rangle  S = S^\top S = I \ ,
\end{align}
for $P \equiv S^\top  Z$ and the identity matrix $I$. This means that the correlation of $P$ or $Z$ gives $I$ as in the case of a normal distribution. Therefore, for a given $C_x$ we can write the noise vector 
\begin{align}
    \eta =  S D^{1/2} Z\ .
\end{align}
Then, using this vector $\eta_{\mathcal{T}/dt\times 1}$, we solve the dynamics~\eqref{dmft-eq-l} for time $\mathcal{T}$ and compute the matrix $C_x(t,t')$. Then, we reiterate the protocol as discussed above until $C_x(t,t')$ becomes independent of the number of iterations.


\end{document}